\documentclass[reqno,12pt]{article}
\usepackage{psfig, amsmath, amstexnb, amsthm}

\theoremstyle{definition}

\makeatletter

\def\enum{\ifnum \@enumdepth >3 \@toodeep\else
        \advance\@enumdepth \@ne 
        \edef\@enumctr{enum\romannumeral\the\@enumdepth}\list
        {\csname label\@enumctr\endcsname}
        {\setlength{\topsep}{1mm}
        \setlength{\parsep}{0mm}
        \setlength{\itemsep}{0mm}
        \setlength{\labelsep}{2mm}
        \settowidth{\leftmargin}{M.}
        \addtolength{\leftmargin}{\labelsep}
        \usecounter{\@enumctr}
        \def\makelabel##1{\hss\llap{##1}}}\fi}

\def\itemiz{\ifnum \@itemdepth >3 \@toodeep\else \advance\@itemdepth \@ne
        \edef\@itemitem{labelitem\romannumeral\the\@itemdepth}%
        \list{\csname\@itemitem\endcsname}{
        \setlength{\topsep}{0mm}
        \setlength{\parsep}{0mm}
        \setlength{\parsep}{0mm}
        \setlength{\itemsep}{0mm}
        \setlength{\labelsep}{2mm}
        \settowidth{\leftmargin}{M.}
        \addtolength{\leftmargin}{\labelsep}
        \def\makelabel##1{\hss\llap{##1}}}\fi}

\def\captionheadfont@{\scshape}
\def\captionfont@{\small}
\long\def\@makecaption#1#2{%
  \setbox\@tempboxa\vbox{\color@setgroup
    \advance\hsize-3pc\noindent
    \captionfont@\captionheadfont@#1\@xp\@ifnotempty\@xp
        {\@cdr#2\@nil}{.\captionfont@\upshape\enspace#2}%
    \unskip\kern-3pc\par
    \global\setbox\@ne\lastbox\color@endgroup}%
  \ifhbox\@ne 
    \setbox\@ne\hbox{\unhbox\@ne\unskip\unskip\unpenalty\unkern}%
  \fi
  \ifdim\wd\@tempboxa=\z@ 
    \setbox\@ne\hbox to\columnwidth{\hss\kern-3pc\box\@ne\hss}%
  \else 
    \setbox\@ne\vbox{\unvbox\@tempboxa\parskip\z@skip
        \noindent\unhbox\@ne\advance\hsize-3pc\par}%
\fi
  \ifnum\@tempcnta<64 
    \addvspace\abovecaptionskip
    \moveright 1.5pc\box\@ne
  \else 
    \moveright 1.5pc\box\@ne
    \nobreak
    \vskip\belowcaptionskip
  \fi
\relax
}

\makeatother

\def\math#1{\ifmmode
\mathchoice{\mbox{$\displaystyle\rm#1$}}
{\mbox{$\textstyle\rm#1$}}
{\mbox{$\scriptstyle\rm#1$}}
{\mbox{$\scriptscriptstyle\rm#1$}}\else
{\mbox{$\rm#1$}}\fi}		

\def\smath#1{\ifmmode
\mathchoice{\mbox{$\textstyle#1$}}
{\mbox{$\scriptstyle#1$}}
{\mbox{$\scriptscriptstyle#1$}}
{\mbox{$\scriptscriptstyle#1$}}\else
{\mbox{$\textstyle#1$}}\fi}	

\def\vec#1{\ifmmode
\mathchoice{\mbox{$\displaystyle\bf#1$}}
{\mbox{$\textstyle\bf#1$}}
{\mbox{$\scriptstyle\bf#1$}}
{\mbox{$\scriptscriptstyle\bf#1$}}\else
{\mbox{$\bf#1$}}\fi}		

\DeclareMathSymbol{\leqsymb}{\mathalpha}{AMSa}{"36}
\def\leqs{\;\leqsymb\;}
\DeclareMathSymbol{\geqsymb}{\mathalpha}{AMSa}{"3E}
\def\geqs{\;\geqsymb\;}
\DeclareMathSymbol{\gtreqqlesssymb}{\mathalpha}{AMSa}{"54}

\newcommand{\field}[1]{\mathbb{#1}}

\newcommand{\Z}{\field{Z}\,}	
\newcommand{\R}{\field{R}\,}	

\newcommand{\cA}{{\mathcal A}}	

\DeclareMathOperator{\e}{e}		
\DeclareMathOperator{\icx}{i}		
\DeclareMathOperator{\re}{Re}		
\DeclareMathOperator{\im}{Im}		
\DeclareMathOperator{\thyp}{th}		
\DeclareMathOperator{\Arccos}{Arccos}	
\DeclareMathOperator{\Argth}{Argth}	
\DeclareMathOperator{\dd}{d}		
\DeclareMathOperator{\defby}{\raisebox{0.35pt}{\math{:}}\!\!=}

\def\sothat{\Rightarrow}	

\def\dx#1{\dd\!#1}		
\def\sdpar#1#2{\partial_{#2}#1}		
\def\dtot#1#2{\frac{\dx{#1}}{\dx{#2}}}	

\def\brak#1{[#1]}			
\def\abs#1{\lvert#1\rvert}		
\def\norm#1{\lVert#1\rVert}		
\def\pscal#1#2{\langle#1|#2\rangle}	
\def\ccint#1#2{[#1,#2]}			

\def\bigbrak#1{\bigl[#1\bigr]}		
\def\bigpar#1{\bigl(#1\bigr)}		
	
\def\Bigbrak#1{\Bigl[#1\Bigr]}		
\def\Bigpar#1{\Bigl(#1\Bigr)}		
	
\def\Order#1{{\mathcal O}(#1)}	
\def\fix#1{#1^{\star}}		

\def\nth#1{\ensuremath{#1^{\math{th}}}}	
\def\defwd#1{{\bf#1}}		

\def\ie{i.e.,\ }
\def\eg{e.g.\ }

\def\ift{implicit function theorem}

\def\nbh{neighborhood}

\def\eps{\varepsilon}

\def\ph{\varphi}
\def\th{\theta}
\def\x{\xi}
\def\y{\eta}
\def\z{\zeta}

\def\one{{\mathchoice {\rm 1\mskip-4mu l} {\rm 1\mskip-4mu l}
{\rm 1\mskip-4.5mu l} {\rm 1\mskip-5mu l}}}	

\def\sub#1{_{\smath{#1}}}	
\def\ssup#1{^{\smath{#1}}}		

\def\delay{\Psi}
\def\W{\Omega}
\def\ctau{\check{\tau}}		
\def\htau{\hat{\tau}}		
\def\supz#1{#1\ssup{0}}		
\def\tom{\tau\ssup{0}_-}	
\def\top{\tau\ssup{0}_+}	
\def\tsm{\fix{\tau}_-}		
\def\tsp{\fix{\tau}_+}		
\def\xic{\xi_{\math c}}		%
\def\taub{\tau_{\math b}}	

\DeclareMathSymbol{\gordsymb}{\mathalpha}{AMSa}{"3C}
\DeclareMathSymbol{\lordsymb}{\mathalpha}{AMSa}{"34}

\def\sord{\approx}		

\def\crit#1{#1_{\math{c}}}	
\def\tprob#1#2{w(#1\vert#2)} 	

\def\figref#1{Fig.\ \ref{#1}}	

\def\writefig#1 #2 #3 {\rlap{\kern #1 truecm
\raise #2 truecm \hbox{\protect{\small #3}}}}
\def\figtext#1{\smash{\hbox{#1}}
\vspace{-5mm}}

\def\bibtitle#1#2{#1, {\em #2}}                         
\def\bibref#1#2#3#4#5{#1 {\bf #2}:#3--#4 (#5)}        
\def\bibarticle#1#2#3#4#5#6#7{\bibtitle{#1}{#2},
\bibref{#3}{#4}{#5}{#6}{#7}.}
\def\bibpreprint#1#2#3#4{#1, {\em #2}, preprint {\tt #3} (#4).}
\def\bibbook#1#2#3#4{#1, {\em #2} (#3, #4).}

\def\DE{Diff.\ Equ.}

\def\DU{Diff.\ Urav.\ }
\def\JPA{J.\ Phys.\ A}

\def\JSP{J.\ Stat.\ Phys.}

\def\PRA{Phys.\ Rev.\ A}
\def\PRB{Phys.\ Rev.\ B}
\def\PRE{Phys.\ Rev.\ E}
\def\PRL{Phys.\ Rev.\ Letters}

\def\SIAM{SIAM J.\ Appl.\ Math.}

\begin{document}


\title{Memory Effects and Scaling Laws\\
in Slowly Driven Systems}

\author{
N. Berglund and H. Kunz \\
{\it Institut de Physique Th\'eorique} \\
{\it Ecole Polytechnique F\'ed\'erale de Lausanne} \\
{\it PHB-Ecublens, CH-1015 Lausanne, Switzerland} \\
{\rm e-mail: }{\tt berglund@iptsg.epfl.ch, kunz@dpmail.epfl.ch} \\
}

\date{July 17, 1998}

\maketitle

\begin{abstract}
This article deals with dynamical systems depending on a slowly varying
parameter. We present several physical examples illustrating memory
effects, such as metastability and hysteresis, which frequently appear in
these systems. A mathematical theory is outlined, which allows to show
existence of hysteresis cycles, and determine related scaling laws.
\end{abstract}

\vspace{5mm}

\vspace{5mm}

\newpage


\section{Introduction}
\label{sec_in}

There exist many instances where the dynamics of a system depends on a
parameter which varies slowly in time. This parameter is often controllable
by the experimentalist, who can modify it at will. A well--known example of
this situation is that of a ferromagnet on which is imposed a low frequency
magnetic field. One can also think of chemical reactions occurring in a
reactor in which the flux of the injected chemical substances is varying
slowly, or the Couette--Taylor experiment in hydrodynamics where the speed
of rotation of the inner cylinder is slowly modulated. In other
circumstances, the parameter is not controllable, but certainly influences
the dynamics of the system of interest. As examples of this situation we
could mention the case of the impact of solar light on the thermal
convection in the atmosphere, or the seasonal (or even climatic) effect
on the dynamics of populations. 

One of the most interesting phenomena observed in systems with an
adiabatically varying parameter is the familiar one of hysteresis.
Recently, there has been a renewal of interest in this old problem, both
from a theoretical and an experimental point of view. Several authors
\cite{RKP,SE,Jung,Hohl} have particularly analysed properties of the
hysteresis cycle, such as its area, which appears to scale in a nontrivial
way with the adiabatic parameter.

In this article, we concentrate on dynamical systems with a finite number
of degrees of freedom, depending on a parameter in such a way that the
system undergoes bifurcations when the parameter is considered to be
static. The static (or ``frozen'') situation corresponds to measurements
made, in principle during very long times, at successive fixed values of
the parameter. We then ask what is happening when the parameter is varying
slowly in time, instead of being kept fixed. This question is closely
related to the opposite one: can the static bifurcation diagram be
determined experimentally by varying the parameter slowly in time (a
possible temptation for the impatient experimentalist)?

We have recently developed a coherent mathematical framework to deal with
adiabatic systems, in particular to show existence of hysteresis cycles
and determine their scaling laws \cite{B}. The purpose of this article is
to explain these methods by illustrating them on a few concrete physical
examples. 

We begin, in Section \ref{sec_1D}, by presenting the most important
features of one--dimensional (1D) systems, which are illustrated by a few
generic examples. We discuss in particular a simple geometric method to
determine scaling laws near bifurcation points. 

In Section \ref{sec_Lz}, we use the Lorenz model to illustrate the
phenomenon of bifurcation delay.  When translated into the language of
Rayleigh--B\'enard (RB) convection, this phenomenon means that the slow and
periodic variation of the temperature gradient in time leads to a delayed
appearance of convection rolls and to hysteresis. The Lorenz model being a
good approximation close to the instability threshold, since it contains
two dominant modes of the bifurcation, this delay should be observable in
the real RB convection. To explain the delay, we introduce two new methods
especially designed for $n$--dimensional ($n$D) systems: dynamic
diagonalization and adiabatic manifolds.  

In Section \ref{sec_mh}, we present a simple mean field model for the
dynamics of a ferromagnet in a slowly oscillating magnetic field. In the 1D
case, we discuss the concept of dynamic phase transition introduced in
\cite{TO}, and derive a scaling law for the hysteresis area. In the 2D
case, we examine the effect of anisotropy on the mechanism of magnetization
reversal and the shape of hysteresis cycles. 

In Section \ref{sec_rp}, we discuss a simple mechanical system (which was
introduced in \cite{BK}), displaying chaotic instead of periodic
hysteresis. This phenomenon depends only on a few qualitative features of
the system, and should be observable in a larger class of nonlinear
oscillators including inertia and involving a symmetry breaking
bifurcation. 

Finally, Section \ref{sec_ec} is dedicated to some examples of
the effect of eigenvalue crossings. These crossing give rise to an effective
interaction between otherwise independent modes, which is essential in the
sense that it cannot be eliminated by a change of variables. The interaction
may, however, be delayed in certain cases. 

Throughout this text, we use the following mathematical setting. The
``frozen'' dynamical system is supposed to be described by a family of
ordinary differential equations
\begin{equation}
\label{in1}
\dtot{x}{t} = F(x,\lambda), \qquad
x \in \R^n, \quad \lambda \in \R.
\end{equation}
The associated adiabatic system is given by
\begin{equation}
\label{in2}
\dtot{x}{t} = F(x,\lambda(\eps t)), 
\end{equation}
where $\lambda(\tau)$ is a given function, and $\eps$ is the small
\defwd{adiabatic parameter}. It is useful to introduce the \defwd{slow
time} $\tau=\eps t$, so that \eqref{in2} can be rewritten as 
\begin{equation}
\label{in3}
\eps\dtot{x}{\tau} = F(x,\lambda(\tau)), 
\end{equation}
or, in short form, $\eps\dot{x} = f(x,\tau)$. We denote by
$\pscal{\cdot}{\cdot}$ the usual scalar product in $\R^n$ and by
$\norm{\cdot}$ the Euclidean norm. 

There is a large literature on singular perturbed problems of this type.
Results on linear systems can be found in \cite{Wasow}. For a review of
results on dynamic bifurcations, see \cite{Benoit}. In particular, the
phenomenon of bifurcation delay has been rigorously analysed in two
important articles by Neishtadt  \cite{Ne1,Ne2}. Certain hysteresis
phenomena in slow--fast systems similar to the Van der Pol equation have
been analysed in \cite{MKKR}. Our geometric method to determine scaling
exponents, as well as the procedures using dynamic diagonalization and
adiabatic manifolds, however, are new to our knowledge. Here we only
outline some essential ideas of these methods, detailed proofs can be found
in \cite{B}.


\section{One--dimensional systems}
\label{sec_1D}

In this section, we will consider one--dimensional (1D) adiabatic equations
of the form 
\begin{equation}
\label{1D1}
\eps\dot{x} = F(x,\lambda(\tau)) = f(x,\tau), 
\qquad x,\tau\in\R,
\end{equation}
where the dot denotes derivation with respect to $\tau$, and $f(x,\tau)$ is
assumed to be an analytic real--valued function (weaker results as those
stated below hold for differentiable functions). 

The static bifurcation diagram of \eqref{1D1} is obtained by determining the
solutions of $f(x,\tau)=0$, which are generically curves subdividing the
plane into regions where $f$ is positive or negative. Let $\fix{x}(\tau)$ be
such an equilibrium curve. The \ift\ tells us that if the linearization
$a(\tau)=\sdpar{f}{x}(\fix{x}(\tau),\tau)$ does not vanish, then
$\fix{x}(\tau)$ is a smooth curve. It corresponds to stable solutions if
$a(\tau)$ is negative, and to unstable ones of $a(\tau)$ is positive.

In such a situation, one can prove the existence of a particular solution
$\bar{x}(\tau)$ of the adiabatic equation \eqref{1D1} tracking the curve
$\fix{x}(\tau)$ at a distance of order $\eps$: 
\begin{equation}
\label{1D2}
\bar{x}(\tau) = \fix{x}(\tau) + \Order{\eps}.
\end{equation}
Moreover, this solution admits an asymptotic power series in $\eps$, which
does not converge in general, but admits, however, an optimal truncation at
exponentially small order: 
\begin{equation}
\label{1D3}
\bar{x}(\tau) = \fix{x}(\tau) + \sum_{j=1}^{N(\eps)} x_j(\tau)\eps^j +
\Order{\e^{-1/C\eps}}, \qquad N(\eps) = \Order{1/\eps}.
\end{equation}
We call $\bar{x}(\tau)$ an \defwd{adiabatic solution} associated with the
equilibrium branch $\fix{x}(\tau)$. Other solutions of \eqref{1D1} are
attracted or repelled exponentially fast by adiabatic ones, and tend to
switch between the \nbh s of different equilibrium branches in a time of
order $\eps\abs{\ln\eps}$. As long as there are no bifurcation points, the
solutions thus remain most of the time close to equilibria, and there is no
hysteresis in the system.

Let us now consider the effect of bifurcations. At these special points,
several equilibrium branches may meet, causing the solutions to choose
between several possible directions, which is the basic mechanism of
hysteresis. Moreover, equilibrium branches are in general no longer tracked
at a distance of order $\eps$, but at a distance scaling in some other,
nontrivial way with $\eps$, which we now show how to compute. 

If the origin is a bifurcation point of $f$, we can write in some \nbh\
\begin{equation}
\label{1D4}
f(x,\tau) = \sum_{n,m\geqs 0} c_{nm}x^n\tau^m, 
\qquad c_{00} = c_{10} = 0.
\end{equation} 
Assume that $f(x,\tau)$ admits an equilibrium branch scaling as
$\fix{x}(\tau) \sord \abs{\tau}^q$ (we use this notation to indicate that
$c_-\abs{\tau}^q \leqs \fix{x}(\tau) \leqs c_+\abs{\tau}^q$, where
$c_{\pm}$ are positive constants independent of $\tau$ and $\eps$). A
standard result of bifurcation theory states that $-q$ is necessarily equal
to the slope of a segment of \defwd{Newton's polygon}. This polygon is
constructed as the convex envelope of the set of points $(n,m)$ such that
$c_{nm}\neq 0$, completed by a horizontal and a vertical
(\figref{fig_1DNewton}). The linearization
$a(\tau)=\sdpar{f}{x}(\fix{x}(\tau),\tau)$ scales generically as
$\abs{\tau}^p$, where $p$ is the ordinate at $1$ of the  tangent to
Newton's polygon with slope $-q$. 

\begin{figure}
 \centerline{
 \psfig{figure=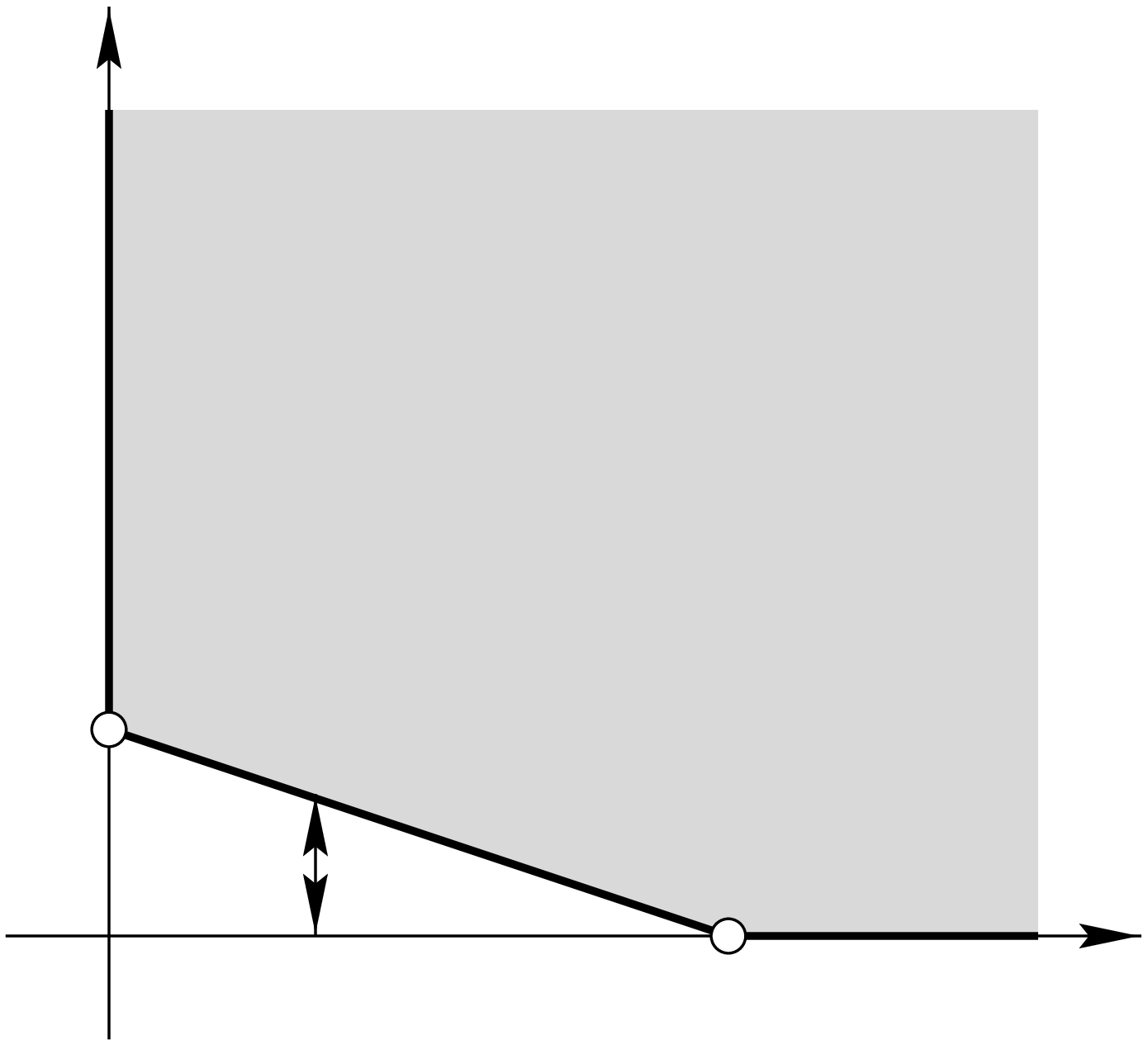,height=40mm,clip=t}
 \hspace{5mm}
 \psfig{figure=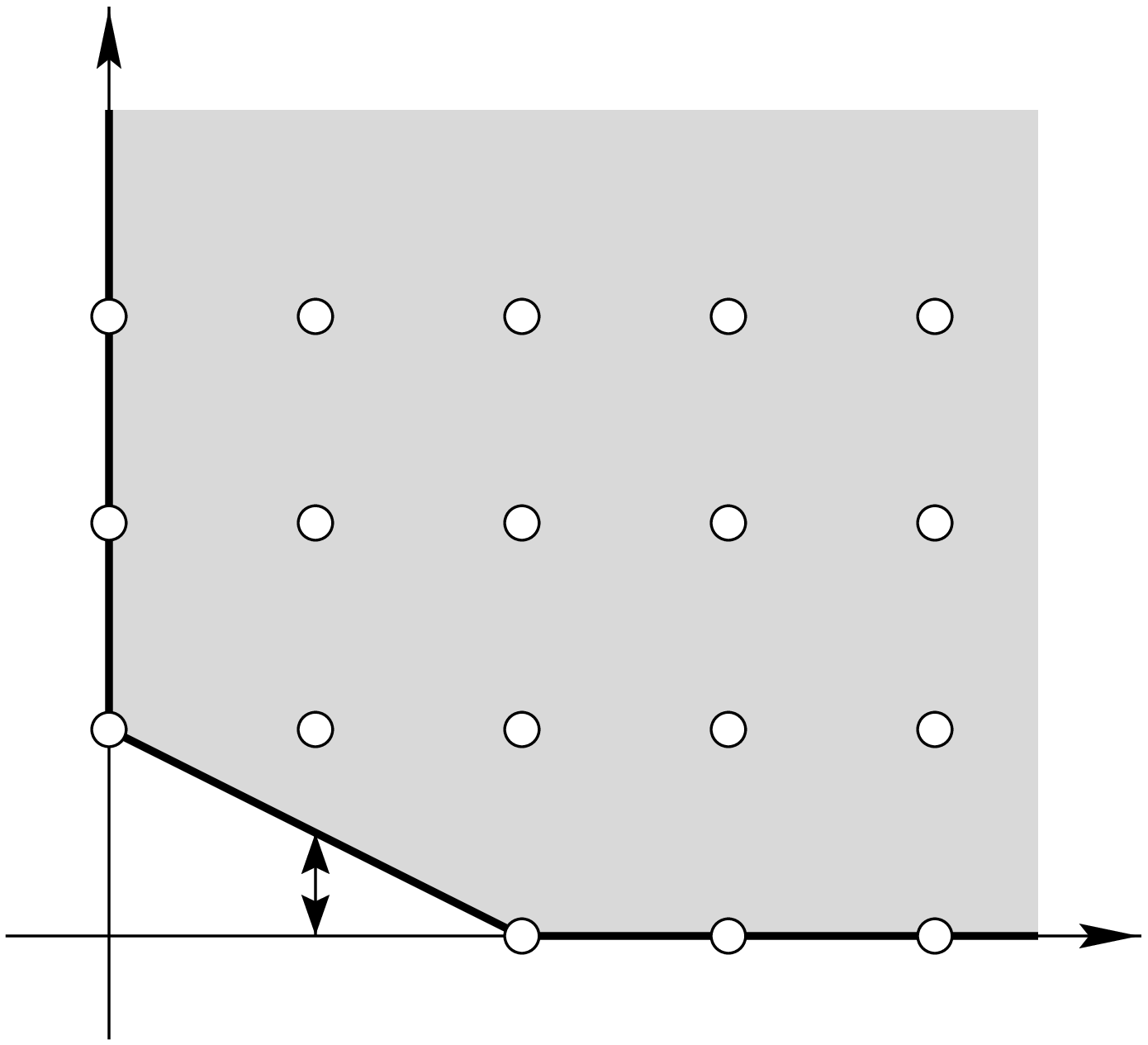,height=40mm,clip=t}
 \hspace{5mm}
 \psfig{figure=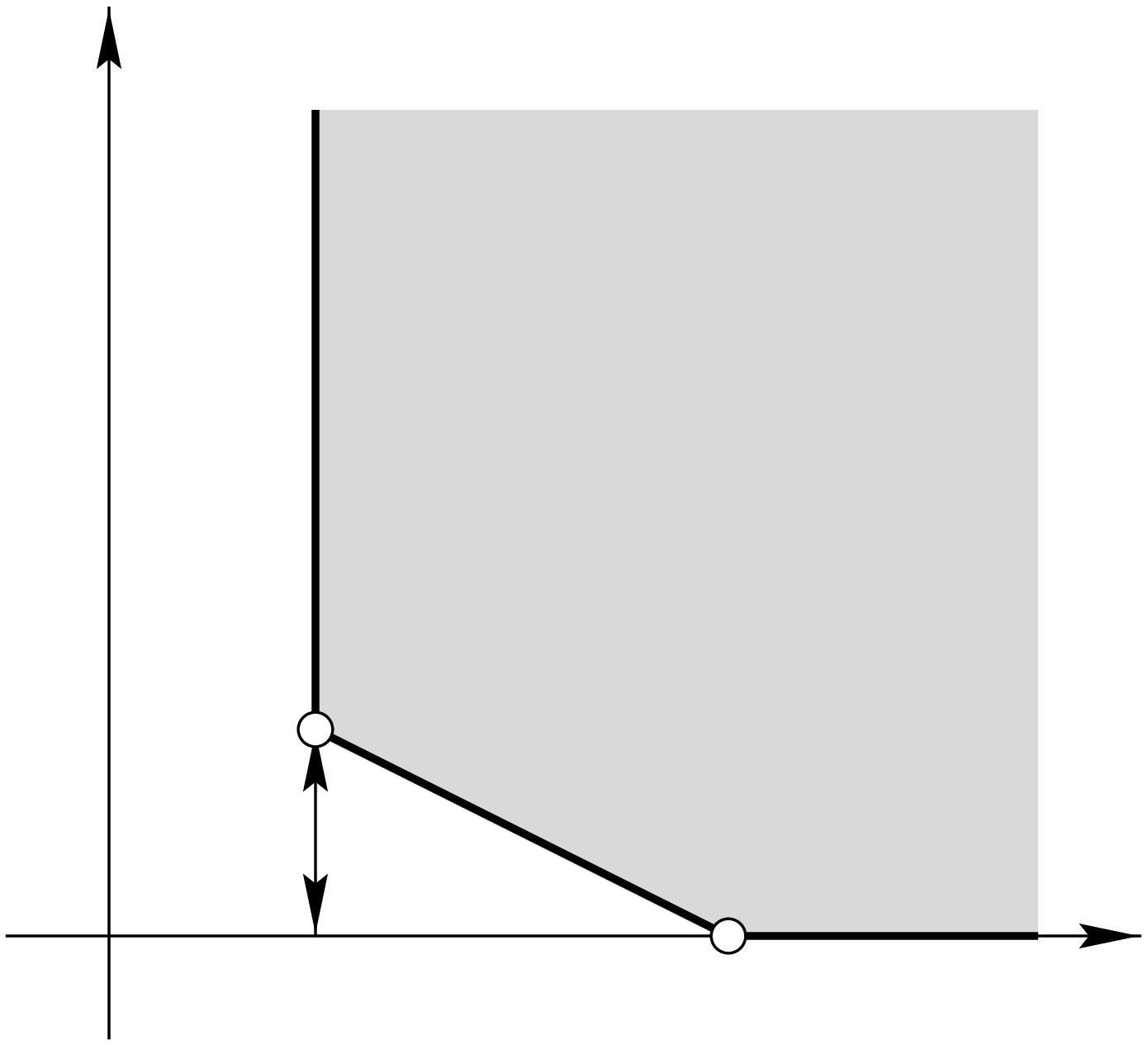,height=40mm,clip=t}
 }
 \figtext{
 	\writefig	0.0	4.2	a
 	\writefig	0.05	1.6	$1$
	\writefig	0.65	4.2	$m$
	\writefig	1.15	0.5	$1$
	\writefig	2.7	0.5	$3$
	\writefig	3.95	0.5	$n$
	\writefig	0.85	1.1	$\frac23$
 	\writefig	5.15	4.2	b
 	\writefig	5.2	1.6	$1$
	\writefig	5.8	4.2	$m$
	\writefig	6.3	0.5	$1$
	\writefig	7.1	0.5	$2$
	\writefig	9.1	0.5	$n$
	\writefig	6.0	1.1	$\frac12$
 	\writefig	10.3	4.2	c
 	\writefig	10.35	1.6	$1$
	\writefig	10.95	4.2	$m$
	\writefig	11.45	0.5	$1$
	\writefig	13.0	0.5	$3$
	\writefig	14.25	0.5	$n$
	\writefig	11.15	1.25	$1$
 }
 \caption[]
 {Newton's polygons for the most generic bifurcations discussed in the
 text. Dots mark points for which $c_{nm}\neq 0$. The slope of segments
 correspond to the possible exponents $q$ of equilibrium branches through the
 bifurcation point. The ordinate at $1$ of these segments is the exponent
 $p$ of the linearization.}
\label{fig_1DNewton}
\end{figure}

These two easily determined numbers $p$ and $q$ are usually sufficient to
characterize the scaling behaviour at leading order in $\eps$. In fact,
different behaviours take place in an \defwd{inner region}
$\abs{\tau}\leqs\eps^{1/p+1}$ and in an \defwd{outer region}
$\abs{\tau}\geqs\eps^{1/p+1}$. In particular, if
$\fix{x}(\tau)\sord\abs{\tau}^q$ is a decreasing stable branch arriving at
the bifurcation point, above which $f$ is negative, one can show that 
\begin{equation}
\label{1D5}
\bar{x}(\tau) - \fix{x}(\tau) \sord
\begin{cases}
\eps\abs{\tau}^{q-p-1} 	& \text{for $\tau\leqs-\eps^{1/p+1}$,} \\
\eps^{q/p+1} 		& \text{for $-\eps^{1/p+1}\leqs\tau\leqs 0$.}
\end{cases}
\end{equation}

Combining a local analysis around bifurcation points with a global
analysis, which is usually easy in 1D, one can determine the qualitative
properties of dynamics. In particular, if $\lambda(\tau)$ is a periodic
function, one can construct the Poincar\'e map (which is necessarily a
monotonous function) in order to prove existence of hysteresis cycles and
determine their scaling laws. 

Let us illustrate this procedure on the simple model equation given by 
\begin{equation}
\label{1D6}
F(x,\lambda,\mu) = -\mu x - x^3 + \lambda.
\end{equation}
Mathematically, this function is a generic two--parameter perturbation of
the vector field $-x^3$ \cite{HK}. Physically, it describes the overdamped
motion of a particle in a Ginzburg--Landau potential
\begin{equation}
\label{1D7}
\Phi(x,\lambda,\mu) = \frac12 \mu x^2 + \frac14 x^4 - \lambda x,
\end{equation}
where $\mu = T - \crit{T}$ represents the difference between the
temperature and its critical value, and $\lambda$ is an external field. 
The quartic potential described by the first two terms is fairly generic in
physical systems presenting the symmetry $x\mapsto -x$, while the linear
term is the simplest possible asymmetric perturbation. 

We begin with the situation where $\mu=1$ is fixed, and
$\lambda(\tau)=\sin\tau$ is slowly oscillating. The equation
$F(x,\lambda,1)=0$ admits a single, stable equilibrium branch
$\fix{x}(\lambda)$, given implicitly by $\fix{x}(\lambda)^3 +
\fix{x}(\lambda) = \lambda$. All solutions are attracted by a periodic
solution $\bar{x}(\tau) = \fix{x}(\lambda(\tau)) + \Order{\eps}$, enclosing
an area of order $\eps$ (\figref{fig_1Dhyst1}a). In the adiabatic limit
$\eps\to 0$, this area vanishes and there is no hysteresis. 

\begin{figure}
 \centerline{\psfig{figure=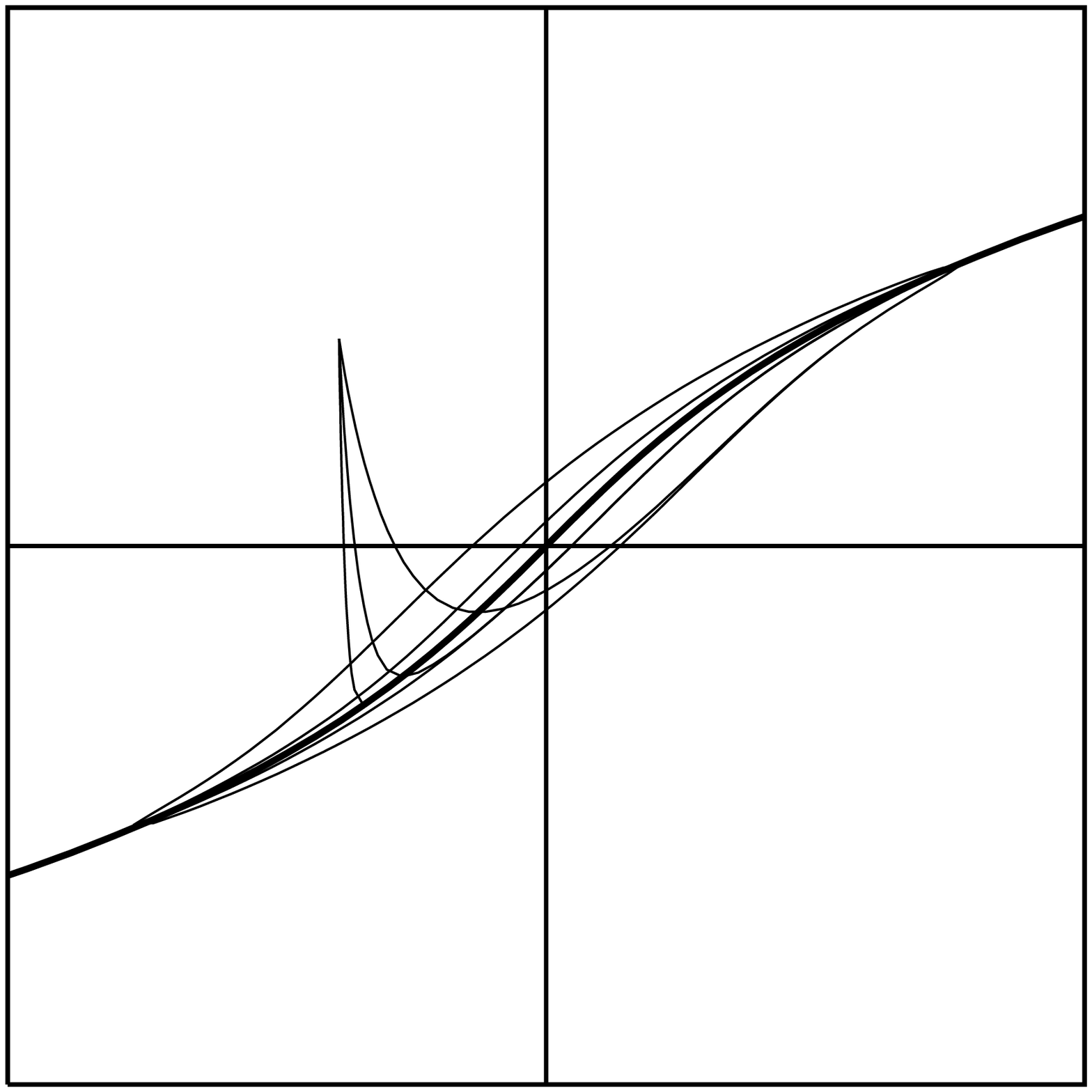,height=50mm,clip=t}
 \hspace{18mm}
 \psfig{figure=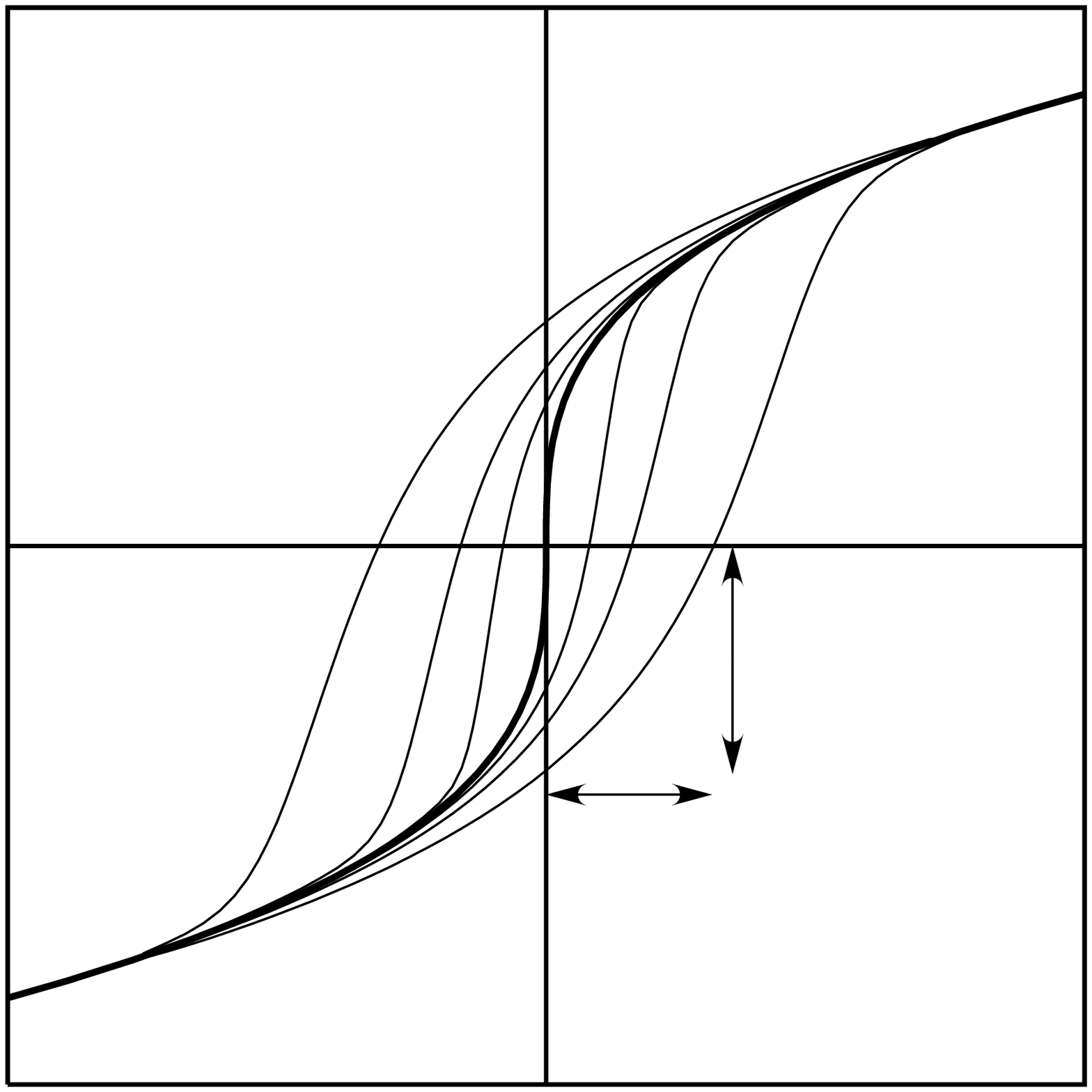,height=50mm,clip=t}}
 \figtext{
 	\writefig	0.9	5.2	a
 	\writefig	4.0	5.2	$x$
 	\writefig	6.0	3.1	$\lambda$
	\writefig	7.9	5.2	b
 	\writefig	11.0	5.2	$x$
 	\writefig	13.0	3.1	$\lambda$
 	\writefig	11.9	2.4	$\eps^{1/5}$
 	\writefig	11.0	1.35	$\eps^{3/5}$
 }
 \caption[]
 {(a) Orbits, in the $(\lambda,x)$--plane, of the system
 $\eps\dot{x} = -\mu x - x^3 + \lambda(\tau)$, for $\mu=1$ and $\eps =
 10^{-3/2}, 10^{-2}$ and $10^{-5/2}$. The solutions are attracted by a
 periodic orbit, enclosing an area of order $\eps$. (b) Same as (a), but for
 $\mu=0$. At $\lambda=0$, the periodic orbit is at a distance of order
 $\eps^{1/5}$ of the equilibrium point, and it encloses an area of order
 $\eps^{4/5}$.}
\label{fig_1Dhyst1}
\end{figure}

If $\mu=0$ and $\lambda(\tau)=\sin\tau$, the unique equilibrium branch
$\fix{x}(\lambda)=\lambda^{1/3}$ admits the origin as a bifurcation point.
Using Newton's polygon or a direct calculation, we find that the exponents
determining the scaling behaviour are $q=\frac13$ and $p=\frac23$
(\figref{fig_1DNewton}a). The orbits are attracted by a periodic one,
crossing the $x$--axis at a distance of order $\eps^{1/5}$ from the origin,
and enclosing an area 
\begin{equation}
\label{1D7b}
\cA(\eps)\sord\eps^{4/5}
\end{equation} 
(\figref{fig_1Dhyst1}b). The cycle still collapses with
$\fix{x}(\lambda(\tau))$ in the adiabatic limit, but with a much slower
rate. These exponents have been found in \cite{Hohl,GBS} using other  
methods (which are less general than ours). 

If $\mu=-1$ and $\lambda(\tau)=\sin\tau$, there are two bifurcation points
at $(\pm\crit{\lambda},\mp\crit{x})$, where $\crit{\lambda}=\sqrt{4/27}$
and $\crit{x}=\sqrt{1/3}$. Two stable branches $\fix{x}_{\pm}(\tau)$ and
one unstable branch $\fix{x}_0(\tau)$ meet at these bifurcation points
(\figref{fig_1Dhyst2}a). Since they are crossed with nonzero velocity,
Newton's polygon shows that the associated exponents are $q=p=\frac12$
(\figref{fig_1DNewton}b). In fact, close to these points, the dynamics in
translated coordinates is governed by the equation 
\begin{equation}
\label{1D8}
\eps\dot{y} = -\tau - y^2 + \text{higher order terms.}
\end{equation} 
Solutions cross the $y$--axis at $y\sord\eps^{1/3}$. Using the scaling
$y=\eps^{1/3}z$ and $\tau=\eps^{2/3}\sigma$, one shows that $y$ remains of
order $\eps^{1/3}$ until a time $\fix{\tau}\sord\eps^{2/3}$, and then
quickly leaves the bifurcation region to jump on the other stable branch.
The orbits are attracted by a hysteresis cycle with area satisfying
\begin{equation}
\label{1D9}
\cA(\eps) - \cA(0) \sord \eps^{2/3}, 
\end{equation}
where $\cA(0)=\frac32$ is the area situated between $\fix{x}_+(\lambda)$
and $\fix{x}_-(\lambda)$ for $-\crit{\lambda}<\lambda<\crit{\lambda}$. This
time, the hysteretic behaviour persists in the adiabatic limit. The main
contribution of order $\eps^{2/3}$ to the excess area comes from the
delayed jump. The scaling law \eqref{1D9} was also obtained in \cite{Jung}
using an exact solution of \eqref{1D8} (without the higher order terms). 

\begin{figure}
 \centerline{\psfig{figure=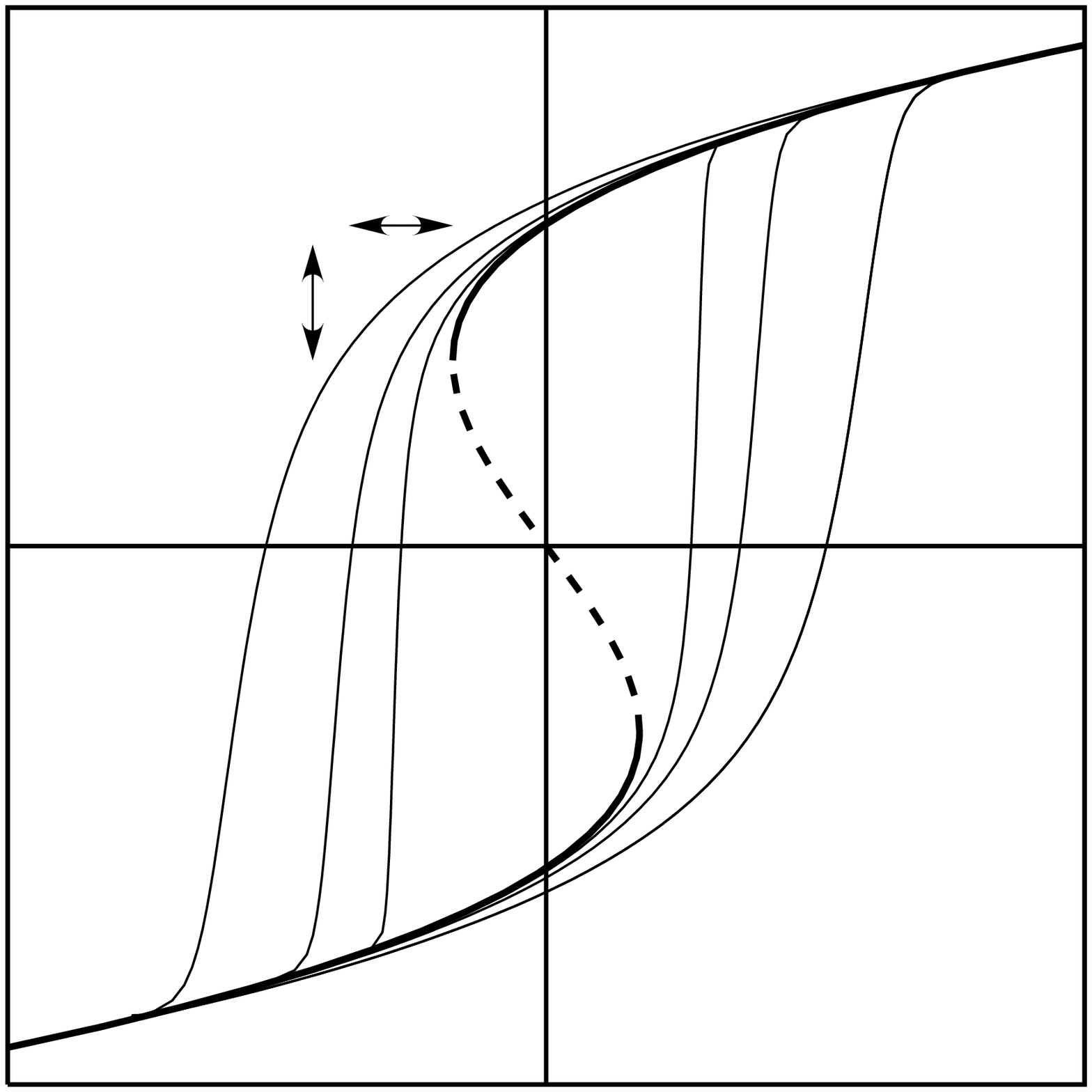,height=50mm,clip=t}
 \hspace{18mm}
 \psfig{figure=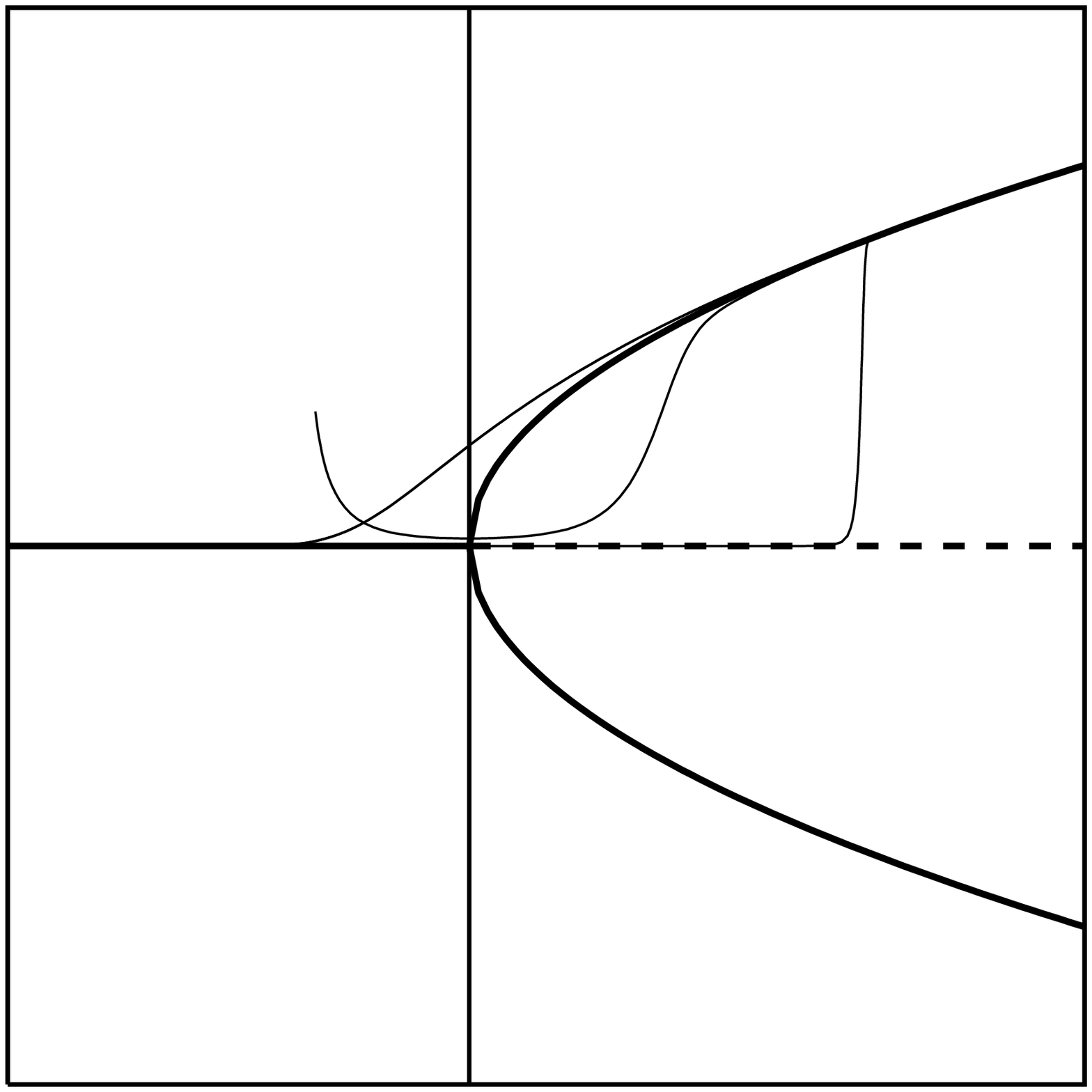,height=50mm,clip=t}}
 \figtext{
 	\writefig	0.9	5.2	a
 	\writefig	4.0	5.2	$x$
 	\writefig	6.0	3.1	$\lambda$
 	\writefig	3.0	4.7	$\eps^{2/3}$
 	\writefig	2.0	4.0	$\eps^{1/3}$
	\writefig	7.9	5.2	b
 	\writefig	10.8	5.2	$x$
 	\writefig	13.0	3.1	$\mu$
 }
 \caption[]
 {(a) Same as \figref{fig_1Dhyst1}, for $\mu=-1$. The periodic
 orbit lies at a distance at most $\Order{\eps^{1/3}}$ from a limiting
 hysteresis cycle, composed of stable equilibrium branches and two vertical
 lines. The enclosed area is $\cA(\eps) = \cA(0) + \Order{\eps^{2/3}}$. 
 (b) Hysteresis cycle of the equation $\eps\dot{x} = \mu(\tau) x -
 x^3$.}
\label{fig_1Dhyst2}
\end{figure}

As a final example, let us consider the situation where $\lambda=0$ and
$\mu(\tau)$ is the varying parameter. The static bifurcation diagram
displays a pitchfork bifurcation at the origin, involving the branches
$x\equiv 0$ and $x = \pm\sqrt{\lambda}$, for which $q=\frac12$ and $p=1$
(\figref{fig_1DNewton}c). An important new phenomenon is \defwd{bifurcation
delay}: near $x=0$, the dynamics is essentially governed by the linearized
equation
\begin{equation}
\label{1D10}
\eps\dot{x} = \mu(\tau)x \quad \sothat \quad 
x(\tau) = \exp\bigbrak{\frac{1}{\eps}\int_{\tau_0}^\tau \mu(s)\dx s}
x(\tau\sub0).
\end{equation}
Starting at a time where $\mu<0$, $x(\tau)$ remains exponentially small as
long as the integral in \eqref{1D10} is negative, which is true for a while
beyond the instant where $\mu$ becomes positive. If the solution finally
jumps on the stable branch and $\mu$ is decreased again, this branch is
followed adiabatically, which leads to hysteresis (\figref{fig_1Dhyst2}b).
The area of the cycle follows the scaling law
\begin{equation}
\label{1D11}
\cA(\eps) - \cA(0) \sord \eps^{3/4},
\end{equation}
where $\cA(0)$ depends on the bifurcation delay time. 

These examples indicate that one-dimensional systems are relatively well
understood. In fact, our method to determine scaling laws is general, and
quite straightforward to apply to other bifurcations than the special cases
described here. Moreover, one can expect that when a mode of a larger
system undergoes bifurcation, the dynamics of the associated adiabatic
system will be governed by an effective 1D equation, describing the motion
of this particular mode, which explains why the same scaling laws are
observed for more complicated systems. The Lorenz model discussed in the
next section illustrates this reduction of variables.  


\section{Bifurcation delay in the Lorenz model}
\label{sec_Lz}

Let us now turn to the behaviour of higher--dimensional systems
\begin{equation}
\label{Lz0}
\eps\dot{x} = F(x,\lambda(\tau)) = f(x,\tau), 
\qquad x\in\R^n,\;\tau\in\R.
\end{equation}
Adiabatic solutions with an expansion of the form \eqref{1D3} still exist in
the vicinity of \defwd{hyperbolic} equilibria, \ie branches of equilibrium
points around which the linearization of $f$ has no purely imaginary
eigenvalues. 

The behaviour of neighbouring solutions is far more complicated to analyse
than in the 1D case. One can, however, obtain valuable informations by
analysing the linearized system first, before dealing with nonlinear terms
using appropriate tools such as invariant manifolds. 

We illustrate these techniques on the Lorenz model with slowly varying
temperature gradient $r(\tau)$:
\begin{equation}
\label{Lz1}
\begin{split}
\eps\dot{x}_1 &= \sigma (x_2-x_1) \\
\eps\dot{x}_2 &= r(\tau)x_1 - x_2 - x_1 x_3 \\
\eps\dot{x}_3 &= -b x_3 + x_1 x_2,
\end{split}
\end{equation}
where we assume that $b,\sigma>0$. This model has been introduced as an
approximation to Rayleigh--B\'enard convection, but also describes other
systems such as lasers. It is well known that if $r\leqs 1$ is fixed, the
origin is a globally asymptotically stable fixed point, whereas when $r>1$,
the origin is hyperbolic and two new equilibria $C_{\pm} =
(\pm\sqrt{b(r-1)},\pm\sqrt{b(r-1)},r-1)$ appear, which correspond
physically to convection rolls. 

We will study this system when $r(\tau)$ is slowly oscillating around $r=1$,
and stays well below the chaotic region. It can be written in compact form 
\begin{equation}
\label{Lz2}
\eps\dot{x} = A(\tau)x + b(x),
\end{equation}
where $b(x)$ is quadratic. Note that the identically zero function is a
particular solution of this equation. The matrix $A(\tau)$ has three
eigenvalues
\begin{equation}
\label{Lz3}
a_{1,2}(\tau) = -\tfrac12(\sigma+1) \pm s(\tau), \quad
a_3 = -b, \quad
s(\tau) \defby \tfrac12 \sqrt{(\sigma+1)^2 + 4\sigma(r(\tau)-1)}. 
\end{equation}
The eigenvalues $a_2$ and $a_3$ are always negative, while $a_1(\tau)$ has
the same sign as $r(\tau)-1$. We thus expect that the motion will
essentially follow the eigenspace of $a_1(\tau)$. For this reason, we will
construct a change of variables isolating this particular direction. To
do this, we begin by searching a linear transformation which should
diagonalize the linear part of \eqref{Lz2}. To this end, observe that if
$S(\tau;\eps)$ is a matrix satisfying
\begin{equation}
\label{Lz4}
\eps\dot{S} = AS - SD 
\end{equation}
where $D(\tau;\eps)$ is diagonal, then the change of variables $x=Sy$
transforms \eqref{Lz2} into
\begin{equation}
\label{Lz5}
 \eps\dot{y} = D(\tau)y + S^{-1}b(Sy).
\end{equation}
The key point is that we can prove the existence of a bounded solution of
\eqref{Lz4}, admitting asymptotic series
\begin{equation}
\label{Lz6}
\begin{split}
S(\tau;\eps) &= S_0(\tau) + \eps S_1(\tau) + \eps^2 S_2(\tau) + \dotsb \\
D(\tau;\eps) &= D_0(\tau) + \eps D_1(\tau) + \eps^2 D_2(\tau) + \dotsb
\end{split}
\end{equation}
which can be truncated to exponentially small order, just as the adiabatic
solution \eqref{1D3}.\footnote{These matrices are not unique, since every
column of $S$ can be multiplied by a function of time, which will of course
affect terms of order $\eps$ in $D$.} In particular, $S_0(\tau)$ is the
matrix diagonalizing $A$ statically, and the entries of $D_0(\tau)$ are
eigenvalues of $A(\tau)$. The proof uses the fact that $a_1(\tau)\neq
a_2(\tau)$. 

In the specific case of the Lorenz equations, a linear transformation given
to leading order by
\begin{equation}
\label{Lz7}
x_1 = \sigma(y+z_1) + \Order{\eps}, \quad 
x_2 = \tfrac{\sigma-1}{2}(y+z_1) + s(y-z_1) + \Order{\eps}, \quad 
x_3 = z_2
\end{equation}
yields the equation 
\begin{equation}
\label{Lz8}
\begin{split}
\eps\dot{y} &= d_1(\tau)y + b_1(y,z,\tau) \\
\eps\dot{z} &= D_2(\tau)z + b_2(y,z,\tau), 
\end{split}
\end{equation}
where $d_1(\tau) = a_1(\tau) + \Order{\eps}$, $D_2(\tau)$ is diagonal with
entries $a_2(\tau) + \Order{\eps}$ and $a_3$, and $b_1$, $b_2$ are
quadratic. 

\begin{figure}
 \centerline{\psfig{figure=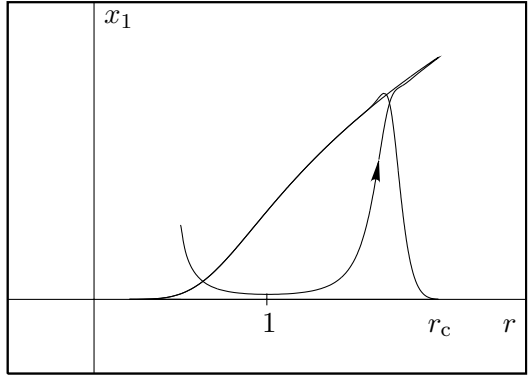,height=50mm,clip=t}}
 \figtext{
 	\writefig	5.2	5.2	$x_1$
 	\writefig	10.5	1.1	$r$
 	\writefig	9.5	1.1	$\crit{r}$
 	\writefig	7.3	1.1	$1$
}
 \caption[]
 {Rotation frequency of convection rolls $x_1$ as a function of the
 periodically varying temperature difference $r$ in the Lorenz model. After
 a first transient cycle, the motion settles on a hysteresis cycle, on which
 $x_1$ increases rapidly and decreases slowly.}
\label{fig_lorenz}
\end{figure}

In order to deal with the nonlinear terms, we introduce invariant manifolds.
Consider the partial differential equation
\begin{equation}
\label{Lz9}
\eps\sdpar{v}{\tau}(y,\tau) = D_2(\tau)v(y,\tau) + b_2(y,v,\tau)  
-\sdpar{v}{y}(y,\tau) \bigbrak{d_1(\tau)y + b_1(y,v,\tau)}.
\end{equation}
It can be shown that this equation admits, in a \nbh\ of $y=0$, a solution
$v(y,\tau) = \Order{y^2}$. When $r>1$, it lies at a distance of order $\eps$
from the instantaneous unstable manifold of the origin, but it can be
continued to times where $r<1$. The change of variables $z = \z + v(y,\tau)$
transforms the second equation of \eqref{Lz8} into 
\begin{equation}
\label{Lz10}
\eps\dot{\z} = \bigbrak{D_2(\tau) + \beta_2(y,\z,\tau)}\z, 
\end{equation}
where $\beta_2$ is of order $\abs{y}+\norm{\z}$. This equation admits
$\z=0$ as invariant manifold. A similar change of variables $y = \y +
u(\z,\tau)$ transforms the first equation into
\begin{equation}
\label{Lz11}
\eps\dot{\y} = \bigbrak{d_1(\tau) + \beta_1(\y,\z,\tau)}\y, 
\end{equation}
defining a stable manifold separating the basins of attraction of $C_+$ and
$C_-$. 

Since $D_2$ has negative eigenvalues, one easily shows that $\z(\tau)$ goes
to zero exponentially fast. Thus the effective dynamics will take place on
the invariant manifold $z=v(y,\tau)$, where it is governed by the scalar
equation
\begin{equation}
\label{Lz12}
\begin{split}
\eps\dot{y} &= d_1(\tau)y + h(\tau)y^3 + \Order{y^5}, \\
h(\tau) &= -\frac{\sigma^2}{2bs} \Bigpar{\frac{\sigma-1}{2}+s} +
\Order{\eps}.
\end{split}
\end{equation}
($y=\y$ on the unstable manifold $\z=0$).
As $r(\tau)$ is varied periodically around $r=1$, $d_1(\tau) = a_1(\tau) +
\Order{\eps}$ changes sign. The situation is thus very similar to the last
example of Section \ref{sec_1D}. Assume that $d_1(\tau)$ becomes negative
at $\tau_0$, positive at $\tau_1$, and negative again at $\tau_0+1$.
Asymptotically, the solution will stay close to the origin for $\tau\sub0 +
n \leqs \tau \leqs \hat{\tau} + n$, where $\hat{\tau}>\tau\sub1$ is the
\defwd{delay time} defined by the relation
\begin{equation}
\label{Lz13}
\int_{\tau_0}^{\hat{\tau}} d_1(\tau)\dx \tau = 0
\end{equation}
(this time exists if the average of $d_1(\tau)$ over one period is
positive, otherwise solutions stay indefinitely close to the origin). For
$\hat{\tau}+n<\tau<\tau_0+n+1$, the asymptotic solution follows $C_+$ or
$C_-$ adiabatically. Thus the bifurcation delay leads once again to
\defwd{hysteresis} (\figref{fig_lorenz}). 

When $r(\tau)$ is varied back and forth (at least when $r$ does not become
too large), the solution always follows the {\em same} equilibrium (which
one it chooses depends on the initial condition). In the case of
Rayleigh--B\'enard convection, $x_1$ measures the rotation frequency of
convection rolls. When $r(\tau)$ is increased, these rolls will appear
suddenly, with a positive frequency, at some $\crit{r}=r(\htau)>1$. When
$r(\tau)$ is decreased again, they slowly decelerate to disappear smoothly
as $r$ becomes smaller than 1. The rolls will always turn in the same
direction. We believe that it would be interesting to try to observe this
delay experimentally.


\newpage
\section{Hysteresis in mean field ferromagnets}
\label{sec_mh}

Hysteresis in ferromagnets has been known and studied experimentally for a
long time. Interest in a microscopic understanding of hysteresis and
associated scaling laws has been renewed by the numerical study of
\cite{RKP}. The internal dynamics of ferromagnets, however, is so
complicated that its modeling by ordinary differential equations is not
obvious. We will consider here a simple Curie--Weiss model, which can be
described by an effective mean field equation.

Consider the Hamiltonian
\begin{equation}
\label{mh1}
H(\sigma) = -\frac{1}{2N} \sum_{i\neq j\in\Lambda}
\pscal{\sigma_i}{J\sigma_j} - \sum_{i\in\Lambda} \pscal{h}{\sigma_i},
\end{equation}
where $\Lambda$ is a subset of $\Z^d$ with $N$ sites, the spins $\sigma_i$
are unit vectors in $\R^n$, $J$ is a fixed coupling matrix and $h$ the
magnetic field. We introduce a stochastic spin flip dynamics of Glauber
type. The detailed balance condition \cite{Ka} is satisfied by a transition
probability by unit time of the form
\begin{equation}
\label{mh2}
\begin{split}
\tprob{\sigma'}{\sigma} &= \sum_{i\in\Lambda} \prod_{j\neq i}
\delta(\sigma'_j-\sigma_j) \e^{\beta\pscal{\sigma'_i}{h_i(\sigma)}}
g(h_i(\sigma)),\\
h_i(\sigma) &\defby h + \frac{1}{N}J\sum_{j\neq i} \sigma_j,
\end{split}
\end{equation}
where $\beta=T^{-1}$ is the inverse temperature, and $g(h)$ an arbitrary
function.

To derive a deterministic equation of motion, we consider a sequence of
systems with $N$ sites, $N\to\infty$. Under appropriate assumptions on the
initial probability distribution, one can derive in the thermodynamic limit
a deterministic equation for the magnetization $m$ of the
form\footnote{This is the simplest equation, obtained for a particular
choice of $g$. Other choices yield a multiplicative factor in the
right--hand--side.}
\begin{equation}
\label{mh3}
\dtot{m}{t} = -m + \beta(Jm+h) F_n(\beta\norm{Jm+h}), 
\end{equation}
where $F_n(x)$ depends on the dimension $n$ of the spins. In particular, 
\begin{equation}
\label{mh4}
F_1(x) = \frac{\thyp x}{x}, \qquad
F_2(x) = \frac12 - \frac{1}{16}x^2 + \Order{x^4}.
\end{equation}
It can be shown that corrections to $m(t)$ resulting from a finite $N$ are
of order $N^{-1/2}$, and obey a Langevin equation \cite{Mar1,Mar2}.

We will now describe some interesting properties of the 1D and 2D models in
a slowly oscillating magnetic field, paying attention in particular to the
mechanism of magnetization reversal and its influence on the shape of
hysteresis cycles.

\subsection{One--dimensional case: dynamic phase transition}

\begin{figure}
 \centerline{\psfig{figure=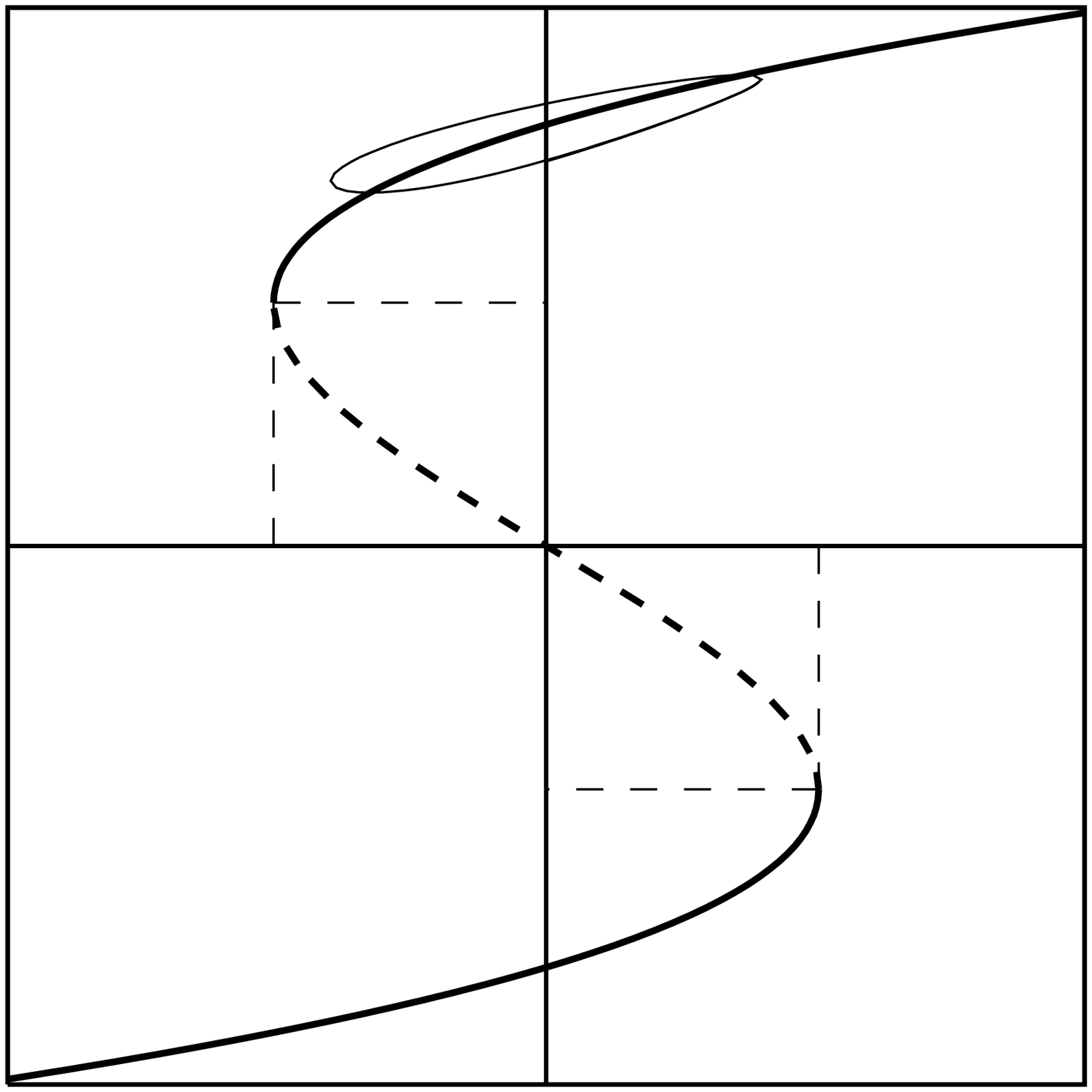,height=50mm,clip=t}
 \hspace{18mm}
 \psfig{figure=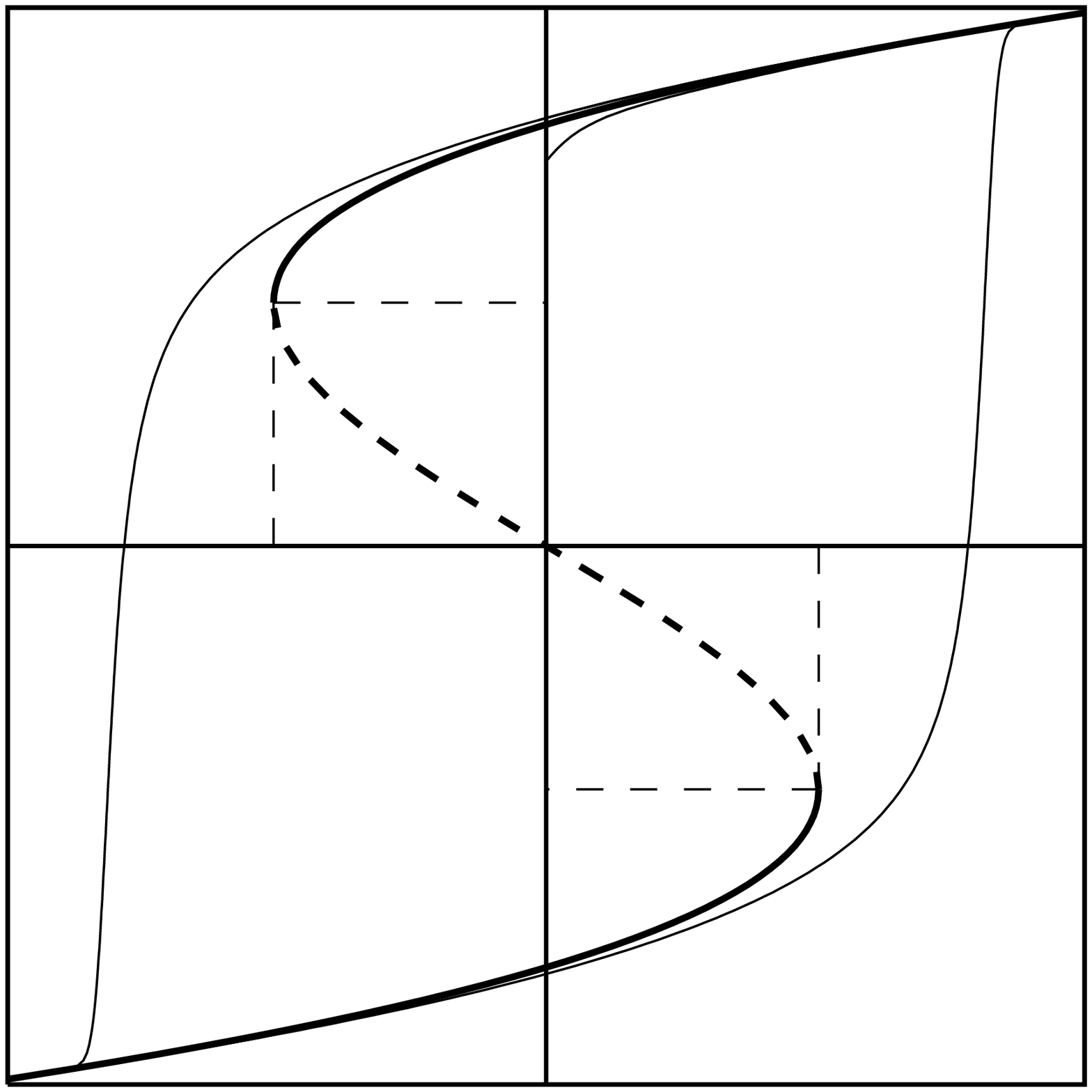,height=50mm,clip=t}}
 \figtext{
 	\writefig	0.9	5.2	a
 	\writefig	4.0	5.2	$m$
 	\writefig	4.0	4.05	$\crit{m}$
 	\writefig	3.05	1.75	$-\crit{m}$
 	\writefig	6.0	3.15	$h$
 	\writefig	4.9	3.15	$\crit{h}$
 	\writefig	2.3	2.65	$-\crit{h}$
	\writefig	7.9	5.2	b
	\writefig	11.0	5.2	$m$
	\writefig	11.0	4.05	$\crit{m}$
	\writefig	10.05	1.75	$-\crit{m}$
	\writefig	13.0	3.15	$h$
	\writefig	11.9	3.15	$\crit{h}$
	\writefig	9.3	2.65	$-\crit{h}$
 }
 \caption[]
 {Solutions of \eqref{m1D1} illustrating the phenomenon of ``dynamic phase
 transition''. When the amplitude $h\sub0$ of the magnetic field is smaller
 than the critical field $\crit{h}$, the magnetization oscillates around a
 nonzero average, and encloses an area of order $\eps h\sub0$ (a). When
 $h\sub0$ is larger than $\crit{h}$, the average magnetization is zero, and
 the periodic solution encloses an area $\cA(\eps) = \cA(0) +
 \Order{\eps^{2/3}}$ (b).}
\label{fig_mh1}
\end{figure}

For 1D spins, the adiabatic equation of motion can be written as\footnote{We
neglect here terms of order $\eps$ stemming from the slow time dependence of
$h$ in the derivation of the equation of motion.}
\begin{equation}
\label{m1D1}
\eps\dot{m} = -m + \thyp\beta(Jm+h(\tau)), 
\end{equation}
where we will consider a periodic magnetic field of the form
\begin{equation}
\label{m1D2}
h(\tau) = h_0 \sin(2\pi\tau).
\end{equation}
For positive inverse temperature $\beta=T^{-1}$, we may rescale the
variables in such a way that $J=1$. If $\beta<1$, there is no static
hysteresis. If $\beta>1$, the static bifurcation diagram is similar to the
equation $\eps\dot{x} = -x+x^3+\lambda(\tau)$ discussed in Section
\ref{sec_1D}: it contains two saddle--node bifurcations $(\pm\crit{h},
\mp\crit{m})$, where 
\begin{equation}
\label{m1D3}
\crit{m}(T) = \sqrt{1-T}, \qquad
\crit{h}(T) = \crit{m} - T \Argth\crit{m}.
\end{equation}
As we have seen in Section \ref{sec_1D}, when $h_0\gg\crit{h}$ orbits are
attracted by a hysteresis cycle with zero average magnetization
(\figref{fig_mh1}b) and area $\cA(0) + \Order{\eps^{2/3}}$. This scaling
law was already obtained in \cite{Jung}. When $h_0<\crit{h}$, the
magnetization never sees any bifurcation point, and it follows
asymptotically a cycle of width $\Order{\eps}$ with nonzero average
magnetization (\figref{fig_mh1}a).  

This phenomenon was observed numerically and called ``dynamic phase
transition'' by \cite{TO}, who also called ``ferromagnetic'' or
\defwd{F--region} the domain of $(T,h_0)$--plane where the asymptotic cycle
has a nonzero magnetization, and ``paramagnetic'' or \defwd{P--region} the
domain where it has zero average magnetization. In the adiabatic limit,
these regions are delimited by the line $h=\crit{h}(T)$. For positive
$\eps$, there may be a small overlap between these regions, where a
symmetric P--cycle and an asymmetric F--cycle coexist. 

We claim that for small $\eps$ and $T=\beta^{-1}<1$, the F--region grows by
a distance of order $\eps$, the P--region shrinks by an amount of the same
order (but may overlap the F--region), and the corresponding cycles obey
the scaling laws
\begin{equation}
\label{m1D4}
\begin{split}
\text{F--cycle:}\quad & \quad
\text{$\cA(\eps,h_0)\sord h_0\eps$ if $h\sub0<\crit{h}$}, 
\quad\text{$\eps\ln\abs{\eps}$ if $h\sub0=\crit{h}$}, \\
\text{P--cycle:}\quad & \quad
\cA(\eps,h_0) \sord \cA_0 + \eps^{2/3}(h_0-\crit{h})^{1/3}.
\end{split}
\end{equation} 

\begin{figure}
 \centerline{\psfig{figure=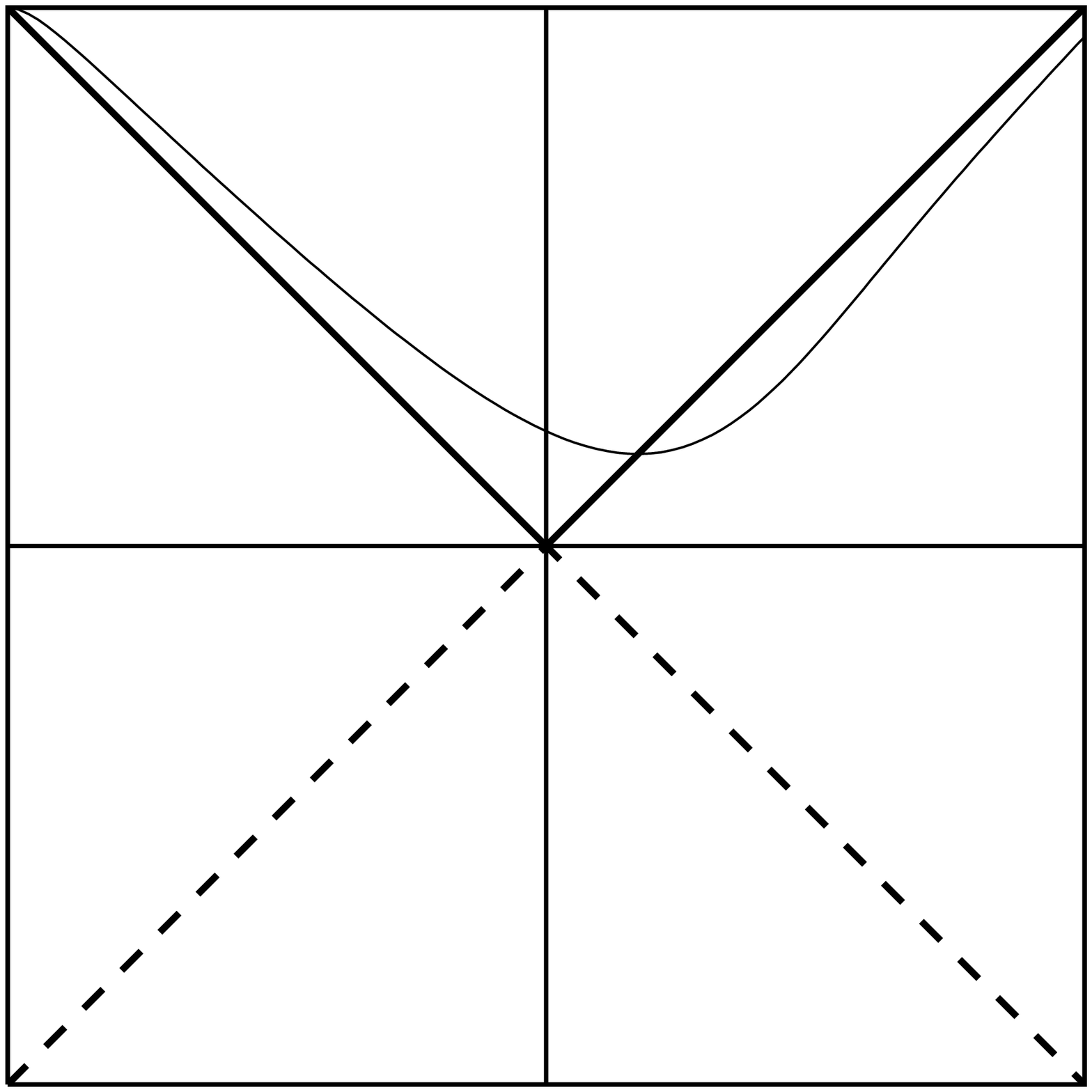,height=50mm,clip=t}
 \hspace{18mm}
 \psfig{figure=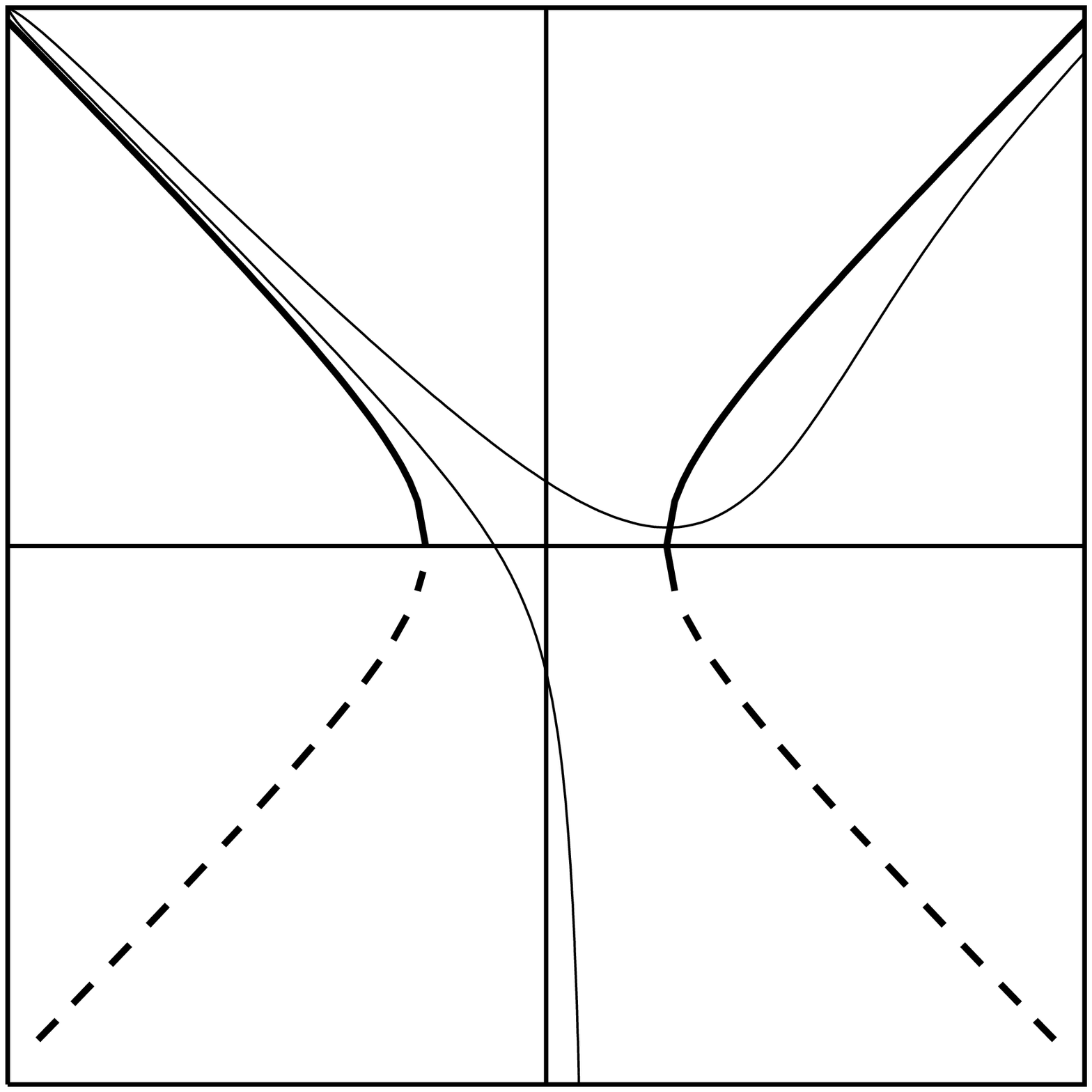,height=50mm,clip=t}}
 \figtext{
 	\writefig	0.9	5.2	a
 	\writefig	4.0	5.2	$y$
 	\writefig	6.0	3.15	$\tau$
	\writefig	7.9	5.2	b
	\writefig	11.0	5.2	$y$
	\writefig	13.0	3.15	$\tau$
	\writefig	9.4	2.6	$-\sqrt{\delta}$
	\writefig	11.6	2.6	$\sqrt{\delta}$
 }
 \caption[]
 {If the amplitude of the magnetic field is equal to $h\sub0 = \crit{h} +
 \delta$, the motion near the turning point is governed by the Ricatti
 equation \eqref{m1D5}. If $\delta=0$, it describes a transcritical
 bifurcation, and adiabatic solutions follow the upper branch (a). This
 means that we are still in the F--region. For positive $\delta$, we show
 that this behaviour subsists as long as $\delta = \Order{\eps}$. In
 (b), trajectories are shown for two different values of $\eps$. If $\eps <
 \delta$, the solution escapes from below after a delay of order
 $\eps^{2/3}\delta^{-1/6}$, and we have reached the P--region.}
\label{fig_mh2}
\end{figure}

Let us indicate how we obtain these scaling laws. Assume that $h_0 =
\crit{h}+\delta$. After translating the coordinates to the bifurcation point
$(-\crit{h},\crit{m})$ and scaling them in a proper way, equation
\eqref{m1D1} becomes
\begin{equation}
\label{m1D5}
\eps\dot{y} = -y^2 - \delta + \tau^2 + \text{higher order terms.}
\end{equation}
For $\delta=0$, this equation displays a transcritical bifurcation at the
origin. For positive $\delta$, it splits up into two saddle--node
bifurcations, with a gap of width $2\sqrt{\delta}$ (\figref{fig_mh2}). 
If $\delta<\Order{\eps}$, the transformation $\tau=\sqrt{\delta}\sigma$,
$y=\sqrt{\eps}z$ yields the equation
\begin{equation}
\label{m1D6}
\dtot{z}{\sigma} = -\tilde{\eps}z^2 + \tilde{\eps}^3(\sigma^2-1), 
\qquad \tilde{\eps} = \sqrt{\delta/\eps}, 
\end{equation}
which can be used to show that $z$ cannot move enough to slip through the
gap, so that we are in the P--region. 

If $\delta>\Order{\eps}$, the transformation $\tau=\sqrt{\delta}(\sigma-1)$,
$y=\sqrt{\delta}z$ gives 
\begin{equation}
\label{m1D7}
(\eps\delta^{-1}) \dtot{z}{\sigma} = -z^2 - \sigma + \sigma^2, 
\end{equation}
which is exactly the equation studied in Section \ref{sec_1D}. In
particular, the trajectory slips through the gap after a time delay of order
$\sqrt{\delta}(\eps\delta^{-1})^{2/3} = \eps^{2/3}\delta^{-1/6}$. During
this time, the magnetic field has reached a value $h_0 +
\Order{\eps^{2/3}\delta^{1/3}}$, which implies the scaling relation
\eqref{m1D4}. 

Finally, when $\delta\sord\eps$, the trajectory may behave in either way. A
more careful analysis of the Poincar\'e map shows that even though there is
a small region where stable F-- and P--cycles can coexist, the transition
is sharp, in the sense that the average magnetization jumps discontinuously
from one cycle to another. In \cite{TO}, a smooth transition, where the
magnetization goes to $0$ continuously, has been observed for larger values
of $\eps$. 

\subsection{Two--dimensional case: effect of anisotropy}

If we retain only the leading terms in \eqref{mh3}, we obtain the
Ginzburg--Landau equation
\begin{equation}
\label{m2D1}
\eps\dot{m} = (\beta J-\one)m - \tfrac12\beta J m \norm{\beta J m}^2 +
\beta h , 
\end{equation}
which describes the linearly driven, overdamped motion of a particle in a
sombrero--shaped potential. We shall assume that $h$ is parallel to an
eigenvector of the symmetric matrix $J$. In the subspace of this
eigenvector, the equation reduces to the previously studied case, and a
minimal field amplitude $\crit{h}$ is necessary to reverse magnetization.
When the magnetization has a transverse component, however, it can also
turn around the potential maximum, for a much smaller field amplitude. We
are thus going to focus on this situation. 

\begin{figure}
 \centerline{\psfig{figure=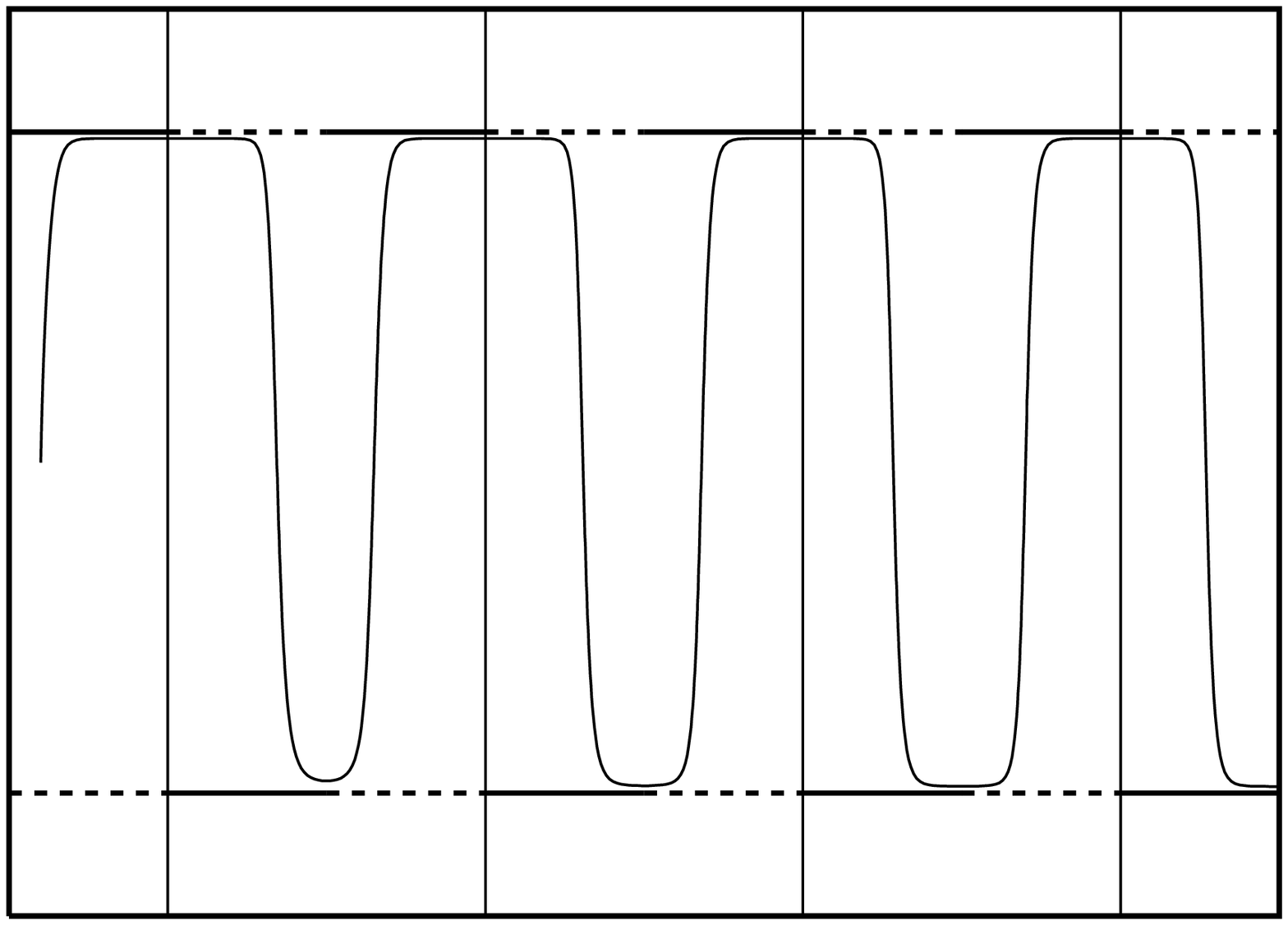,height=45mm,clip=t}
 \hspace{18mm}
 \psfig{figure=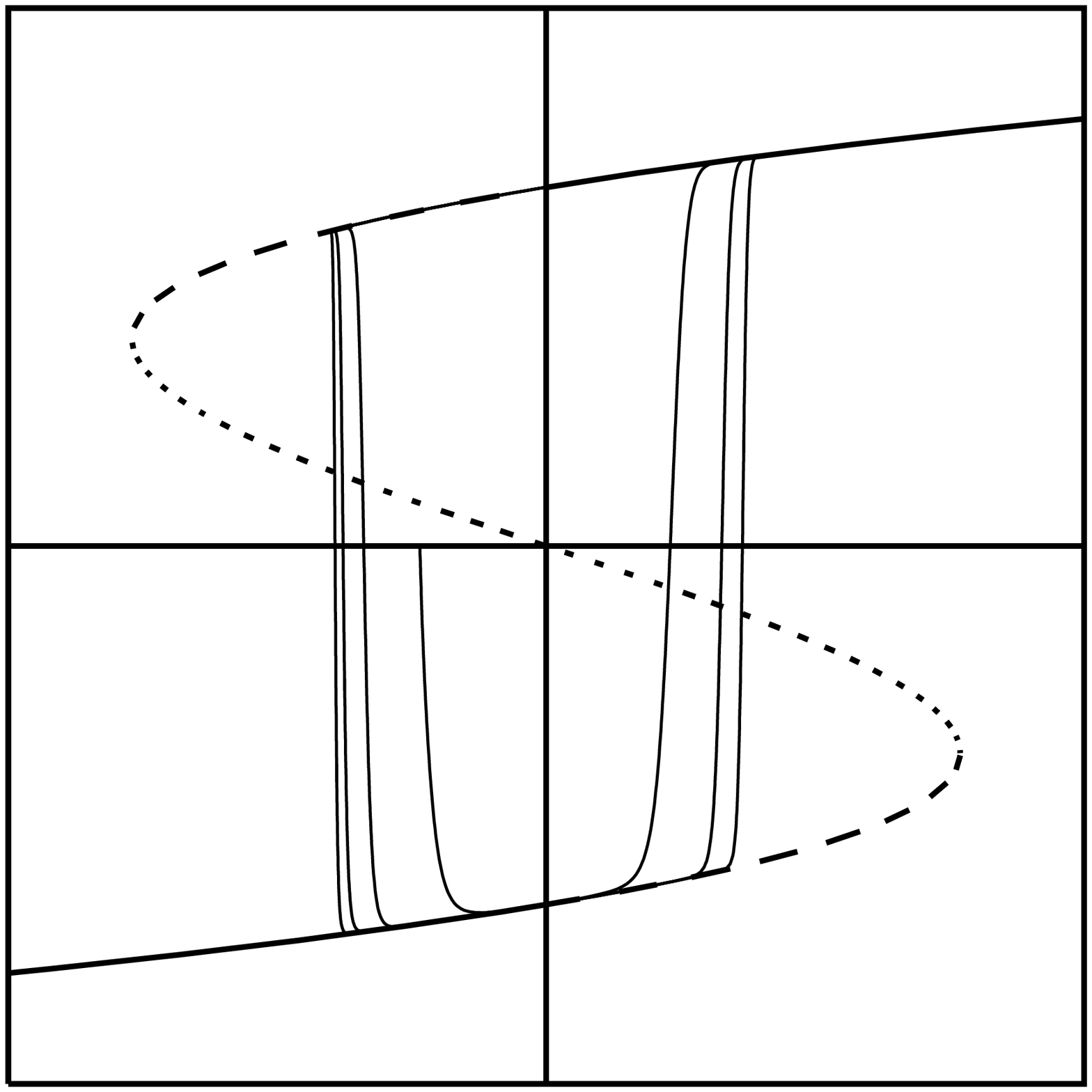,height=45mm,clip=t}}
 \figtext{
 	\writefig	0.6	4.7	a
	\writefig	8.9	4.7	b
 	\writefig	1.2	4.7	$\ph$
 	\writefig	1.45	4.1	$\pi$
	\writefig	11.65	4.7	$m_1$
 	\writefig	6.95	0.8	$\tau$
	\writefig	13.3	2.85	$h_1$
 	\writefig	1.2	0.8	$\tau\sub0$
 	\writefig	2.2	0.8	$\tau\sub1$
 	\writefig	2.65	0.8	$\tau\sub2$
 	\writefig	3.35	0.8	$1$
 }
 \caption[] 
 {(a) Evolution of $\ph(\tau)$, solution of equation \eqref{m2D3}. The
 magnetization quickly rotates at times $\tau\sub{n}$, determined
 recursively by the relation $\tau\sub{n+1} = \frac{n}{2} +
 \delay(\tau\sub{n}-\frac{n}{2})$. (b) The plot of $m_1$ as a function of
 $h_1$ shows the asymptotic hysteresis cycle, which is determined solely
 by the delay times. Thick solid lines, dashed and dotted lines represent
 respectively sinks, saddles and sources of the static system. Due to
 bifurcation delay, the magnetization follows the hyperbolic branch for
 some time, but ultimately rotates around the unstable origin.}
\label{fig_mh3}
\end{figure}

In the isotropic case, we may choose $J=\one$ to obtain the equation
\begin{equation}
\label{m2D2}
\eps\dot{m} = (\beta-1)m - \tfrac12\beta^3m\norm{m}^2 + \beta h(\tau),
\end{equation}
where we take a magnetic field $h(\tau)=(h_1(\tau),0)$, with
$h_1(\tau)=h_0\sin(2\pi\tau)$. It is useful to write this equation in polar
coordinates, with $m=(r\cos\ph,r\sin\ph)$, to get
\begin{equation}
\label{m2D3}
\begin{split}
\eps\dot{r} &= (\beta-1)r - \tfrac12\beta^3r^3 + \beta h_1(\tau)\cos\ph \\
\eps\dot{\ph} &= -\frac{\beta}{r} h_1(\tau)\sin\ph.
\end{split}
\end{equation}
If $h_1(\tau_0)<0$, the magnetization settles near the left equilibrium,
determined by $\ph=\pi$ and $r=r_-(\tau_0)$, the largest solution of
$(\beta-1)r - \tfrac12\beta^3r^3 - \beta h_1(\tau_0)=0$. When the field
becomes positive, the phenomenon of bifurcation delay causes $\ph$ to remain
for some time in unstable equilibrium near $\pi$, until it switches to $0$
at a time $\tau_1=\Psi(\tau_0)$, defined by
\begin{equation}
\label{m2D4}
\int_{\tau_0}^{\Psi(\tau_0)} \frac{h(\tau)}{r_-(\tau)} \dx\tau = 0.
\end{equation}
Because of the symmetry,  the next time of delayed magnetization reversal
is then given by $\tau_2 = \frac12 + \Psi(\tau_1-\frac12)$
(\figref{fig_mh3}). Subsequent reversal times are determined by the
recursive formula $\tau_n = \frac{n}{2} + \Psi(\tau_n-\frac{n}{2})$. It
turns out that this \defwd{self--determined bifurcation delay} finally
settles at a fixed point of the map $\tau\mapsto\Psi(\tau)-\frac12$. 
 
\begin{figure}
 \centerline{\psfig{figure=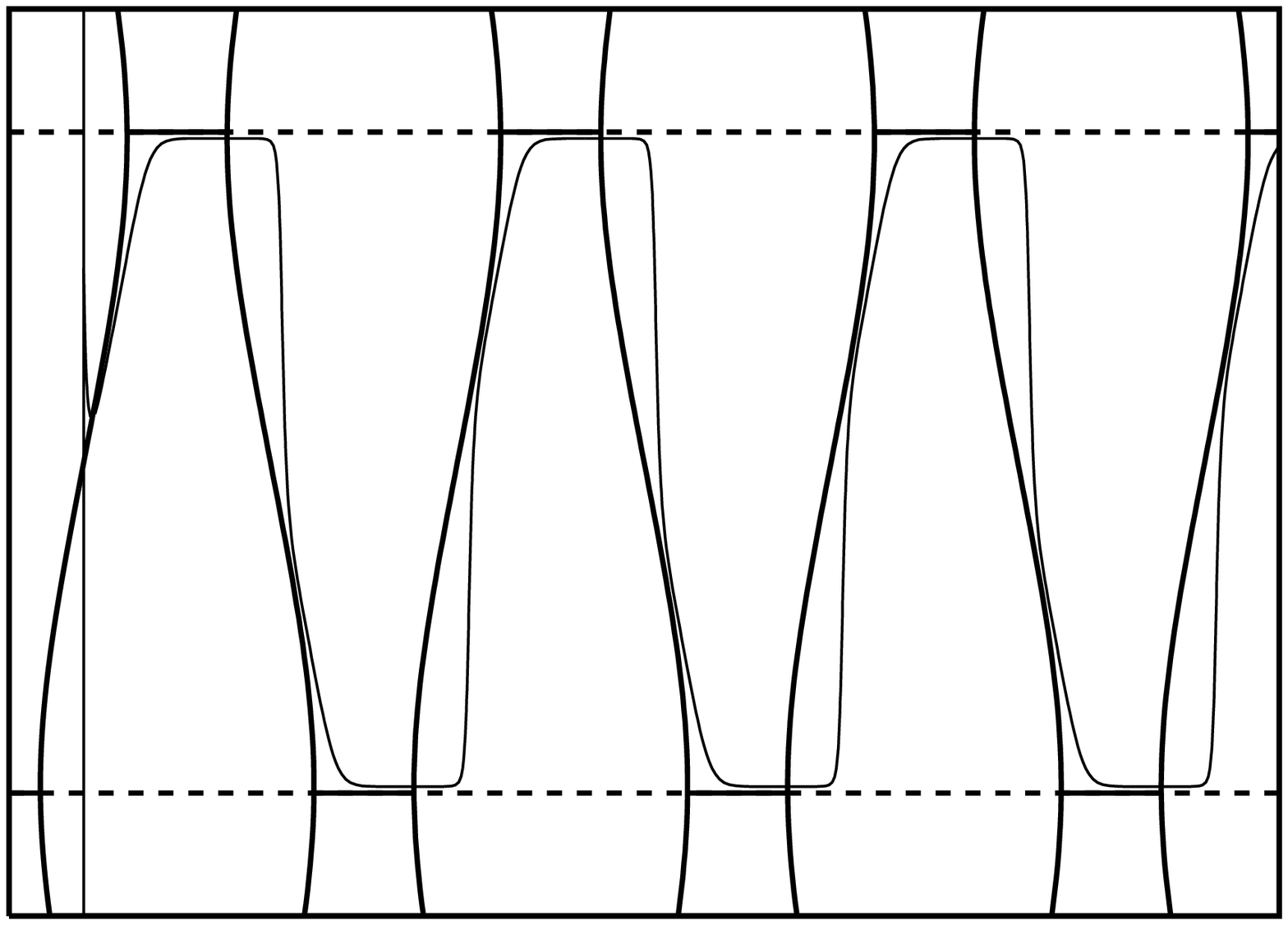,height=45mm,clip=t}
 \hspace{18mm}
 \psfig{figure=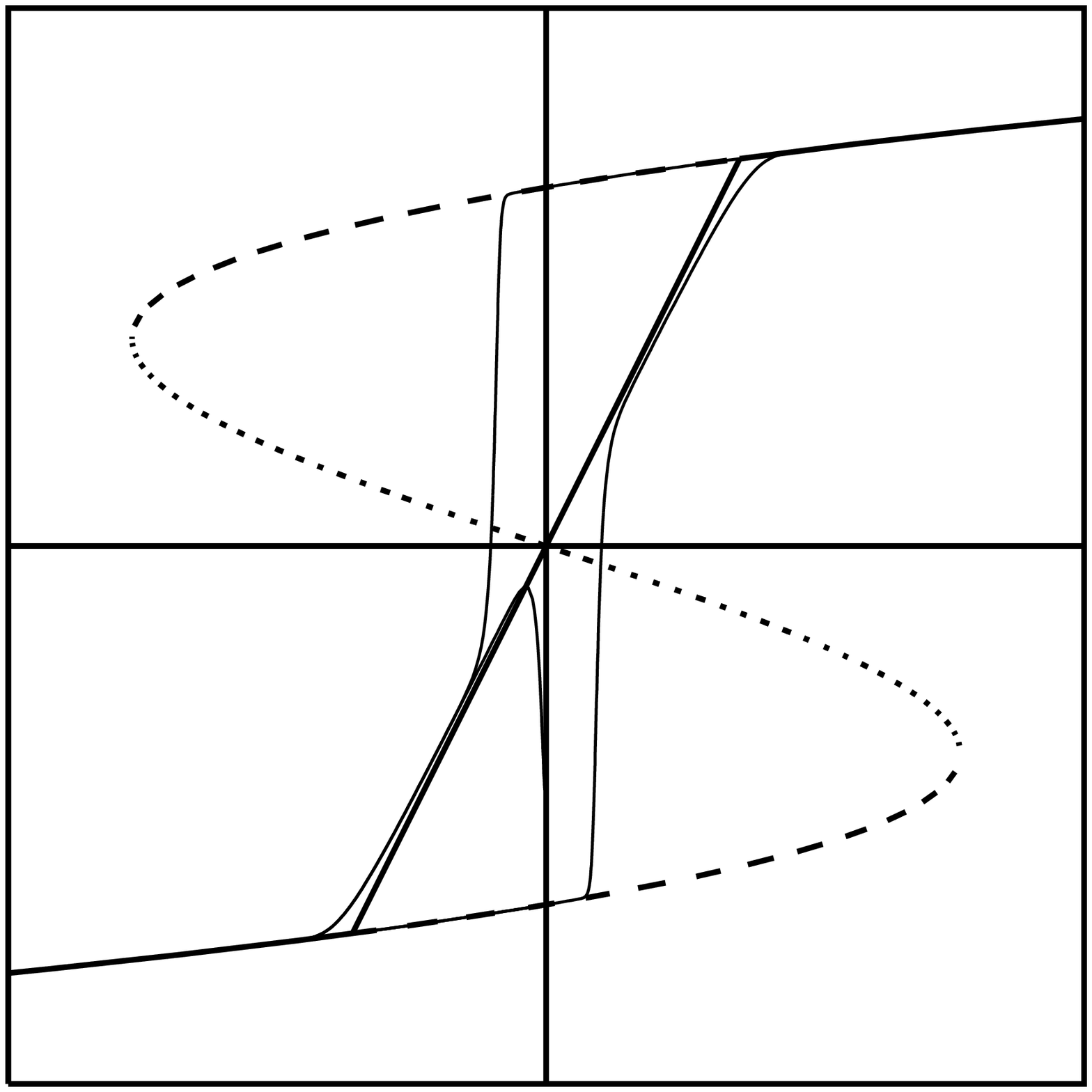,height=45mm,clip=t}}
 \figtext{
 	\writefig	0.6	4.7	a
	\writefig	8.9	4.7	b
 	\writefig	1.05	4.7	$\ph$
 	\writefig	1.05	4.1	$\pi$
	\writefig	11.65	4.7	$m_1$
 	\writefig	6.95	0.8	$\tau$
 	\writefig	3.1	0.8	$1$
	\writefig	13.3	2.85	$h_1$
 }
 \caption[]
 {(a) Evolution of $\ph$ in the anisotropic case with $\gamma>1$. Due to
 bifurcation delay, $\ph$ spends some time near $0$ or $\pi$, even when
 these points are unstable. It always drops back, however, to the
 transverse stable position. (b) The resulting hysteresis loop looks
 triangular. Curved lines are longitudinal equilibria, and the straight
 line represents transverse branches.}
\label{fig_mh4}
\end{figure}

We now turn to the anisotropic case where 
$J=\bigpar{\begin{smallmatrix}1&0\\0&\gamma\end{smallmatrix}}$.  In the
coordinates $m=(r\cos\ph,\gamma^{-1}r\sin\ph)$, the second equation of
\eqref{m2D3} becomes
\begin{equation}
\label{m2D5}
\eps\dot{\ph} = \beta(1-\tfrac12\beta^2r^2)(\gamma-1)\sin\ph\cos\ph -
\frac{\beta}{r} h_1(\tau)\sin\ph. 
\end{equation}
The case $\gamma<1$ is not very interesting, since the anisotropy enhances
the effect of the magnetic field, and tends to align the magnetization with
it.  If $\gamma>1$, a new stable transversal equilibrium exists for small
magnetic field. Its coordinates are determined by the relations
$\norm{\beta Jm}^2 = 2(1-\beta^{-1}\gamma^{-1})$ and
$(1-\gamma^{-1})m_1=\beta h_1$. The resulting hysteresis cycle is composed
of two triangular loops (\figref{fig_mh4}), since after leaving the
unstable position $\ph=0$ or $\pi$, the magnetization drops to the
transverse branch, which it follows until merging with the longitudinal
branch. 

We point out that if the magnetic field is slightly tilted with respect to
the eigenvectors of $J$, the pitchfork bifurcations in \figref{fig_mh4}
transform into saddle--nodes, which suppresses the bifurcation delay. The
result is that instead of oscillating back and forth, the magnetization
performs full circles, always in the same direction. 


\section{Chaotic hysteresis of a rotating pendulum}
\label{sec_rp}

The examples we have considered up to now all described an overdamped,
effectively one--dimensional motion, which displayed hysteretic, but not
chaotic properties. We present here an example taking into account inertia,
which turns out to have far more complicated dynamics. 

\begin{figure}
 \centerline{\psfig{figure=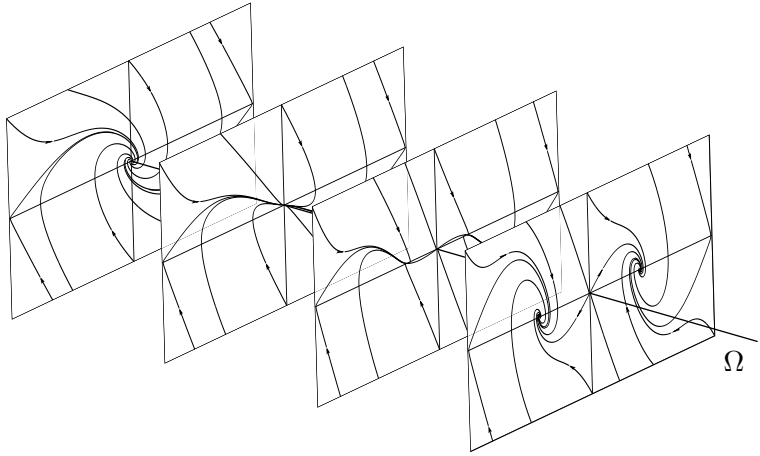,height=60mm,clip=t}}
 \figtext{
 	\writefig	12.0	1.6	$\W$
 }
 \caption[]
 {Phase portraits of the rotating pendulum for different values of the
 rotation frequency $\W$.}
\label{fig_rp1}
\end{figure}

Consider a mathematical pendulum mounted on a rotating table, turning with
angular frequency $\W$. The pendulum is subject to weight, friction and
a centrifugal torque, so that its equation of motion can be written in
dimensionless variables 
\begin{equation}
\label{rp1}
\begin{split}
\dot{q} &= p \\
\dot{p} &= -2\gamma p - \sin q + \W^2\sin q\cos q, 
\end{split}
\end{equation}
where $q$ is the angle between pendulum and vertical, and $\gamma>0$ is a
friction coefficient. This equation also describes the motion of a particle
in a symmetric potential, shaped as a single well when $\W<1$ and as a
double well when $\W>1$. The origin $O$ is always an equilibrium, while
for $\W>1$, two new stable equilibria appear at 
\begin{equation}
\label{rp2}
Q_{\pm} = (\pm\fix{q}(\W),0), \qquad
\fix{q}(\W) = \Arccos \W^{-2}. 
\end{equation}
The eigenvalues of the linearization of \eqref{rp1} around $O$ and $Q_{\pm}$
are given, respectively, by 
\begin{equation}
\label{rp3}
a_{\pm}^o = -\gamma \pm \sqrt{\gamma^2+\W^2-1}, \qquad
a_{\pm}^{\star} = -\gamma \pm
\sqrt{\gamma^2-\W^2+\W^{-2}}. 
\end{equation}
There are four qualitatively different phase portraits, delimited by the
values $\W=1$ and $\W=\W_{\pm}(\gamma)$, where
\begin{equation}
\label{rp4}
\W_-(\gamma)^2 = 1-\gamma^2, \qquad
\W_+(\gamma)^2 = \tfrac12\bigbrak{\gamma^2+\sqrt{\gamma^2+4}},  
\end{equation}
namely (see \figref{fig_rp1}):
\begin{itemiz}
\item	when $0<\W<\W_-(\gamma)$, $O$ is a stable focus; 
\item	when $\W_-(\gamma)<\W<1$, $O$ is a stable node; 
\item	when $1<\W<\W_+(\gamma)$, $O$ is a saddle and $Q_{\pm}$ are stable
	nodes;
\item	when $\W>\W_+(\gamma)$, $O$ is a saddle and $Q_{\pm}$ are stable
	focuses.
\end{itemiz}

\begin{figure}
 \centerline{\psfig{figure=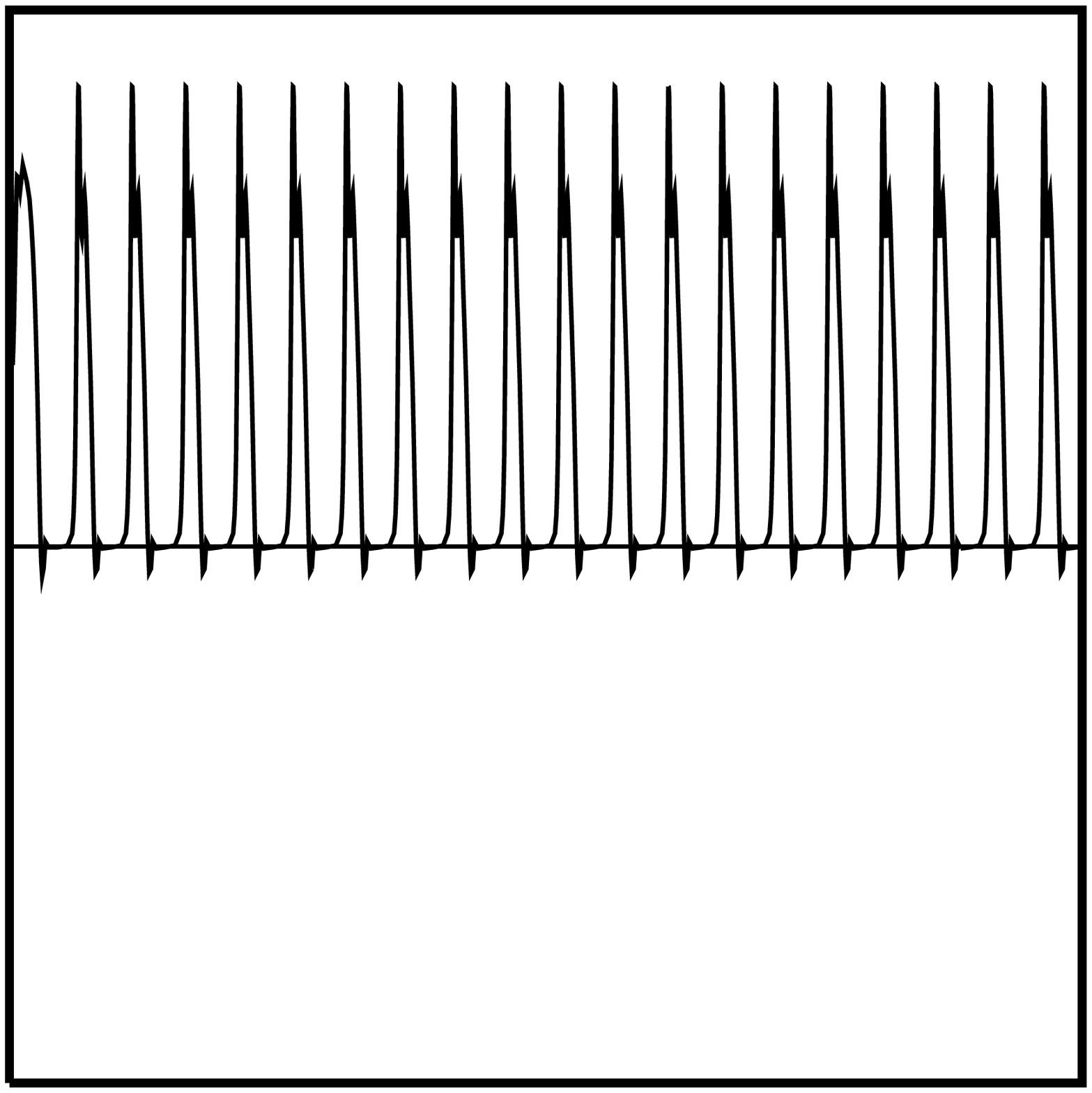,height=35mm,clip=t}
 \hspace{1mm}
 \psfig{figure=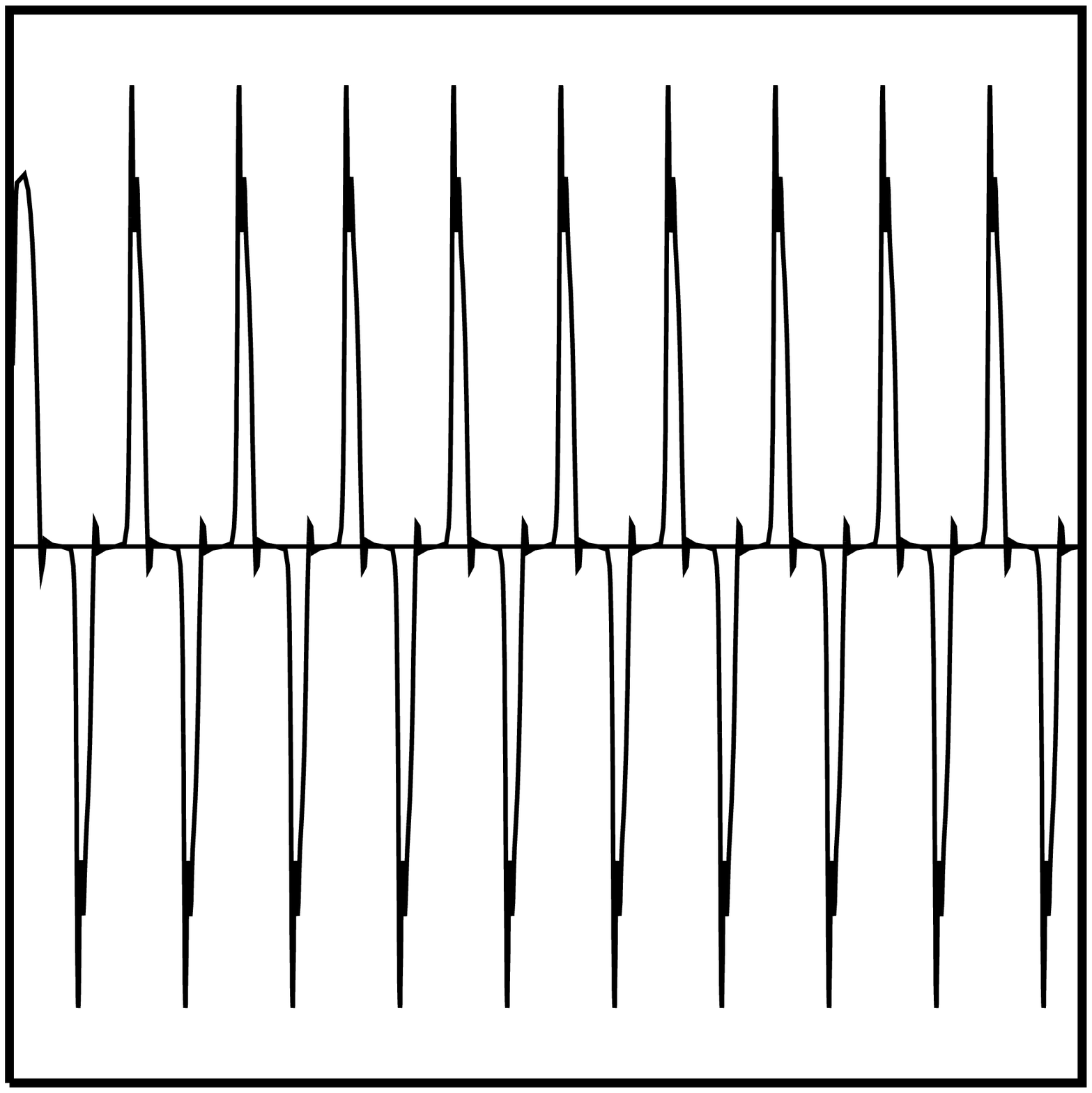,height=35mm,clip=t}
 \hspace{1mm}
 \psfig{figure=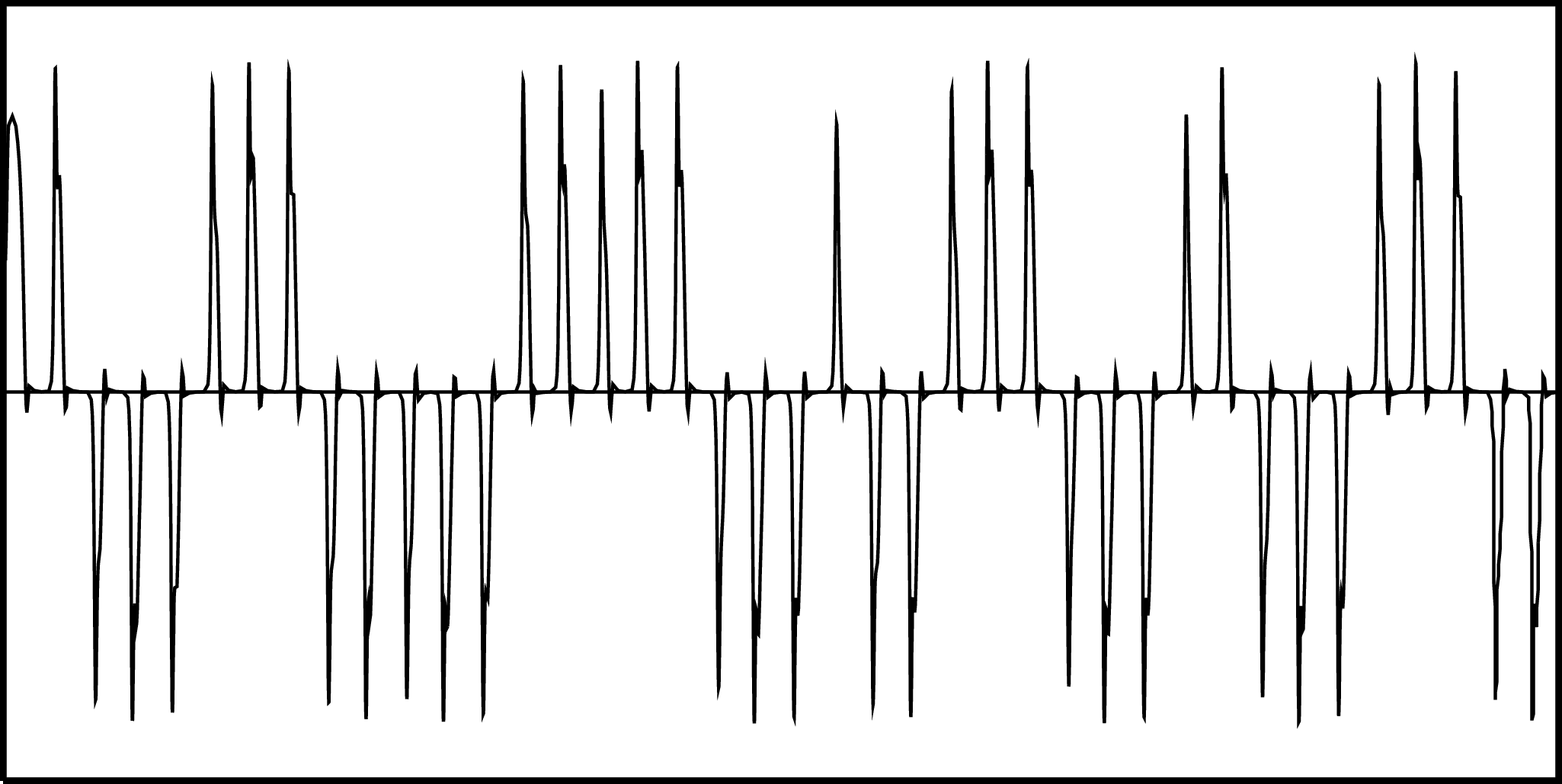,height=35mm,clip=t}}
 \figtext{
 	\writefig	0.1	0.58	a
	\writefig	3.95	0.58	b
	\writefig	7.85	0.58	c
  }
 \caption[]
 {Solutions $q(\tau)$ for slightly different values of the
 adiabatic parameter $\eps$. The time scale has been contracted in such a
 way as to show 20, resp.\ 40 periods of $\W$. One observes (a)
 solutions with the same period then $\W(\tau)$, (b) solutions with
 twice the period of $\W(\tau)$, going alternatively to one side and
 the other one, and (c), if $\eps$ is carefully adjusted, solutions which
 have no apparent period.}
\label{fig_rp2}
\end{figure}

If $\W=\W(\eps t)$ is made slowly and periodically time--dependent, we
obtain the adiabatic system
\begin{equation}
\label{rp5}
\begin{split}
\eps\dot{q} &= p \\
\eps\dot{p} &= -2\gamma p - \sin q + \Omega(\tau)^2\sin q\cos q.
\end{split}
\end{equation}
This system displays two interesting phenomena. The first one is a
bifurcation delay similar to the one already observed in previous examples:
when $\W$ is increased beyond 1, the pendulum remains for some time in
unstable equilibrium close to the origin, before joining one of the stable
equilibria $Q_+$ or $Q_-$. When $\W$ is decreased again below 1, the
pendulum follows this equilibrium until it joins the origin again, leading
to hysteresis. The second interesting phenomenon is related to the sequence
of visited equilibria, which depends on the value of the adiabatic
parameter (\figref{fig_rp2}). For some values, the pendulum always chooses
the same equilibrium, just as the Lorenz system always chooses the same
direction of rotation for the convection rolls. For other values of $\eps$,
however, one observes a sequence with twice the driving period, in which the
pendulum visits alternatively the equilibria $Q_+$ and $Q_-$. Between these
periodic behaviours, it is even possible to observe apparently random
sequences, which we called \defwd{chaotic hysteresis} \cite{BK}.  

In order to explain this behaviour, we now compute an asymptotic expression
for the Poincar\'e map in the $(q,p)$--plane, during one period of
$\W(\tau)$. If $\W(\tau)$ remains within the interval $\ccint{\W_-}{\W_+}$,
the system can be reduced to 1D as in Section \ref{sec_Lz}, and there is no
possibility for chaotic motion. We thus consider the case where $\W(\tau)$
has a larger amplitude (\figref{fig_rp3}). It is useful to introduce the
notations 
\begin{equation}
\label{rp6}
\alpha^o(\tau_2,\tau_1) = \re \int_{\tau_1}^{\tau_2} a_+^o(\tau)\dx\tau, 
\qquad
\phi^o(\tau_2,\tau_1) = \im \int_{\tau_1}^{\tau_2} a_+^o(\tau)\dx\tau. 
\end{equation}
Similar functions $\fix{\alpha}$ and $\fix{\phi}$ are defined for the
linearizations around $Q_{\pm}$. 

For $\ctau<\tau<1$, orbits are attracted by the stable origin. They remain
close to it until a bifurcation delay time $\htau+1$, defined by the
relation $\alpha^o(\htau+1,\ctau)=0$. During this part of motion, the system
can be essentially described by its linearization around the origin. Except
in a \nbh\ of $\tau^o_{\pm}$, where the eigenvalues $a^o_{\pm}$ cross, we
can carry out a dynamic diagonalization as in Section \ref{sec_Lz}. The
actual crossings are described by a local analysis, using Airy's equation.
Combining these steps, we obtain that
\begin{equation}
\label{rp7}
x(\htau+1) = S(\htau+1) 
\begin{pmatrix}
\cos\bigpar{\tfrac{\phi^o}{\eps}} & 
\e^{-\delta^o_2/\eps}\sin\bigpar{\tfrac{\phi^o}{\eps} + \th^o_2} \\ 
-\e^{-\delta^o_1/\eps}\sin\bigpar{\tfrac{\phi^o}{\eps} + \th^o_1} & 
\e^{-\delta^o_3/\eps}\cos\bigpar{\tfrac{\phi^o}{\eps} + \th^o_3}  
\end{pmatrix}
S(\ctau) x(\ctau),
\end{equation}
where $\phi^o = \phi^o(\top,\tom)+\Order{\eps}$ is the dynamic phase of
oscillations around the origin, and the columns of $S(\tau)$ are close to
the eigenvectors associated with the origin. The positive factors
$\delta^o_j$ describe the asymmetric contraction due to the difference
between $a^o_+$ and $a^o_-$, and the $\th^o_j$ are geometric phase shifts.
It can be shown that the effect of nonlinear terms can be absorbed in these
small geometric corrections. 

\begin{figure}
 \centerline{\psfig{figure=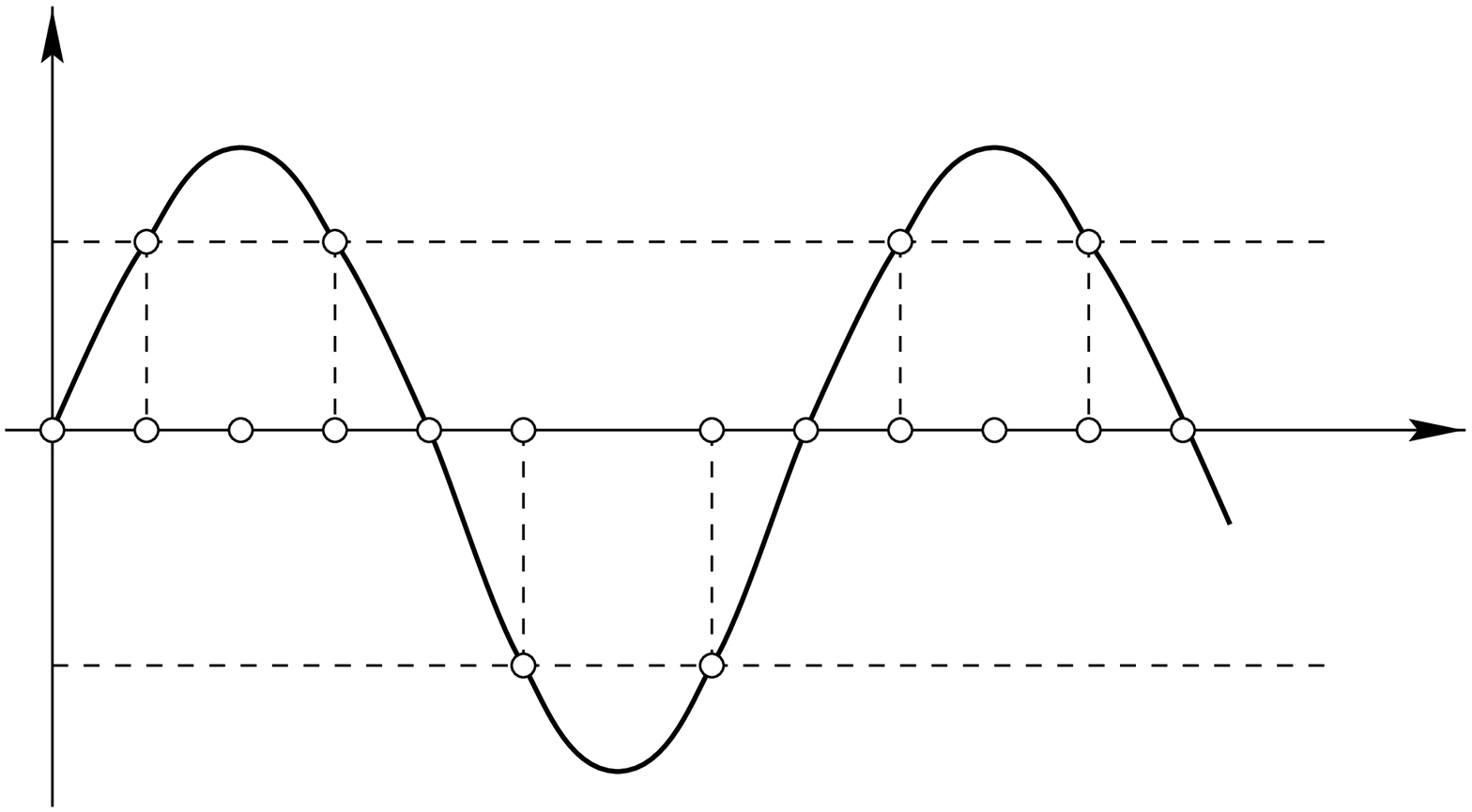,height=40mm,clip=t}}
 \figtext{
	\writefig	4.15	4.2	$\W(\tau)$
 	\writefig	2.95	3.3	$\W_+(\gamma)$
 	\writefig	2.95	1.15	$\W_-(\gamma)$
  	\writefig	3.4	2.3	$1$
	\writefig	10.4	2.6	$\tau$
 	\writefig	7.6	2.6	$1$
  	\writefig	7.1	2.6	$\top$
 	\writefig	6.2	2.6	$\tom$
	\writefig	5.8	2.6	$\ctau$
	\writefig	5.2	2.05	$\tsp$
	\writefig	4.8	2.05	$\htau$
	\writefig	4.3	2.05	$\tsm$
	\writefig	8.2	2.05	$\htau+1$
  }
 \caption[]
 {Function $\W(\tau)$ considered in the analysis. The instants when $\W$
 crosses the values $\W_{\pm}$ and $1$ delimit the different phases of the
 motion.}
\label{fig_rp3}
\end{figure}

\begin{figure}
 \centerline{\psfig{figure=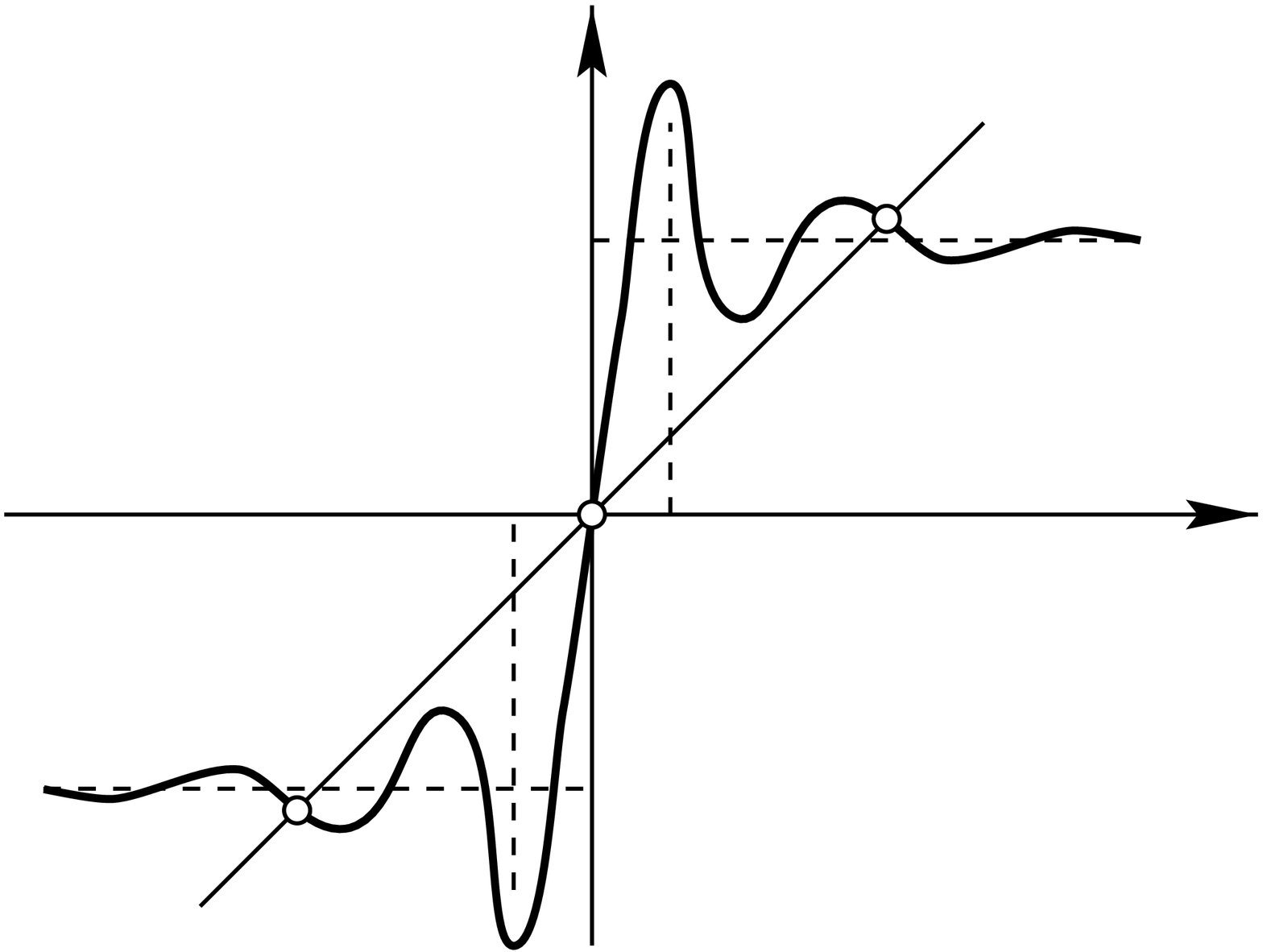,height=45mm,clip=t}
 \hspace{8mm}
 \psfig{figure=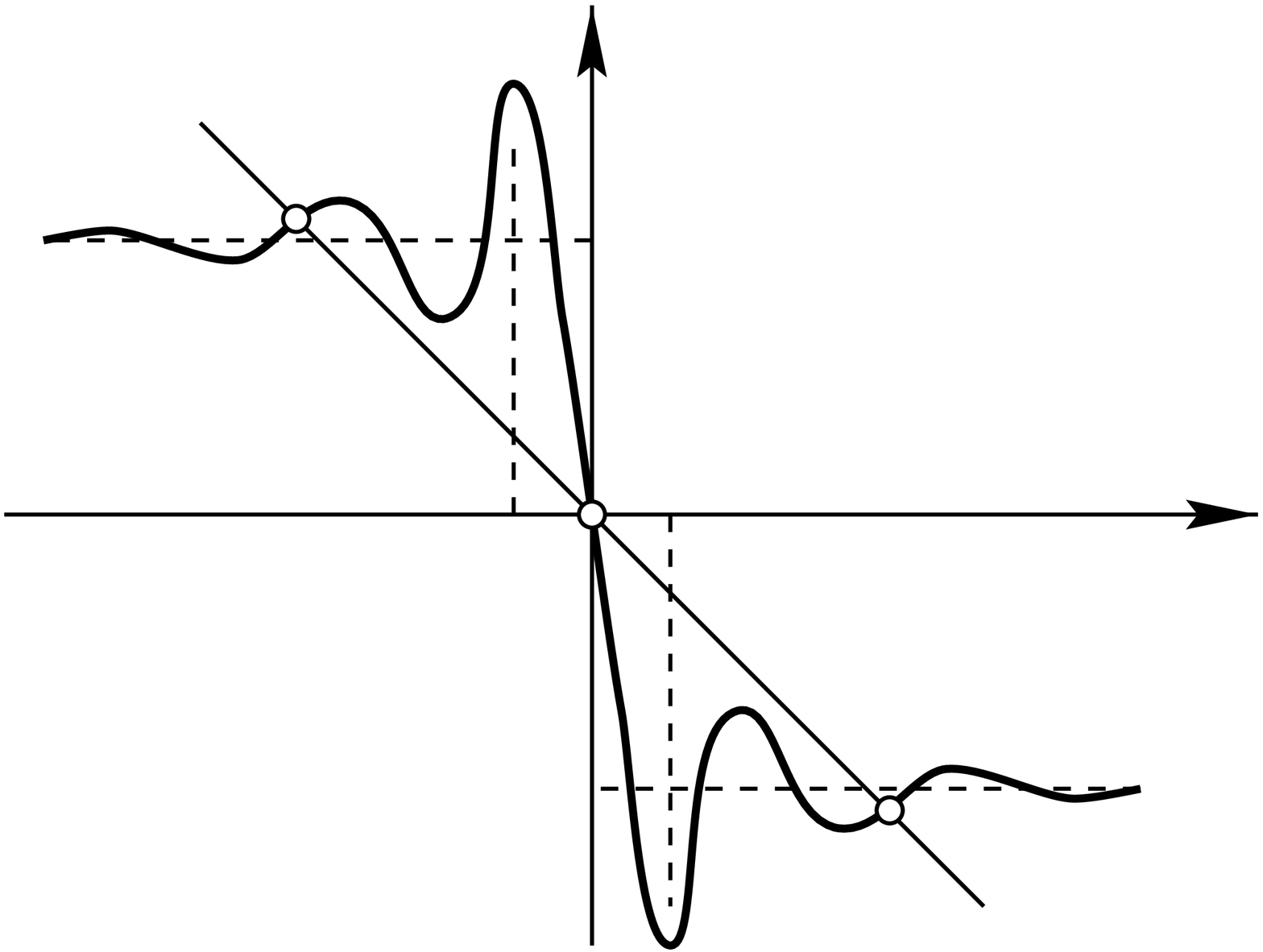,height=45mm,clip=t}}
 \figtext{
 	\writefig	0.55	4.6	a
	\writefig	2.6	4.5	$T\sub1(\x)$
 	\writefig	6.5	2.2	$\x$
	\writefig	1.0	3.8	$C\eps^{1/4}\cos(\supz{\phi}/\eps)$
	\writefig	3.8	1.2	$-C\eps^{1/4}\cos(\supz{\phi}/\eps)$
 	\writefig	4.0	2.2	$\xic$
 	\writefig	2.9	2.8	$-\xic$
	\writefig	7.6	4.6	b
 	\writefig	11.0	4.5	$T\sub1(\x)$
	\writefig	13.5	2.2	$\x$
	\writefig	10.9	3.8	$C\eps^{1/4}\cos(\phi^o/\eps)$
	\writefig	7.75	1.2	$-C\eps^{1/4}\cos(\phi^o/\eps)$
 	\writefig	11.05	2.8	$\xic$
 	\writefig	9.95	2.2	$-\xic$
 }
 \caption[]
 {Schematic shape of the function $T\sub1(\x)$ of equation \eqref{rp11} (a)
 for $\cos(\supz{\phi}/\eps) > 0$ and (b) for $\cos(\supz{\phi}/\eps) < 0$.
 In the first case, there are two symmetric stable fixed points. In the
 second case, there is a stable orbit of period 2.}
\label{fig_rp4}
\end{figure}

The part of motion between $\htau$ and $\ctau$ is essentially nonlinear.
Near the origin, we may use invariant manifolds as in Section \ref{sec_Lz}
to transform \eqref{rp5} into
\begin{equation}
\label{rp8}
\begin{split}
\eps\dot{\x} &= \brak{a^o_+(\tau,\eps) + \beta_+(\x,\y,\tau,\eps)}\x \\
\eps\dot{\y} &= \brak{a^o_-(\tau,\eps) + \beta_-(\x,\y,\tau,\eps)}\y, 
\end{split}
\end{equation}
where $a^o_{\pm}(\tau,\eps)=a^o_{\pm}(\tau)+\Order{\eps}$ and $\beta_{\pm}$
are of order $\abs{\x}+\abs{\y}$. Starting at $\htau$ with a small initial
condition $(\x_0>0,\y_0)$, the second equation in \eqref{rp8} shows that
$\y$ becomes exponentially small. The first one is used to prove that $\x$
reaches a distance $d$ from the origin ($d$ not too large) at a time
$\bar{\tau}(\x_0)+\Order{\eps}$, where 
\begin{equation}
\label{rp9}
\alpha^o(\bar{\tau}(\x_0),\htau) = -\eps\ln(\x_0/d), 
\end{equation}
provided $\x_0>\xic=\e^{-\alpha^o(\tsp,\htau)/\eps}$ (for smaller $\x_0$,
the orbit does not reach $Q_+$ before the time $\tsp$). For
$\tau>\bar{\tau}(\x_0)$, the trajectory is attracted by $Q_+$, around which
we carry out a similar analysis than around the origin, with the result
\begin{equation}
\label{rp10}
\begin{split}
\x(\ctau) &= C\eps^{1/4} +
\e^{\fix{\alpha}/\eps}\cos\bigpar{\tfrac{\fix{\phi}}{\eps}} \\
\y(\ctau) &= 
\e^{(\fix{\alpha}-\fix{\delta})/\eps}
\sin\bigpar{\tfrac{\fix{\phi}}{\eps}+\fix{\th}}, 
\end{split}
\end{equation}
where $\fix{\alpha} = \fix{\alpha}(\ctau,\bar{\tau}(\x_0)) +
\Order{\sqrt{\eps}}$ and  $\fix{\phi} = \fix{\phi}(\tsp,\bar{\tau}(\x_0)) +
\Order{\sqrt{\eps}}$, while $C$, $\fix{\delta}$ and $\fix{\th}$ are constant
at lowest order in $\eps$. The position at $\ctau$ thus depends essentially
on $\x_0$ through the delayed bifurcation time $\bar{\tau}(\x_0)$, in such a
way that \eqref{rp10} is the parametric equation of a squeezed spiral (which
is essentially the image of the unstable manifold of the origin under the
flow from $\htau$ to $\ctau$). 

\begin{figure}
 \centerline{
 \psfig{figure=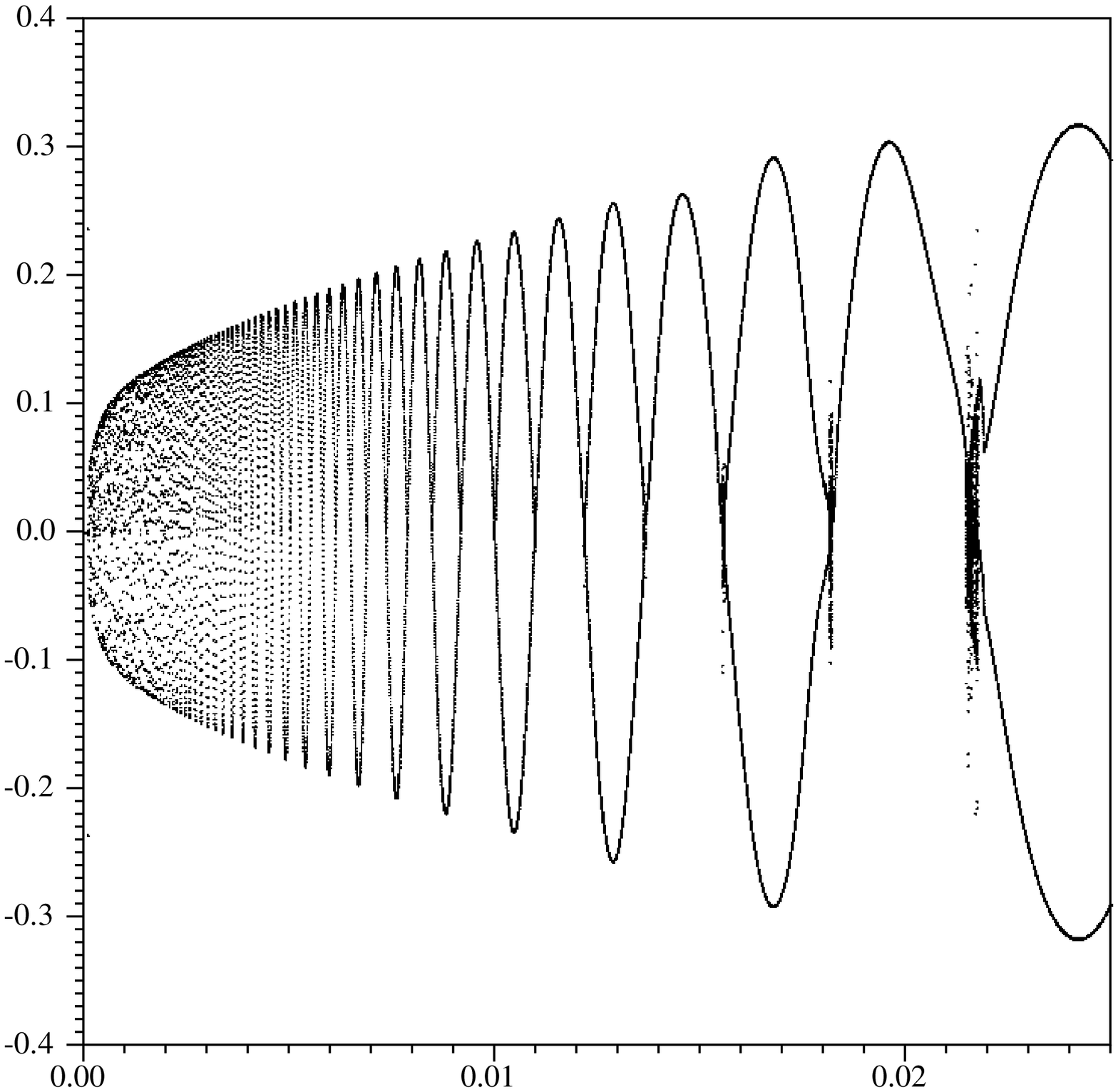,height=70mm,clip=t}
 \psfig{figure=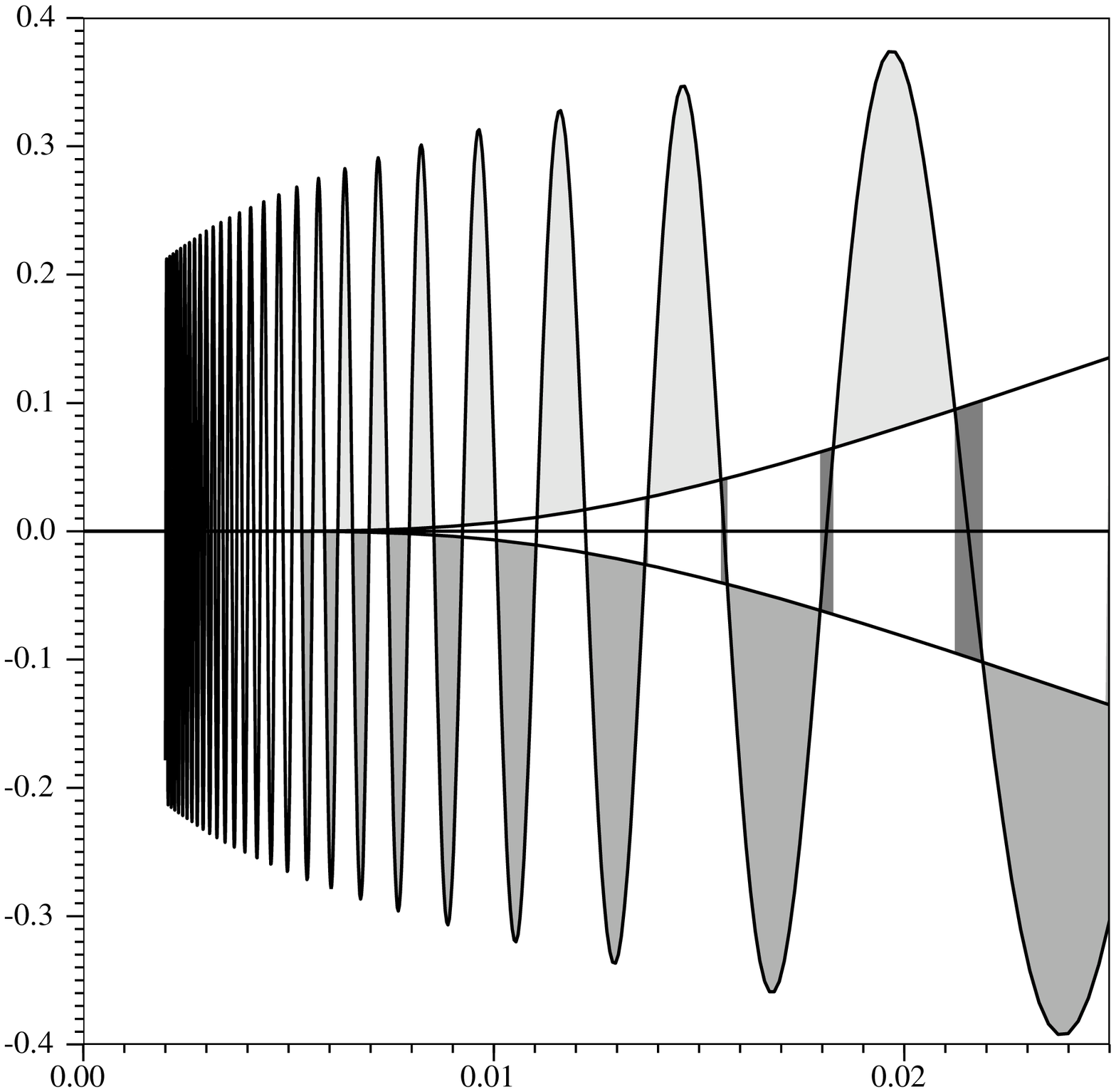,height=70mm,clip=t}}
 \figtext{
 	\writefig	0.6	7.0	a
 	\writefig	6.8	0.4	$\eps$
 	\writefig	8.4	7.0	b
 	\writefig	14.6	0.4	$\eps$
 }
 \caption[]
 {(a) Numerically computed bifurcation diagrams of the Poincar\'e map. For
 each value of $\eps$, we have plotted the asymptotic value of
 $q(\htau+n)$, for {\em one} initial condition. On the domain $0 < \eps <
 0.025$, the diagram clearly shows the alternates of regions with a
 period--1 and a period--2 cycle, separated by small chaotic zones. (b)
 Plots of the functions $\eps^{1/4}\cos(\phi^o/\eps)$ and
 $\pm\e^{-\mu/\eps}$. Light gray zones are those where the theory predicts
 existence of a period--1 cycle, medium gray zones those with a period--2
 cycle. Dark gray zone are those where chaotic hysteresis is possible, and,
 indeed, observed. We point out that in figure (b), the dynamic phase
 $\phi^o(0)$ has been computed analytically. Only the next-to-leading-order
 correction to $\phi^o(\eps)$ (which results in a phase shift) has been
 chosen in order to fit the numerical results.}
\label{fig_rp5}
\end{figure}

Combining this result with \eqref{rp7}, we finally obtain a Poincar\'e map
of the form
\begin{equation}
\label{rp11}
\begin{split}
\x_1 = T_1(\x_0;\y_0,\eps) = &
\cos\bigpar{\tfrac{\phi^o}{\eps}} \Bigbrak{C\eps^{1/4} +
\e^{\fix{\alpha}/\eps} \cos\bigpar{\tfrac{\fix{\phi}}{\eps}}} \\
&+ \e^{(\fix{\alpha}-\delta)/\eps} 
\sin\bigpar{\tfrac{\phi^o}{\eps} + \th^o} 
\sin\bigpar{\tfrac{\fix{\phi}}{\eps} + \fix{\th}},\\
\y_1 = T_2(\x_0;\y_0,\eps) = & \Order{\e^{-\delta^o_2}/\eps},
\end{split}
\end{equation}
where $\x_j = \x(\htau+j)$ measures the distance to the stable manifold of
the origin (it is close to $q$), and $\y_j = \y(\htau+j)$ measures the
distance to the unstable manifold.  This expression is valid for $\x_0>0$,
but is easily extended to negative $\x_0$, since the Poincar\'e map is odd.
The dynamics is thus essentially determined by the 1D map $\x_0\mapsto
T_1(\x_0;0,\eps)$, which is oscillating around $\pm
C\eps^{1/4}\cos(\tfrac{\phi^o}{\eps})$ (\figref{fig_rp4}). One easily shows
the existence of a positive constant $\mu$ such that, if
$\eps^{1/4}\cos(\phi^o/\eps)>\e^{-\mu/\eps}$, this map admits stable fixed
points at $\pm\fix{\x}\sord\pm\eps^{1/4}\cos(\phi^o/\eps)$, which
correspond to cycles of period 1. When
$\eps^{1/4}\cos(\phi^o/\eps)<-\e^{-\mu/\eps}$, there is an orbit of period
2, for which the pendulum alternatively visits the left and right
equilibrium (\figref{fig_rp4}). These properties can be shown to hold for
the 2D map, which is confirmed by numerical simulations
(\figref{fig_rp5}). 

\begin{figure}
 \centerline{
 \psfig{figure=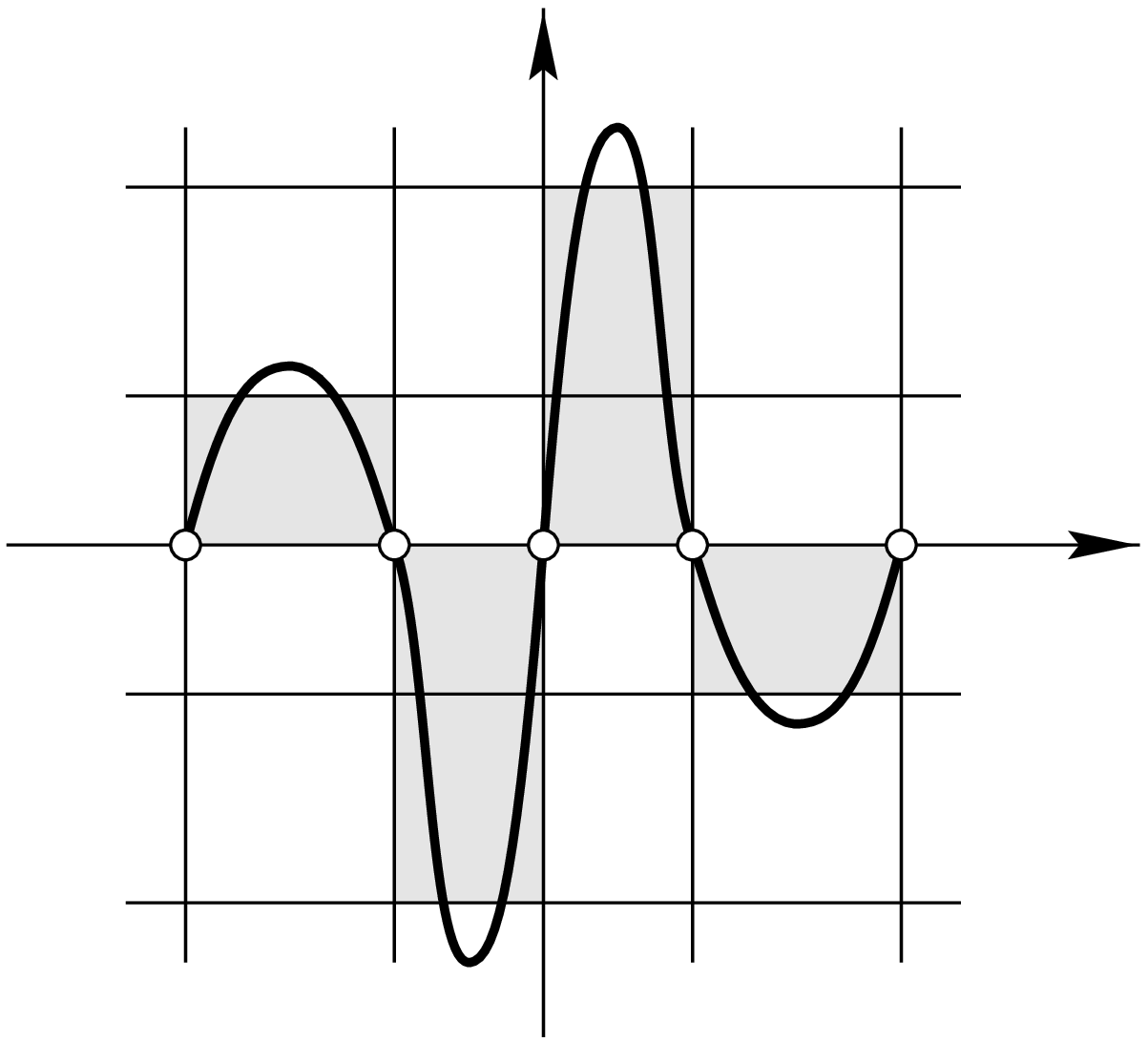,height=45mm,clip=t}
 \hspace{15mm}
 \psfig{figure=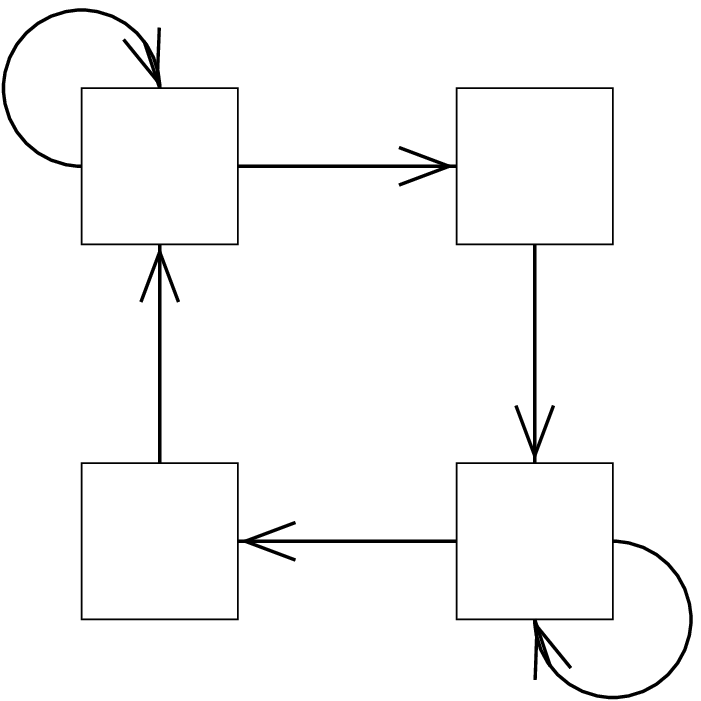,height=42mm,clip=t}}
 \vspace{-1mm}
 \figtext{
 	\writefig	1.5	4.5	a
 	\writefig	4.5	4.6	$T_1(\x)$
 	\writefig	2.15	2.25	$-\x_2$
 	\writefig	3.05	2.25	$-\x_1$
  	\writefig	5.05	2.75	$\x_1$
 	\writefig	5.95	2.75	$\x_2$
 	\writefig	6.6	2.2	$\x$
 	\writefig	2.9	0.4	$I_{-2}$
 	\writefig	3.7	0.4	$I_{-1}$
  	\writefig	4.5	0.4	$I_1$
 	\writefig	5.3	0.4	$I_2$
	\writefig	8.8	4.5	b
 	\writefig	9.8	1.7	$I_{-2}$
 	\writefig	11.25	1.7	$I_{-1}$
  	\writefig	9.9	3.12	$I_1$
 	\writefig	11.35	3.12	$I_2$
  }
 \caption[]
 {The interval map (a) admits the Markov subgraph (b), which allows for
 periodic orbits of all periods, except possibly 3. }
\label{fig_rp6}
\end{figure}

Chaotic motion is possible in the intermediate regions, where
$\abs{\eps^{1/4}\cos(\phi^o/\eps)}<\e^{-\mu/\eps}$. For the 2D map, it is
difficult to prove existence of such a motion, but one can do more for the
simplified 1D map, using symbolic dynamics. In fact, when
$\cos(\phi^o/\eps)=0$ and under certain conditions on $\Omega(\tau)$, one
finds that $T_1(\x)$ behaves as in \figref{fig_rp6}a: it vanishes at two
points $\x_1$ and $\x_2$, and, being odd, also at $-\x_1$ and $-\x_2$. These
points define four intervals $I_{-2}$, $I_{-1}$, $I_1$ and $I_2$, and the
maximum of $T_1$ on $I_1$ is larger than $\x_2$, while its minimum on $I_2$
is smaller than $-\x_1$. The \defwd{Markov graph} of $T_1$ is defined as the
graph with sites $I_j$, admitting an oriented edge $I_j\to I_k$ whenever
$T_1(I_j)\supset I_k$. It is known \cite{MS} that for every path in the
Markov graph, there exists an orbit visiting the corresponding sequence of
intervals. In particular, using Sarkovskii's theorem \cite{MS}, it is
possible to prove the existence of periodic orbits of every period except
possibly 3. 

The mathematical pendulum analysed here has been realized experimentally,
and all the phenomena predicted by the equations have been observed. They
depend, in fact, only on a few qualitative features of the system. The
origin should be a focus for some values of the parameter, in order to
allow the orbits to wind around it. It should be hyperbolic for other
values of the parameter, for which two new stable equilibria should exist.
Chaotic motion requires, in addition, these asymmetric equilibria to be
sometimes focuses, in order to create the oscillations in the Poincar\'e
map. Under these conditions, it should be possible to observe chaotic
hysteresis for other nonlinear oscillators (see Section
\ref{ssec_cruise}). 
 

\section{Examples of eigenvalue crossings}
\label{sec_ec}

We mentioned in Section \ref{sec_Lz} that adiabatic linear systems of the
form $\eps\dot{x} = A(\tau)x$ could be diagonalized (and thus solved) by the
change of variables $x=S(\tau;\eps)y$, where $S$ is a matrix satisfying the
equation
\begin{equation}
\label{ec1}
\eps\dot{S} = AS - SD,
\end{equation}
and $D$ is a suitable diagonal matrix. This procedure is only useful,
however, if we manage to control the transformation matrix $S$, which should
be bounded (\eg close to the matrix $S_0(\tau)$ which diagonalizes $A$
statically). Such a control turns out to be possible at least in two cases:
when the eigenvalues of $A(\tau)$ have different real parts, or, (in a more
restricted sense) when they  have the same real part but different imaginary
parts.\footnote{More generally, the system can be bloc--diagonalized when
the eigenvalues can be split into two groups with non--crossing real parts.}

This leaves open the question of the effect of different types of eigenvalue
crossings. The most generic case, when $A$ is not diagonalizable at the
crossing time, has been mentioned in Section \ref{sec_rp}. It can be studied
using the properties of Airy functions (see also \cite{Wasow}). 
In the present section, we illustrate the effect of two other types of
crossing. The first one occurs when $A(\tau)$ is symmetric, and can thus be
diagonalized even when it has identical eigenvalues. The second one arises
when the eigenvalues' real parts cross, but their imaginary parts are
different. We call this situation \defwd{eigenvalue cruising}; it is closely
related to properties of dynamic Hopf bifurcations discussed in
\cite{Ne1,Ne2}.

\subsection{Symmetric crossing}

Let us consider the overdamped motion of a particle in the 2D potential
\begin{equation}
\label{ecs1}
\Phi(x,\eps t) = -\frac12 \pscal{x}{A(\eps t)x} + \frac14 \norm{x}^4, 
\end{equation}
where $A$ is a symmetric matrix. The equation of motion can be written 
\begin{equation}
\label{ecs2}
\eps\dot{x} = A(\tau)x - \norm{x}^2 x.
\end{equation}
We assume that the matrix $A(\tau)$ is given by
\begin{equation}
\label{ecs3}
A(\tau) = a(\tau) 
\begin{pmatrix}
\cos 2\th(\tau) & \phantom{-}\sin 2\th(\tau) \\
\sin 2\th(\tau) & - \cos 2\th(\tau)
\end{pmatrix},
\end{equation}
so that it admits eigenvalues $\pm a(\tau)$ and eigenvectors $v_1 =
(\cos\th,\sin\th)$ and $v_2 = (-\sin\th,\cos\th)$. Thus, the potential
$\Phi$ has minima at $\pm\sqrt{a}\,v_1$ if $a$ is positive, and at
$\pm\sqrt{-a}\,v_2$ if $a$ is negative. To diagonalize the linearized
equation $\eps\dot{x} = A(\tau)x$, we may try to solve equation \eqref{ec1}
with matrices $S$ and $D$ of the form
\begin{equation}
\label{ecs4}
S = 
\begin{pmatrix}
\cos\th_1(\tau) & -\sin\th_2(\tau) \\
\sin\th_1(\tau) & \phantom{-}\cos\th_2(\tau)
\end{pmatrix}, \qquad 
D = 
\begin{pmatrix}
d_1(\tau) & 0 \\ 0 & d_2(\tau)
\end{pmatrix}.
\end{equation}
Substitution of this Ansatz in \eqref{ec1} yields the relations 
\begin{align}
\label{ecs5a}
\eps\dot{\th}_1 &= -a(\tau) \sin 2(\th_1-\th(\tau)), & 
d_1(\tau) &= a(\tau)\cos 2(\th_1-\th(\tau)), \\
\label{ecs5b}
\eps\dot{\th}_2 &= a(\tau) \sin 2(\th_2-\th(\tau)), & 
d_2(\tau) &= -a(\tau)\cos 2(\th_2-\th(\tau)).
\end{align}
If $a(\tau)$ does not vanish (\ie when there is no eigenvalue crossing),
these equations admit equilibrium branches at $\th_1 = \th_2 = \th(\tau)$,
of opposite stability. By the results of Section \ref{sec_1D}, we know
that they admit particular adiabatic solutions $\th_1(\tau) = \th(\tau) +
\Order{\eps}$ and $\th_2(\tau) = \th(\tau) + \Order{\eps}$. The evolution
operator of the linearized system can thus be written
\begin{equation}
\label{ecs6}
\begin{split}
U(\tau,\tau_0) &= S(\tau) 
\begin{pmatrix}
\e^{\delta_1(\tau,\tau_0)/\eps} & 0 \\ 
0 & \e^{\delta_2(\tau,\tau_0)/\eps}
\end{pmatrix}
S(\tau_0)^{-1}, \\
\delta_{1,2}(\tau,\tau_0) &= \int_{\tau_0}^{\tau} d_{1,2}(s) \dx s 
= \pm \int_{\tau_0}^{\tau} a(s) \dx s + \Order{\eps^2}. 
\end{split}
\end{equation}
The columns of $S(\tau)$ can be considered as \defwd{dynamic eigenvectors}
which are close to the static eigenvectors $v_{1,2}$. They define invariant
subspaces (depending on $\tau$), in which the motion is expanding, resp.\
contracting.

When $a(\tau)$ is allowed to vanish, new phenomena occur because the
equations \eqref{ecs5a}, \eqref{ecs5b} undergo bifurcation. It is
instructive to consider the case $a(\tau)=-\cos\tau$, for three different
functions $\th(\tau):$ (1) $\th(\tau)=0$, (2) $\th(\tau)=-\cos\tau$ and (3)
$\th(\tau)=\tau$. 

\begin{figure}
 \centerline{\psfig{figure=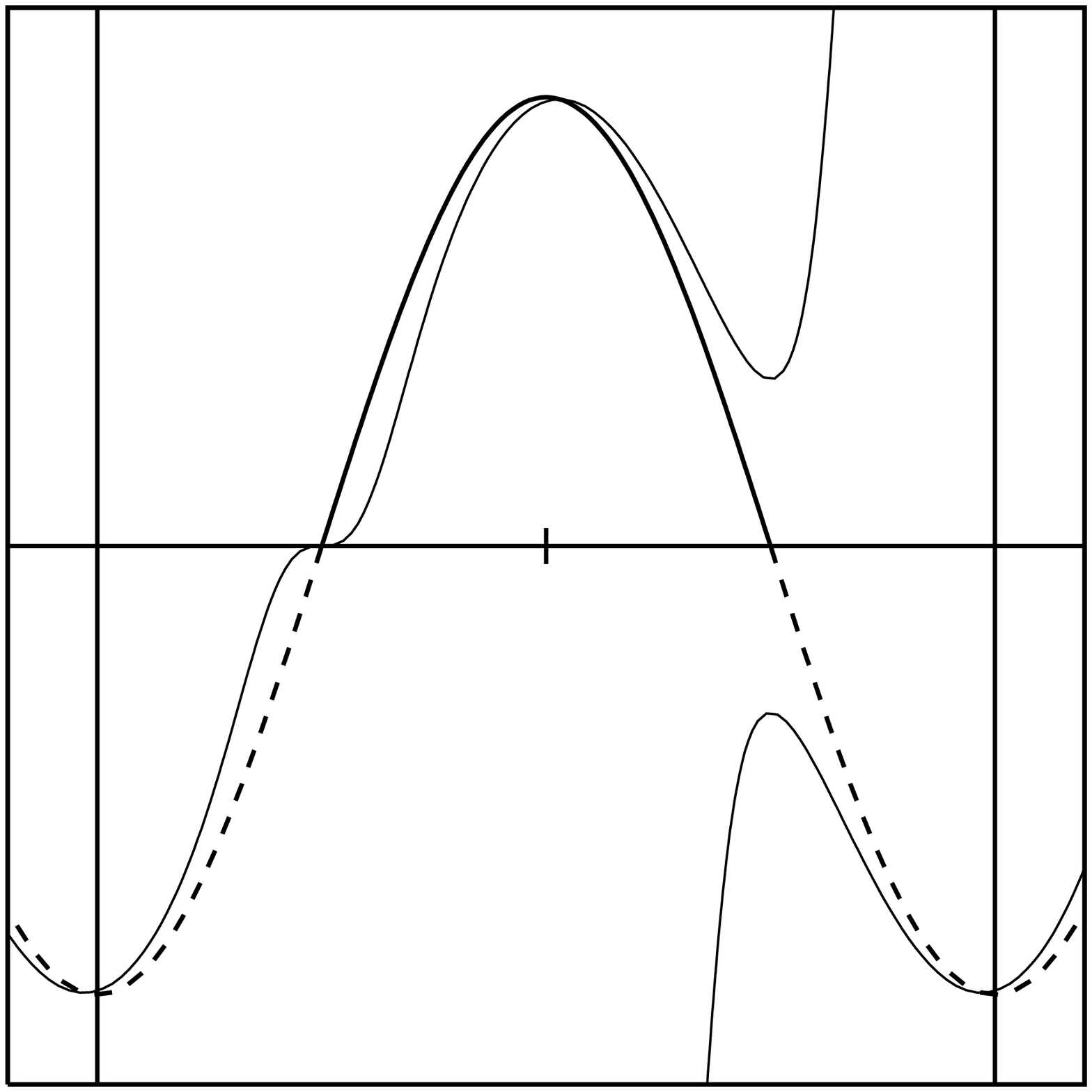,height=50mm,clip=t}
 \hspace{18mm}
 \psfig{figure=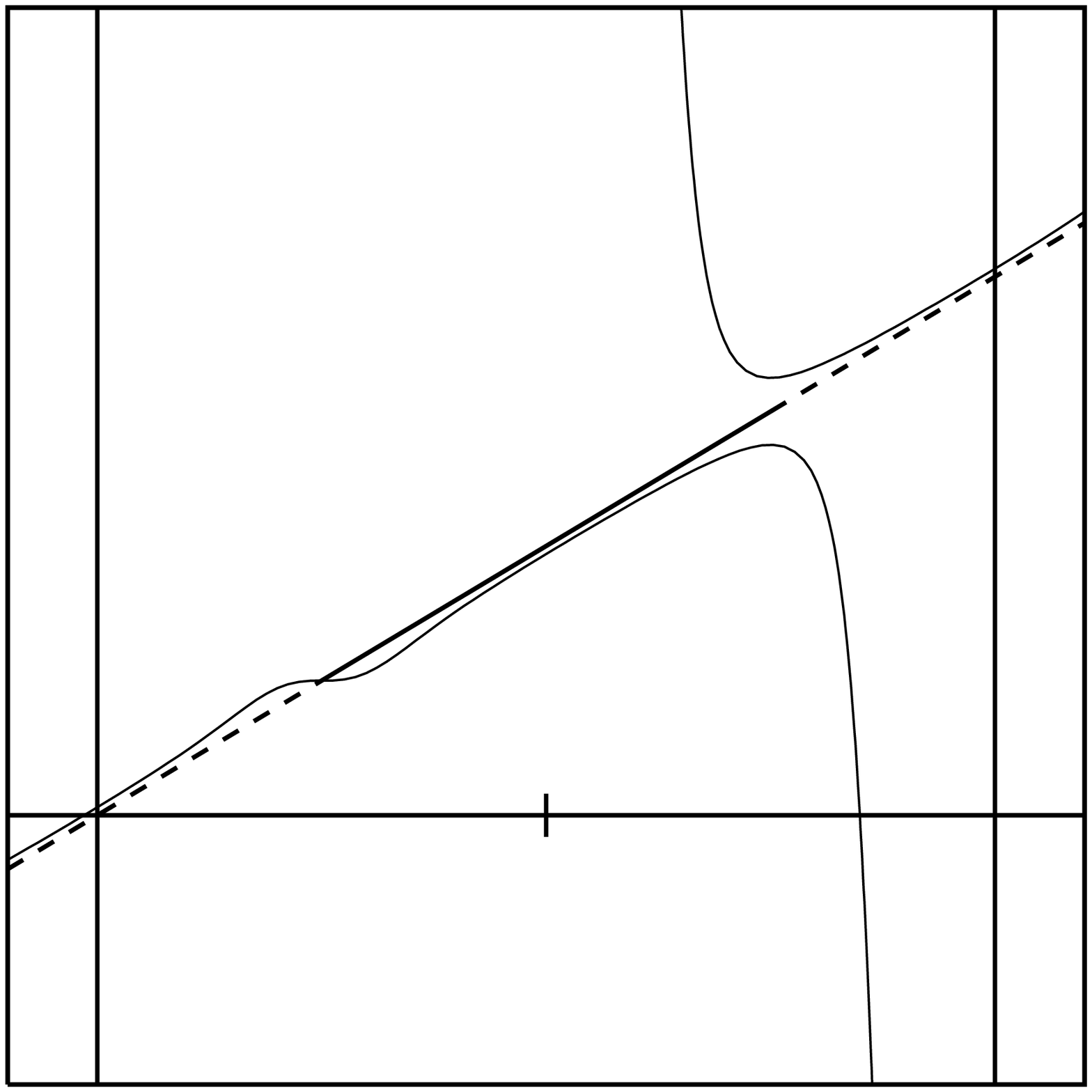,height=50mm,clip=t}}
 \figtext{
 	\writefig	0.9	5.2	a
	\writefig	7.9	5.2	b
 	\writefig	1.95	5.2	$x$
	\writefig	9.0	5.2	$x$
 	\writefig	1.9	2.65	$0$
 	\writefig	3.75	2.65	$\pi$
 	\writefig	6.0	2.75	$\tau$
	\writefig	8.95	1.4	$0$
	\writefig	10.8	1.4	$\pi$
	\writefig	13.0	1.5	$\tau$
  }
 \caption[]
 {Solutions (thin lines) of \eqref{ecs5a} when $a(\tau)=-\cos\tau$: (a) in
 the case $\th(\tau)=-\cos\tau$, (b) in the case $\th(\tau)=\tau$. In both
 cases, one can construct a particular solution remaining close to the
 static equilibrium $\th(\tau)$ (thick lines, where the solid lines indicate
 stable branches and the dashed lines unstable ones), admitting a
 discontinuity at $\tau=\frac{3\pi}{2}$. Solutions of \eqref{ecs5b} behave
 in a similar way, with a discontinuity at $\tau=\frac{\pi}{2}$.}
\label{fig_ecs1}
\end{figure}

If $\th(\tau)=0$, \eqref{ecs5a} admits the solution $\th_1=\th_2\equiv 0$,
and $d_{1,2} = \pm a(\tau)$. The evolution operator can thus be written 
\begin{equation}
\label{ecs7}
U(\tau,0) = 
\begin{pmatrix}
\e^{-\sin(\tau)/\eps} & 0 \\ 0 & \e^{\sin(\tau)/\eps}
\end{pmatrix}.
\end{equation}
The subspaces $x_1=0$ and $x_2=0$ are invariant. If $x_2(0)=0$, $x_1(\tau)$
remains exponentially small until $\tau=\pi$, which is the standard
bifurcation delay.

If $\th(\tau)=-\cos\tau$, it is not possible to construct solutions of
\eqref{ecs5a} remaining indefinitely close to $\th(\tau)$. The best one can
do is to construct periodic solutions $\th_1(\tau)$ and $\th_2(\tau)$
admitting a discontinuity of order $\sqrt{\eps}$, respectively at times
$\frac{3\pi}{2}$ and $\frac{\pi}{2}$ (\figref{fig_ecs1}a). As a result, for
$\frac{\pi}{2}<\tau<\frac{3\pi}{2}$ we have 
\begin{equation}
\label{ecs8}
U(\tau,0) = S(\tau) 
\begin{pmatrix}
\e^{\delta_1(\tau,\pi/2)/\eps} & 0 \\ 
0 & \e^{\delta_2(\tau,\pi/2)/\eps} 
\end{pmatrix}
T 
\begin{pmatrix}
\e^{\delta_1(\pi/2,0)/\eps} & 0 \\ 
0 & \e^{\delta_2(\pi/2,0)/\eps} 
\end{pmatrix}
S(0)^{-1},
\end{equation}
where
\begin{equation}
\label{ecs9}
T = S(\tfrac{\pi}{2}+)^{-1} S(\tfrac{\pi}{2}-) = 
\begin{pmatrix}
1 + \Order{\sqrt{\eps}} & \sin(\th_2^+-\th_2^-) + \Order{\eps} \\ 0 & 1 
\end{pmatrix},
\end{equation}
with $\th_2^{\pm} = \th_2(\frac{\pi}{2} \pm)$. The off-diagonal term of
this matrix induces a transition between the directions which were
invariant before $\tau=\frac{\pi}{2}$. In particular, when $\tau=\pi$, we
have to leading order in $\eps$
\begin{equation}
\label{ecs10}
U(\pi,0) \sord 
\begin{pmatrix}
1 & \sin(\th_2^+-\th_2^-)\e^{2/\eps} \\ 0 & 1 
\end{pmatrix}. 
\end{equation}
This transformation rotates the vertical axis by almost $\pi/2$. The matrix
$U(2\pi,\pi)$ is found to rotate the horizontal axis by almost $-\pi/2$. This
means that there exists no invariant subspace in which the particle performs
an independent motion. The eigenvalue crossing thus results in an
interaction between both modes, with the particle always following the most
unstable direction. The sign of the discontinuities of $\th_{1,2}$ is
important: in this case it induces a back--and--forth oscillation of the
particle between two wells (\figref{fig_ecs2}a). 

\begin{figure}
 \centerline{\psfig{figure=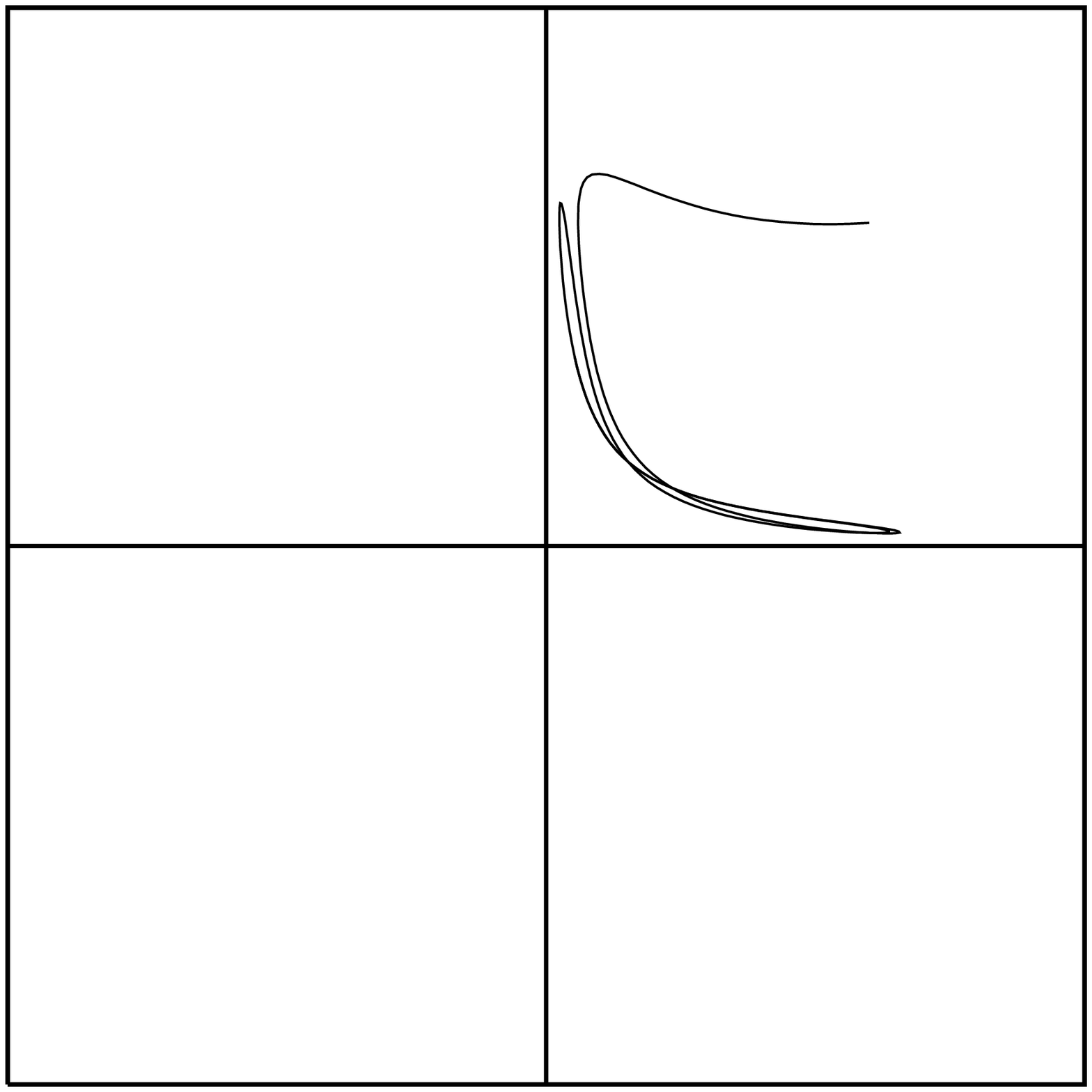,height=50mm,clip=t}
 \hspace{18mm}
 \psfig{figure=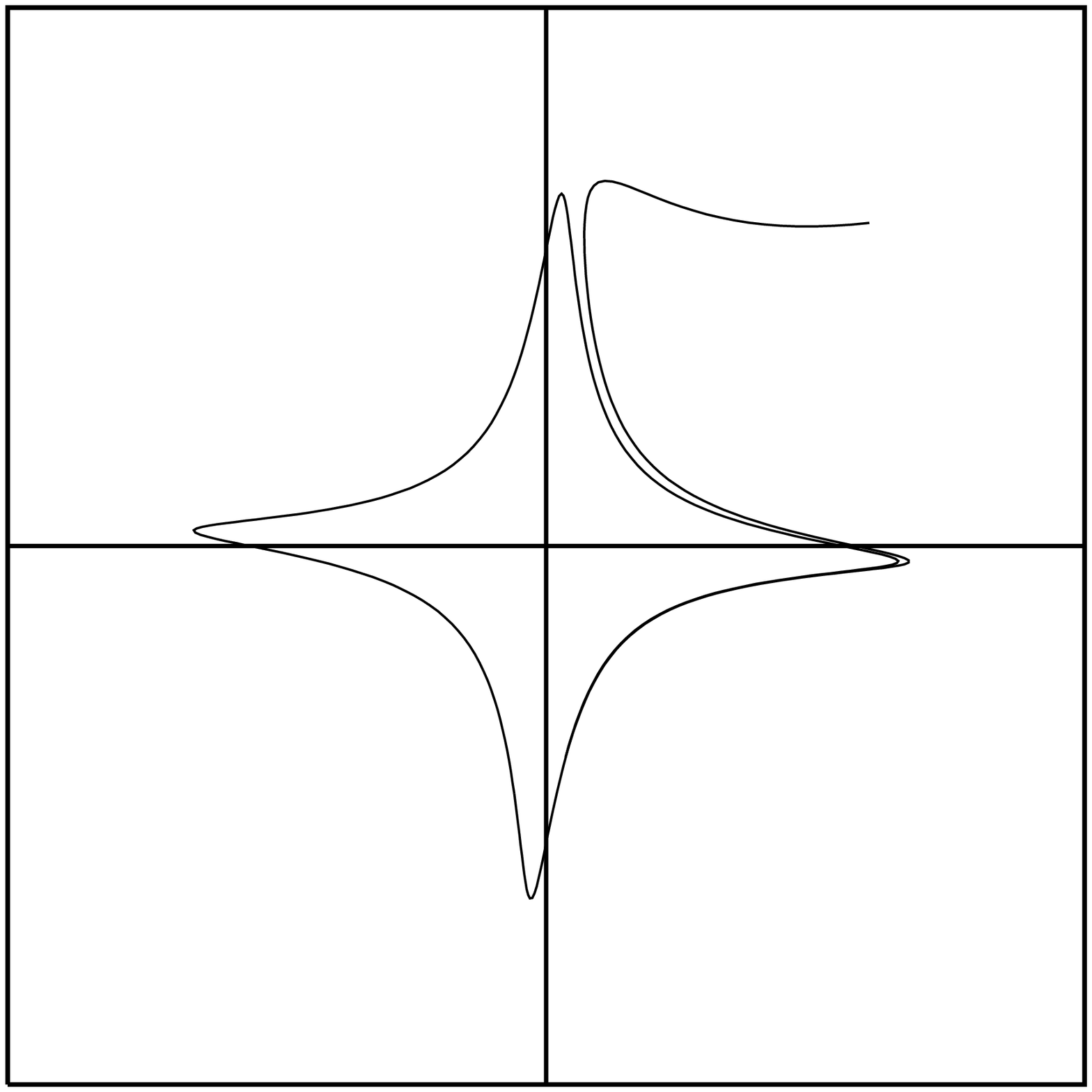,height=50mm,clip=t}}
 \figtext{
 	\writefig	0.9	5.2	a
	\writefig	7.9	5.2	b
 	\writefig	4.0	5.2	$v_2$
	\writefig	11.0	5.2	$v_2$
 	\writefig	5.8	2.7	$v_1$
	\writefig	12.8	2.7	$v_1$
  }
 \caption[]
 {Solutions of \eqref{ecs1} when $a(\tau)=-\cos\tau$, plotted with respect
 to the rotating reference frame $(v_1,v_2)$: (a) in the case
 $\th(\tau)=-\cos\tau$, the particle oscillates back and forth between two
 potential wells, while (b) in the case $\th(\tau)=\tau$ it visits all four
 wells in a row.}
\label{fig_ecs2}
\end{figure}

In the case $\th(\tau) = \tau$, the situation is similar, but with a
discontinuity of $\th_1$ of opposite sign (\figref{fig_ecs1}b). As a
result, the  coordinate axes are always rotated in the same direction, and
the particle visits all four wells in a row (\figref{fig_ecs2}b). 

\subsection{Coupled oscillators and eigenvalue cruising}
\label{ssec_cruise}

We call \defwd{eigenvalue cruising} the situation arising when some
eigenvalues of the matrix $A$ drift past one another at some imaginary
distance. This cruising also leads to an interaction between the modes;
however, unlike in the case of diagonal crossing, this interaction is
\defwd{delayed}. 

Eigenvalue cruisings appear in particular in coupled oscillators. Consider
for instance the system
\begin{equation}
\label{ecc1}
\begin{split}
\ddot{q}_1 + 2\gamma_1\dot{q}_1 + (1-\lambda+q_1^2+4q_2^2)q_1 
- \mu q_2 &= 0, \\ 
\ddot{q}_2 + 2\gamma_2\dot{q}_2 + 4(4-\lambda+q_1^2+4q_2^2)q_2 
+ \mu q_1 &= 0,
\end{split}
\end{equation}
which was introduced by Kobayashi \cite{Ko} to describe the vibrations of a
buckled plate with supersonic flow on one side of the plate. The variables
$q_1$ and $q_2$ are amplitudes of the two dominant Fourier modes of the
deflection, $\lambda$ is the in--plane compressive stress, $\mu$ the
dynamic fluid pressure of the supersonic flow, and $\gamma_{1,2}$ are
friction coefficients (which were taken equal in \cite{Ko}). 

Introducing $p_1 = \dot{q}_1$ and $p_2 = \dot{q}_2$, \eqref{ecc1} can be
written as a 4D first order system for the variables $(q_1,p_1,q_2,p_2)$,
which admits the origin as an equilibrium. The linearization around the
origin is a $4\times 4$ matrix with eigenvalues
\begin{equation}
\label{ecc2}
\begin{split}
a_{1,\pm} &= -\gamma_1 \pm \sqrt{\gamma_1^2 + \lambda - 1} 
+ \Order{\mu^2},\\
a_{2,\pm} &= -\gamma_2 \pm \sqrt{\gamma_2^2 + 4(\lambda - 4)} 
+ \Order{\mu^2}. 
\end{split}
\end{equation}
An eigenvalue cruising arises for instance in the following situation:
assume $\gamma_1=2$, $\gamma_2=1$ and $\mu=0$, so that $a_{1,\pm} = -2
\pm\sqrt{\lambda+3}$ and $a_{2,\pm} = -1 \pm\icx\sqrt{15-4\lambda}$. As
$\lambda$ increases from $-3$ to $\frac{15}{4}$, the complex eigenvalues
$a_{2,\pm}$ correspond to oscillations, while the real eigenvalues
$a_{1,\pm}$ describe an overdamped motion. There is a cruising at
$\lambda=-2$ and the origin becomes unstable at $\lambda=1$
(\figref{fig_ecc1}). The same qualitative features hold for small positive
coupling $\mu$. 

\begin{figure}
 \centerline{\psfig{figure=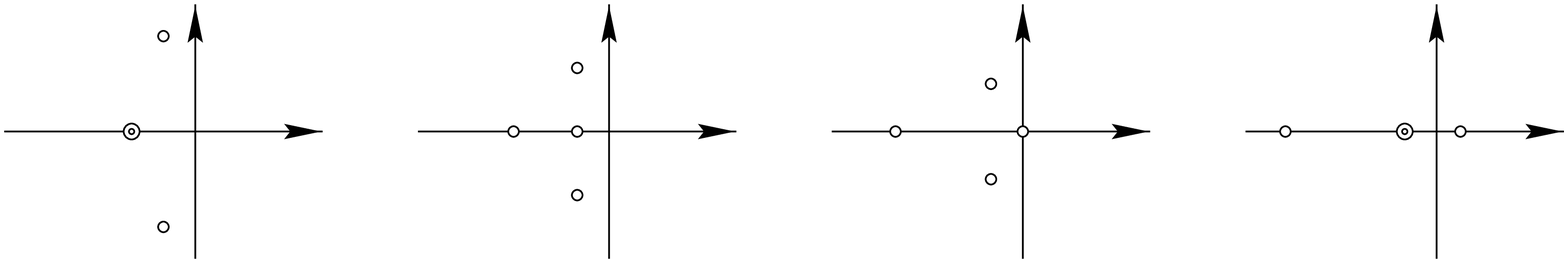,width=150mm,clip=t}}
 \figtext{
 	\writefig	-0.1	0.5	$\lambda=-3$
	\writefig	3.9	0.5	$\lambda=-2$
 	\writefig	7.9	0.5	$\lambda=1$
	\writefig	11.9	0.5	$\lambda=\frac{15}{4}$
  }
 \caption[]
 {Eigenvalues of the linearized Kobayashi equations for $\gamma_1=2$,
 $\gamma_2=1$ and $\mu=0$. There is an eigenvalue cruising for
 $\lambda=-2$.}
\label{fig_ecc1}
\end{figure}

We are interested in the following question. Assume that $\lambda$ is
increased monotonically and adiabatically, starting with a generic initial
condition at a time $\tau_0$ where $\lambda$ is smaller than 1. For what
value of $\lambda$ does the trajectory depart from the origin? The answer
turns out to be related in a rather subtle way to bifurcation delay and
eigenvalue cruising. It is easier to explain this phenomenon on the simple
model equation
\begin{equation}
\label{ecc3}
\eps\dot{x} = A(\tau)x, \qquad 
A(\tau) = 
\begin{pmatrix}
a_1(\tau) & \mu \\ -\mu & a_2(\tau)
\end{pmatrix}, \qquad
\begin{array}{l}
a_1(\tau) = -1+\tau, \\
a_2(\tau) = -1+\icx.
\end{array}
\end{equation}
As in Kobayashi's equations, $a_1$ represents the overdamped mode, and
$a_2$ the oscillating one (in complex notation). The cruising occurs at
$\tau=0$. To diagonalize this equation, we try to solve the equation
$\eps\dot{S}=AS-SD$ with the Ansatz
\begin{equation}
\label{ecc4}
S(\tau) = 
\begin{pmatrix}
1 & s_2(\tau) \\ s_1(\tau) & 1
\end{pmatrix}, 
\qquad
D(\tau) = 
\begin{pmatrix}
d_1(\tau) & 0 \\ 0 & d_2(\tau)
\end{pmatrix}.
\end{equation}
Substitution in the equation for $S$ yields the relations
\begin{align}
\label{ecc5a}
\eps\dot{s}_1 &= -\mu - (\tau-\icx)s_1 - \mu s_1^2, & 
d_1 &= a_1 + \mu s_1, \\
\label{ecc5b}
\eps\dot{s}_2 &=  \mu + (\tau-\icx)s_2 + \mu s_2^2, & 
d_2 &= a_2 - \mu s_2.
\end{align}
The first equation has a static equilibrium at $\fix{s}_1(\tau) =
-\mu/(\tau-i) + \Order{\mu^2}$, which is unstable for $\tau<0$ and stable
for $\tau>0$. One can show that the solution of \eqref{ecc5a} with initial
condition $s_1(0) = \fix{s}_1(0)$ tracks the branch $\fix{s}_1(\tau)$ at a
distance at most $\Order{\eps}$, for both negative and positive
times. 

\begin{figure}
 \centerline{\psfig{figure=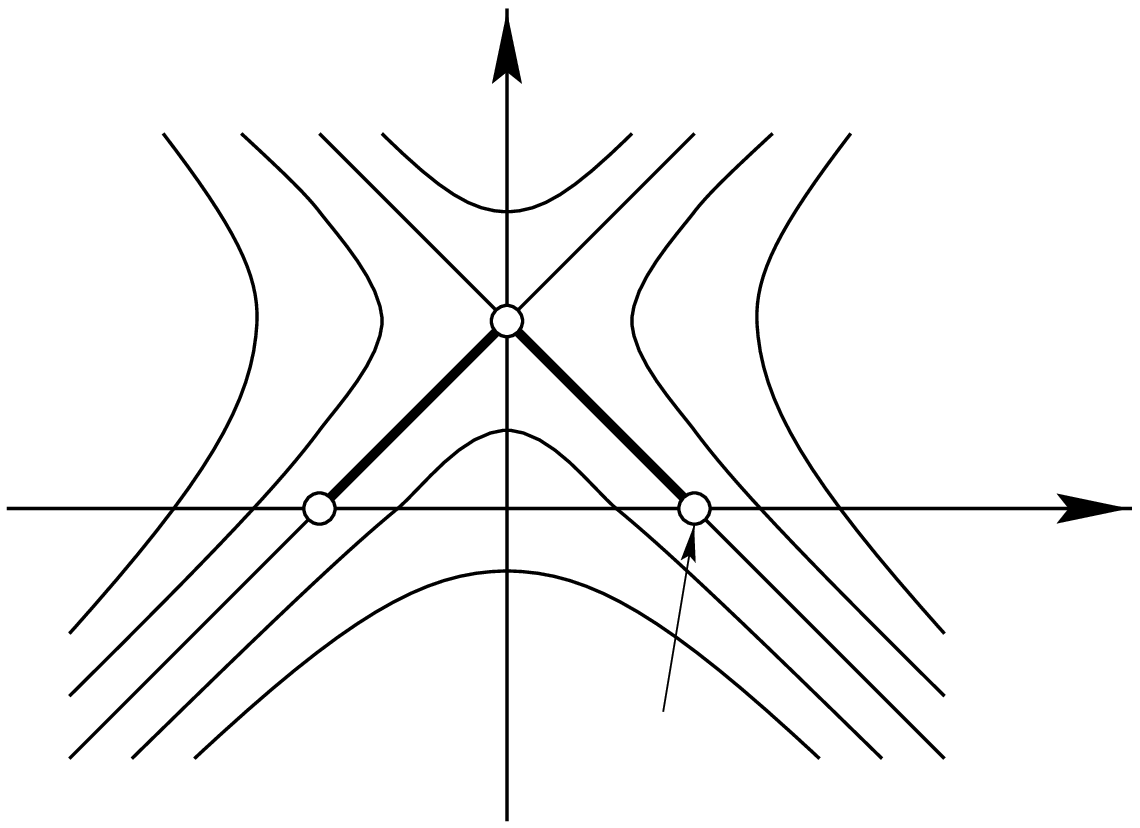,height=45mm,clip=t}
 \hspace{18mm}
 \psfig{figure=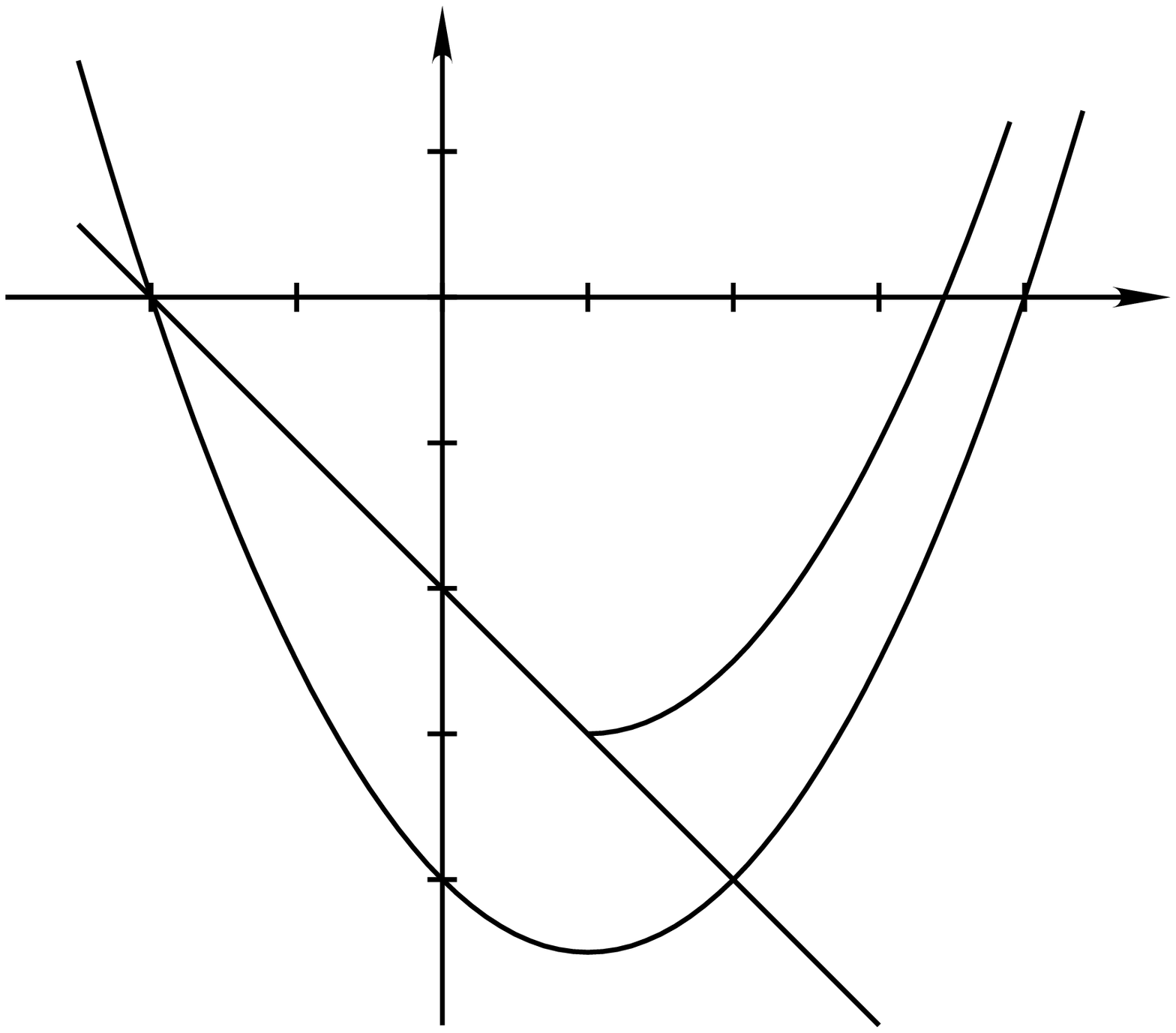,height=50mm,clip=t}}
 \figtext{
 	\writefig	0.5	5.2	a
 	\writefig	3.5	4.7	$\im\tau$
 	\writefig	5.5	2.5	$\re\tau$
 	\writefig	3.4	3.2	$\icx$
 	\writefig	3.7	0.9	$\taub$
	\writefig	7.9	5.2	b
	\writefig	9.3	4.25	$-2$
	\writefig	11.4	3.65	$1$
	\writefig	13.8	3.8	$\tau$
	\writefig	13.0	2.2	$\delta_1(\tau,-2)$
	\writefig	13.0	0.8	$\re\delta_2(\tau,-2)$
	\writefig	11.8	5.0	$\delta_1(\tau,1)+$
	\writefig	11.3	4.6	$\re\delta_2(1,-2)$
  }
 \caption[]
 {(a) The level lines of the function $\re\alpha(\tau)$ of equation
 \eqref{ecc6} are hyperbolas centered at $\tau=\icx$. The largest positive
 time which can be connected to the negative real axis by such a line is the
 buffer time $\taub=1$. (b) The origin becomes unstable when the largest
 exponent of a matrix element of the evolution operator \eqref{ecc8} becomes
 positive. This may happen earlier when $\mu\neq0$, because the oscillators
 interact at the buffer time.}
\label{fig_ecc2}
\end{figure}

The second equation has a more subtle behaviour. It admits an equilibrium
branch at $\fix{s}_2(\tau) = \mu/(\tau-\icx) +  \Order{\mu^2}$, which is
stable for $\tau<0$ and unstable for $\tau>0$; in fact, it undergoes Hopf
bifurcation. Such bifurcations have been studied by Neishtadt
\cite{Ne1,Ne2,Ne3}. The interesting fact is that there exist solutions
tracking the equilibrium branch beyond the bifurcation point, but only
until a time called \defwd{maximal delay} or \defwd{buffer point}. This
point is obtained in the following way: let $a(\tau) = \tau - \icx +
\Order{\mu^2}$ be the linearization of \eqref{ecc5b} around the equilibrium
$\fix{s}_2(\tau)$. Define the function
\begin{equation}
\label{ecc6}
\re\alpha(\tau) = \re\int_0^{\tau} a(s)\dx s = 
\tfrac12\bigbrak{(\re\tau)^2 - (\im\tau-1)^2 + 1} + \Order{\mu^2}.
\end{equation} 
The buffer time $\taub$ is the largest real time which can be connected to
the negative real axis by a path of constant $\re\alpha$ (and with some
additional properties given in \cite{Ne3}). In the present situation,
$\taub=1$ (\figref{fig_ecc2}a). As a consequence, for $\tau\leqs 1$, we can
construct a solution of \eqref{ecc5b} which is close to $\fix{s}_2(\tau)$,
and the evolution operator of \eqref{ecc3} is given by
\begin{equation}
\label{ecc7}
\begin{split}
U(\tau,\tau_0) &= S(\tau)
\begin{pmatrix}
\e^{\delta_1(\tau,\tau_0)/\eps} & 0 \\
0 & \e^{\delta_2(\tau,\tau_0)/\eps} 
\end{pmatrix}
S(\tau_0)^{-1},\\
\delta_1(\tau,\tau_0) &= \int_{\tau_0}^{\tau} d_1(s)\dx s = 
\tfrac12 (\tau^2-\tau_0^2) - (\tau-\tau_0) + \Order{\mu^2}, \\
\delta_2(\tau,\tau_0) &= \int_{\tau_0}^{\tau} d_2(s)\dx s = 
(\icx-1)(\tau-\tau_0) + \Order{\mu^2}. \\
\end{split}
\end{equation}
For $\tau>1>\tau_0$, however, the solution $s_2(\tau)$ necessarily admits a
discontinuity of order $\eps\mu$ at the buffer time. A similar calculation
as in the previous subsection yields the evolution operator
\begin{equation}
\label{ecc8}
U(\tau,\tau_0) = S(\tau)
\begin{pmatrix}
\e^{\delta_1(\tau,\tau_0)/\eps} & 
\Order{\eps\mu}\e^{\brak{\delta_1(\tau,1)+\delta_2(1,\tau_0)}/\eps} \\
0 & \e^{\delta_2(\tau,\tau_0)/\eps} 
\end{pmatrix}
S(\tau_0)^{-1}
\end{equation}
If the system starts away from the origin at $\tau_0<0$, it will follow the
origin exponentially closely until a delay time $\htau$, which is the first
time at which one of the matrix elements becomes of order $1$ again. If
$\mu=0$, we simply have $\htau=2-\tau_0$. When $\mu\neq 0$, however, we
cannot overlook the interaction between the overdamped and the oscillating
mode, which takes effectively place at the buffer time $\taub=1$, and may
cause the system to become unstable at an earlier time. For instance, when
$\tau_0=-2$, the usual delay time for $\mu=0$ would be $\htau=4$, while the
effective delay time for $\mu>0$ is $1+\sqrt{6}+\Order{\mu^2}$
(\figref{fig_ecc2}b). 

A similar phenomenon is observed for the Kobayashi equations \eqref{ecc1},
only with different values of the cruising, buffer and delay times. The
effective value of the bifurcation delay can be shown to depend at leading
order in $\eps$ only on the linearization around the origin. It is
important, together with nonlinear terms, for the global structure of
motion, since it influences the choice of the asymmetric equilibrium the
system follows after leaving the origin. In fact, for large amplitude
oscillations of the form $\lambda(\tau)=8\sin(\tau)$, we observed
numerically that the Kobayashi equations display chaotic hysteresis just as
the rotating pendulum in section \ref{sec_rp}. This is not really
surprising, since even when $\mu=0$, each oscillator is similar to the
rotating pendulum of the amplitude of $\lambda$ is large enough. A positive
$\mu$, however, will modify the bifurcation delay and the dynamic phases
and amplitudes which determine the structure of the Poincar\'e map. 


\section*{Acknowledgments}

This work is supported by the Fonds National Suisse de la Recherche
Scientifique. 

\begin{thebibliography}{MMMM}
\bibitem[{\bf Ben}]{Benoit}	
			\bibbook{E.\ Beno\^{\i}t (Ed.)}
			{Dynamic Bifurcations, Proceeding, Luminy 1990}
			{Sprin\-ger--Verlag, Lecture Notes in Mathematics
			1493}
			{Berlin, 1991}
			
\bibitem[Ber]{B}	\bibtitle{N.\ Berglund}
			{Adiabatic Dynamical Systems and Hysteresis}, 
			Thesis EPFL no 1800. 
			Available at \\
			{\tt
			http://dpwww.epfl.ch/instituts/ipt/berglund/these.html}

\bibitem[BK]{BK}	\bibarticle{N.\ Berglund, H.\ Kunz}
			{Chaotic Hysteresis in an Adiabatically Oscillating
			Double Well}
			{\PRL}{78}{1692}{1694}{1997}
			
\bibitem[GBS]{GBS}	\bibarticle{G.H.\ Goldsztein, F.\ Broner, S.H.\
			Strogatz}
			{Dynamical Hysteresis without Static Hysteresis:
			Scaling Laws and Asymptotic Expansions}
			{\SIAM}{57}{1163}{1187}{1997}
			
\bibitem[{\bf HK}]{HK}		
			\bibbook{J.\ Hale, H.\ Ko\c cak}
			{Dynamics and Bifurcations}
			{Springer--Verlag}		
			{New York, 1991}
			
\bibitem[HL\&]{Hohl}	\bibarticle{A.\ Hohl, H.J.C.\ van der Linden, R.\
			Roy,  G.\ Goldsztein, F.\ Broner, S.H.\ Strogatz}
			{Scaling Laws for Dynamical Hysteresis in a 
			Multidimensional Laser System}
			{\PRL}{74}{2220}{2223}{1995}			

\bibitem[JGRM]{Jung}	\bibarticle{P.\ Jung, G.\ Gray, R.\ Roy, P.\ Mandel}
			{Scaling Law for Dynamical Hysteresis}
			{\PRL}{65}{1873}{1876}{1990}
			
\bibitem[Ka]{Ka}	\bibtitle{K.\ Kawasaki}
			{Kinetics of Ising Models}, in
			\bibbook{C.\ Domb, M.S.\ Green}
			{Phase Transitions and Critical Phenomena, Vol.2}
			{Academic Press}{London, 1972}
			
\bibitem[Ko]{Ko}	\bibarticle{S.\ Kobayashi}
			{Two--dimensional Panel Flutter 1. Simply Supported
			Panel}
			{Trans. Jpn. Soc. Aeronaut. Space Sci.}
			{5}{90}{102}{1962}
			
\bibitem[{\bf Ma1}]{Mar1}	
			\bibbook{P.A.\ Martin}
			{Mod\`eles en M\'ecanique Statistique des Processus
			Irr\'e\-ver\-sibles}
			{Springer--Verlag}{Berlin, 1979}
			
\bibitem[Ma2]{Mar2}	\bibarticle{Ph.A.\ Martin}
			{On the Stochastic Dynamics of Ising Models}
			{\JSP}{16}{149}{168}{1977}
			
\bibitem[{\bf MS}]{MS}	
			\bibbook{W.\ de Melo, S.\ van Strien}
			{One--Dimensional Dynamics}
			{Springer}{Berlin, 1993}
			
\bibitem[{\bf MK\&}]{MKKR}	
			\bibbook{E.F.\ Mishchenko, Yu.S.\ Kolesov, 
			A.Yu.\ Kolesov, N.Kh.\ Rozov}
			{Asymptotic Methods in Singularly Perturbed Systems}
			{Consultants Bureau}{New York, 1994}
			
\bibitem[Ne1]{Ne1}	\bibarticle{A.I.\ Neishtadt}
			{Persistence of stability loss for dynamical 
			bifurcations I}
			{\DE}{23}{1385}{1391}{1987}
			Transl. from \bibref{\DU}{23}{2060}{2067}{1987}.
			
\bibitem[Ne2]{Ne2}	\bibarticle{A.I.\ Neishtadt}
			{Persistence of stability loss for dynamical 
			bifurcations II}
			{\DE}{24}{171}{176}{1988}
			Transl. from \bibref{\DU}{24}{226}{233}{1988}.
			
\bibitem[Ne3]{Ne3}	\bibtitle{A.I.\ Neishtadt}
			{On Calculation of Stability Loss Delay Time for
			Dynamical Bifurcations}
			in \bibbook{D.\ Jacobnitzer Ed.}{\nth{X\!I}
			International Congress of Mathematical
			Physics}{International Press, Boston}{1995}
			
\bibitem[RKP]{RKP}	\bibarticle{M.\ Rao, H.K.\ Krishnamurthy, R.\
			Pandit}
			{Magnetic hysteresis in two model spin systems}
			{\PRB}{42}{856}{884}{1990}

\bibitem[SE]{SE}	\bibarticle{J.-S. Suen, J.L. Erskine}
			{Magnetic Hysteresis Dynamics: Thin $p(1\times 1)$
			Fe Films on Flat and Stepped W(110)}
			{\PRL}{78}{3567}{3570}{1997}
			
\bibitem[TO]{TO}	\bibarticle{T.\ Tom\'e, M.J.\ de Oliveira}
			{Dynamic phase transition in the kinetic Ising model
			under a time--dependent oscillating field}
			{\PRA}{41}{4251}{4254}{1990}
			
\bibitem[{\bf Wa}]{Wasow}		
			\bibbook{W.\ Wasow}
			{Asymptotic expansions for ordinary differential 
			equations}
			{Krieger}		
			{New York, 1965, 1976}
					
\end{thebibliography}

\begin{thebibliography}{MMMM}

\bibitem[BER]{BER}	\bibarticle{S.M.\ Baer, T.\ Erneux, J.\ Rinzel}
			{The slow passage through a Hopf bifurcation:
			delay, memory effects, and resonance}
			{\SIAM}{49}{55}{71}{1989}
			
\bibitem[DT]{DT}	\bibarticle{D.\ Dhar, P.B.\ Thomas}
			{Hysteresis and self--organized criticality in the 
			$O(N)$ model in the limit $N\to\infty$}
			{\JPA}{25}{4967}{4984}{1992}
			
\bibitem[EM]{EM}	\bibarticle{T.\ Erneux, P.\ Mandel}
			{Imperfect bifurcation with a slowly--varying 
			control parameter}
			{\SIAM}{46}{1}{15}{1986}

\bibitem[{\bf GH}]{GH}		
			\bibbook{J.\ Guckenheimer, P.\ Holmes}
			{Nonlinear Oscillations, Dynamical Systems, 
			and Bifurcations of Vector Fields}
			{Springer--Verlag}{New York, 1983}
			
\bibitem[HW]{HeWa}	\bibarticle{Y.-L.\ He, G.-C.\ Wang}
			{Observation of Dynamic Scaling of Magnetic
			Hysteresis in Ultrathin Ferromagnetic Fe/Au(001)
			Films}
			{\PRL}{70}{2236}{2239}{1993}
			
\bibitem[JYW]{JYW}	\bibarticle{Q.\ Jiang, H.-N. Yang, G.-C. Wang}
			{Scaling and dynamics of low--frequency hysteresis
			loops in ultrathin Co films on a Cu(001) surface}
			{\PRB}{52}{14911}{14916}{1995}
			
\bibitem[LP]{LP}	\bibarticle{W.S.\ Lo, R.A.\ Pelcovits}
			{Ising model in a time--dependent magnetic field}
			{\PRA}{42}{7471}{7474}{1990}
			
\bibitem[LZ]{LZ}	\bibarticle{C.N.\ Luse, A.\ Zangwill}
			{Discontinuous scaling of hysteresis losses}
			{\PRE}{50}{224}{226}{1994}
			
\bibitem[ME1]{ME1}	\bibarticle{P.\ Mandel, T.\ Erneux}
			{Laser Lorenz Equations with a Time--Dependent 
			Parameter}
			{\PRL}{53}{1818}{1820}{1984}
			
\bibitem[ME2]{ME2}	\bibarticle{P.\ Mandel, T.\ Erneux}
			{The Slow Passage through a Steady Bifurcation:
			Delay and Memory Effects}
			{\JSP}{48}{1059}{1070}{1987}
			
\bibitem[{\bf MR}]{MR}	
			\bibbook{E.F.\ Mishchenko, N.Kh.\ Rozov}
			{Differential Equations with Small Parameters 
			and Relaxations Oscillations}
			{Plenum}{New York, 1980}
			
\bibitem[NST]{NST}	\bibtitle{A.I.\ Neishtadt, C.\ Sim\`o, D.V.\
			Treschev}
			{On stability loss delay for a periodic trajectory}
			in \bibbook{H.W.\ Broer {\it et al} Eds.}{Nonlinear
			Dynamical Systems and Chaos}{Birkh\"auser,
			Basel}{1996}
			
\bibitem[Ra]{Ra}	\bibarticle{M.\ Rao}
			{Comment on ``Scaling Law for Dynamical
			Hysteresis''}
			{\PRL}{68}{1436}{1437}{1992}
			
\bibitem[RTMS]{RTMS}	\bibarticle{P.A.\ Rikvold, H.\ Tomita, 
			S.\ Miyashita, S.W.\ Sides}
			{Metastable lifetimes in a kinetic Ising model: 
			Dependence on field and system size}
			{\PRE}{49}{5080}{5090}{1994}
			
\bibitem[SRN]{SRN}	\bibpreprint{S.W.\ Sides, P.A.\ Rikvold, M.A.\
			Novotny}
			{Stochastic Hysteresis and Resonance in a Kinetic
			Ising System}
			{cond-mat/9712021}{1997}
			
	
\end{thebibliography}
\end{document}